\documentclass[11pt]{article}
\usepackage{amsfonts}
\usepackage{amssymb}
\usepackage{amsbsy}

\input epsf.sty

\newcommand{\BEQ}{\begin{equation}}     
\newcommand{\BEA}{\begin{eqnarray}}
\newcommand{\BD}{\begin{displaymath}}
\newcommand{\EEQ}{\end{equation}}       
\newcommand{\EEA}{\end{eqnarray}}
\newcommand{\ED}{\end{displaymath}}
\newcommand{\al}{\alpha}                
\newcommand{\cech}{\check}

\newcommand{\del}{\delta}
\newcommand{\Del}{\Delta}
\newcommand{\Id}{{\mathrm{Id}}}
\newcommand{\eps}{\varepsilon}          

\newcommand{\g}{{\mathfrak{g}}}
\newcommand{\h}{{\mathfrak{h}}}

\newcommand{\sch}{{\mathfrak{sch}}}
\newcommand{\sv}{{\mathfrak{sv}}}
\newcommand{\wsv}{{\widetilde{\mathfrak{sv}}}}
\newcommand{\wfsv}{{\widetilde{\mathfrak{fsv}}}}
\newcommand{\age}{{\mathfrak{age}}}

\newcommand{\slin}{{\mathfrak{sl}}}

\newcommand{\so}{{\mathfrak{so}}}

\newcommand{\gal}{{\mathfrak{gal}}}

\newcommand{\conf}{{\mathfrak{conf}}}

\newcommand{\vir}{{\mathfrak{vir}}}
\newcommand{\Phia}{\,{}_{\alpha}\Phi}
\newcommand{\phia}{\,{}_{\alpha}\phi}
\newcommand{\psia}{\,{}_{\alpha}\psi}
\newcommand{\Phima}{\,{}_{-\alpha}\Phi}
\newcommand{\phima}{\,{}_{-\alpha}\phi}
\newcommand{\psima}{\,{}_{-\alpha}\psi}

\newcommand{\R}{\mathbb{R}}
\newcommand{\C}{\mathbb{C}}
\newcommand{\Z}{\mathbb{Z}}
\newcommand{\N}{\mathbb{N}}

\newcommand{\eop}{\hfill $\Box$}        

\newcommand{\II}{{\rm i}}               
\renewcommand{\Re}{{\rm Re\ }}          
\newcommand{\half}{{1\over 2}}          
\newcommand{\ad}{{\mathrm{ad\,}}}       
\newcommand{\wit}[1]{\widetilde{#1}}    
\renewcommand{\vec}[1]{\boldsymbol{#1}} 
\newcommand{\zeile}[1]{\vskip #1 \baselineskip} 

\newcommand{\appsection}[2]{\setcounter{equation}{0} \section*{Appendix #1. #2}
\renewcommand{\theequation}{#1\arabic{equation}}
              \renewcommand{\thesection}{#1} }

\catcode`\@=11
\def\numberbysection{\@addtoreset{equation}{section}
        \def\theequation{\thesection.\arabic{equation}}}
\numberbysection

\begin{document}

\begin{titlepage}

\vskip 1.5 cm
\begin{center}
{\Large \bf On vertex algebra representations of the  Schr\"odinger-Virasoro Lie algebra}
\end{center}

\vskip 2.0 cm
   
\centerline{  {\bf J\'er\'emie Unterberger}$^a$}
\vskip 0.5 cm

\centerline {$^a$Institut Elie Cartan,\footnote{Laboratoire 
associ\'e au CNRS UMR 7502} Universit\'e Henri Poincar\'e Nancy I,} 
\centerline{ B.P. 239, 
F -- 54506 Vand{\oe}uvre-l\`es-Nancy Cedex, France}

\begin{abstract}
The Schr\"odinger-Virasoro Lie algebra $\sv$ is an  extension
of the Virasoro Lie algebra by a nilpotent Lie algebra formed with a bosonic
current of weight $\frac{3}{2}$ and a bosonic current of weight $1$. It is also a natural
infinite-dimensional extension of the Schr\"odinger Lie algebra, which -- leaving aside the invariance under
time-translation --  has been proved to be
a symmetry algebra for many  statistical physics models undergoing a dynamics with dynamical exponent $z=2$ ; it should consequently play a role akin to that of
the Virasoro Lie algebra in two-dimensional equilibrium statistical physics.

We define in this article general Schr\"odinger-Virasoro primary fields by analogy
with conformal field theory, characterized by a 'spin' index and a (non-relativistic) mass, 
and construct vertex algebra representations of $\sv$ out of a charged symplectic  boson  and
a free boson and its associated vertex operators. We also compute two- and
three-point functions of  still conjectural massive fields that are defined
by an analytic continuation with respect to a formal parameter.  
\end{abstract}

\zeile{2} \noindent 
\underline{PACS:} 02.20Tw, 05.70Fh, 11.25Hf, 11.30.Pb\\
\underline{Keywords:} conformal field-theory, correlation functions, 
algebraic structure of integrable models, \\
 Schr\"odinger-invariance, 
supersymmetry, non-equilibrium statistical physics, infinite-dimensional Lie algebras  
\end{titlepage}


\section{Introduction}

The Schr\"odinger-Virasoro algebra $\sv$  is defined in \cite{Hen94,RogUnt06} as the infinite-dimensional Lie algebra generated by $L_n,Y_m,M_p$, $n,p\in\Z$, $m\in\half+\Z$, with
Lie brackets
$$
[L_n,L_p]=(n-p)L_{n+p},\ [L_n,Y_m]=({n\over 2}-m)Y_{n+m},\ [L_n,M_p]=-pM_{n+p}
$$
\BEQ
[Y_m,Y_{m'}]=(m-m')M_{m+m'},\ [Y_m,M_p]=0,\ [M_n,M_p]=0  \label{gl:sv}
\EEQ
where $n,p\in\Z,m,m'\in\half+\Z$. It is a semi-direct product of the non centrally extended
Virasoro
algebra  
 \BEQ \g=\vir_0:=\langle L_n\rangle_{n\in\Z} \EEQ
 by the
two-step nilpotent infinite dimensional Lie algebra 
\BEQ \h=\langle Y_m\rangle_{m\in\half+\Z}\oplus \langle M_p\rangle_{p\in\Z}.
\EEQ
 The $Y_m$ ($m\in\Z+\half$), resp. $M_p$
$(p\in\Z)$, may be seen as the components of $L$-conformal currents with
conformal weight $\frac 32$, resp. $1$. Note that the current $Y$ is {\it bosonic} although its weight is a half-integer. The supersymmetric partner $G$ of the Virasoro field appearing in the Neveu-Schwarz algebra (see
\cite{NS71} or \cite{Kac97}, \textsection 5.9) is also of weight $\frac{3}{2}$, but it is {\it odd}, which changes drastically the representation theory and the range of applications, the 'bosonicity' of $Y$ accounting for the appearance of a {\it space-dependence} which is absent from usual (super)conformal field theory. 

This infinite-dimensional Lie algebra was originally introduced in \cite{Hen94} by looking at the
invariance of the free Schr\"odinger equation in (1+1)-dimensions
\BEQ (2{\cal M}\partial_t-\partial_r^2)\psi=0. \EEQ

Its maximal subalgebra of Lie symmetries (acting projectively on the wave function $\psi$) is known
under the name of Schr\"odinger Lie algebra, $\sch_1$ (see \cite{Lie82,Nied72,Nied74}), and can be embedded into $\sv$ as
$$\sch_1=  \langle L_{-1},L_0,
L_1\rangle\ltimes\langle Y_{-\half},Y_{\half},M_0\rangle=\slin(2,\R)\ltimes\gal,$$
where $\gal$ - isomorphic to the three-dimensional nilpotent Heisenberg Lie algebra - contains the generators
of Galilei transformations;
the generators of $\sch_1$ act on $\psi$ as follows:
\BEQ
 L_{-1}=-\partial_t,\quad L_0=-t\partial_t-\half r\partial_r-\lambda,\quad
L_1=-t^2\partial_t-tr\partial_r-\frac{{\cal M}}{2} r^2-2\lambda t  \label{massM1}
\EEQ
(generators of time translation, scaling transformation - with {\it scaling exponent} $\lambda=\frac{1}{4}$ in
this case -  and 'special' transformation);
\BEQ
Y_{-\half}=-\partial_r,\quad Y_{\half}=-t\partial_r-{\cal M} r,\quad M_0=-{\cal M} \label{massM2}
\EEQ
(generators of space translation, special Galilei transformation and phase shift). All together,
these generate the following finite transformations \cite{HenUnt03}:
\BEQ \psi(t,r)\to \dot{\beta}(t')^{-\lambda} \exp\left(-\frac{{\cal M}}{4} \frac{\ddot{\beta}(t')}{\dot{\beta}(t')} r'^2\right) \psi'(t',r') \EEQ
where $t=\beta(t')=\frac{at'+b}{ct'+d}$ , $r=r'\sqrt{\dot{\beta}(t')}$
for the M\"obius transformations in $SL(2,\R)$;
\BEQ \psi(t,r)\to \exp\left( {\cal M}\left(\half \alpha(t')\dot{\alpha}(t')-r'\dot{\alpha}(t')\right)\right) \psi'(t',r') \EEQ
where 
\BEQ t=t',\quad  r=r'-\alpha(t')=r'-at'-b
\EEQ
for the Galilei transformations;
\BEQ \psi(t,r)\to \exp({\cal M}\gamma) \psi'(t,r)\EEQ
($\gamma$ constant) for the phase shifts.

By a straightforward extrapolation of these formulas to Lie generators of arbitrary integer of half-integer
indices, or -- in other words --  to arbitrary functions of time $\alpha(t),\beta(t),\gamma(t)$, one finds a realization of the Lie algebra $\sv$ or of the Schr\"odinger-Virasoro group (defined in \cite{RogUnt06}) which exponentiates $\sv$.

The original physical motivation for introducing these algebras is the following. 
In the statistical physics of many-body systems far from equilibrium, it is
well-established that a dynamical, time-dependent scale-invariance
frequently arises, even in cases where  the stationary state does {\em not} have 
a static, time-independent
scale invariance.  The scaling generator $L_0$ describes a dynamics with dynamical exponent $z=2$, characteristic of a diffusion-like evolution; a signature of this behaviour is the existence of scaling functions
${\cal G}_R$, ${\cal G}_C$ for the two-time response and correlation functions defined as
(see \cite{PicHen04})
\BEQ R(t,s):=\frac{\partial \langle \phi(t_2,r_2)\rangle}{\partial h(t_1,r_1)} \big|_{h=0}=
s^{-a-1} {\cal G}_R(\frac{t_2}{t_1}, \frac{(r_2-r_1)^2}{t_2-t_1}),\EEQ
\BEQ C(t,s):=\langle \phi(t_1,r_1)\phi(t_2,r_2)\rangle 
= s^{-b} {\cal G}_C(\frac{t_2}{t_1}, \frac{(r_2-r_1)^2}{t_2-t_1})\EEQ
for some scaling exponents $a,b$ (at least in the scaling limit $t_2\gg t_1$, $r_2-r_1\to \infty$), so that, loosely
speaking, the time coordinate scales as the square of the space coordinate(s). 
  For a simple illustration, consider the phase-ordering 
kinetics of a simple magnet (described in terms of an Ising model) with 
a completely random initial state, which at the initial time $t=0$ is
brought into contact with a thermal bath at a sufficiently low
temperature so that more than one stable thermodynamic state exists. 
Then indeed one observes a $z=2$ dynamical scaling, as reviewed in \cite{Bray94}. This is also the case for
many different models at criticality, described for instance by a stochastic Langevin equation or a master equation, for which an equilibrium state does not even exist, see \cite{Hen94,Hen02}.   
Actually, much more can be said: in all these models, there is evidence for the existence of a dynamical
invariance under the subalgebra $\mathfrak{age}_1=\langle L_0,L_1\rangle\ltimes\langle Y_{\pm\half}, M_0\rangle\subset\sch_1$ where the time-invariance generator has been omitted, allowing for an ageing behaviour.
Note for the sake of completeness that the interest has shifted very recently to the case $z\not=2$, which is the general law
 for systems
quenched exactly onto their critical temperature, or else for
equilibrium critical dynamics, and may also apply to  the physically completely
different situation of Lifschitz points in equilibrium spin systems with 
uniaxial competing interactions (for a recent review on the available
evidence for this, see \cite{Hen02,Hen02,Plei02,Dieh06}); however, the symmetry algebras seem to be much more complicated in this case, and they
are not directly related to the Schr\"odinger algebra.

  Coming back to algebra, let us rephrase the physical consequences of symmetry  in a
mathematical way. Let 
$$\rho: g\to \big(\Phi(t,r)\to \rho(g) (\Phi(t,r))=\Phi_g(g.(t,r))\big)$$
be any realization of the Schr\"odinger Lie group Sch$_1$ exponentiating $\sch_1$ as coordinate
transformations acting projectively on a wave-function $\Phi(t,r)$: the statistical field $\Phi(t,r)$
is called {\it quasi-primary} if its $n$-point functions or correlators $\langle \Phi(t_1,r_1)\ldots
\Phi(t_n,r_n)\rangle$ transform covariantly under $\rho$, namely:
\BEQ \langle \Phi_g(g.(t_1,r_1))\ldots \Phi_g(g.(t_n,r_n))\rangle=\langle \Phi(t_1,r_1)\ldots
\Phi(t_n,r_n)\rangle. \EEQ
The predictions of this invariance principle have been extensively developed for different types of realizations of $\sch_1$, including the {\it mass ${\cal M}$ realization} given by formulas
(\ref{massM1},\ref{massM2}) above which define {\it scalar massive fields}, and tested with success for relevant physical systems - see
for instance \cite{HenPicPleUnt03}, \cite{PicHen04} or \cite{Hen06} for a review. A prominent feature of this 
type of covariance is the Bargmann superselection rule with respect to the mass: $n$-point functions
of fields $\Phi_1,\ldots,\Phi_n$ with respective masses ${\cal M}_1,\ldots,{\cal M}_n$ cancel except
if ${\cal M}_1+\ldots+{\cal M}_n=0$.

 The reader {\em should be aware that  the mass}  plays here a very different role by comparison with
relativistic physics or with  critical phenomena at equilibrium: it is the  central charge of the Galilei
algebra, and massless fields have in general no physical interest. Also, it has absolutely nothing to do with
a parameter measuring the distance away from criticality (actually, some kinetic models  at criticality have been proved to exhibit
an  $\age$-invariance!).

The original project was to build the infinite-dimensional Lie algebra $\sv$ into the cornerstone of a 'Schr\"odinger-field theory' with applications to  $z=2$ dynamical scaling,
by analogy with the role played by the Virasoro algebra in the systematic study of two-dimensional statistical physics at equilibrium near the critical temperature. The 'coinduced' representations of $\sv$ introduced in \cite{RogUnt06} and extensively used here are undoubtedly the natural {\it Schr\"odinger-Virasoro primary} (classical) fields to look at, and extend the tensor-density modules ${\cal F}_{\lambda}$ or classical primary
fields (or weight currents) of $\vir$. However, something fails right from the start since no interesting (even linear!)
wave equation exhibiting this infinite-dimensional Lie algebra of symmetries has been found. It seems difficult or impossible to find such wave equations (at least scalar wave equations), see \cite{CherHen04}. There may be a way to escape this problem, see \cite{Unt06}, but it requires the use of a doubly-infinite Lie algebra of invariance (actually,
a 'double' extension of the pseudodifferential algebra on the line) of which $\sv$ appears to be a quotient. This complementary approach is currently under investigation.

The purpose of this paper is to construct explicit non-trivial vertex algebra representations of $\sv$.
We hope that this is only a first step towards a deeper understanding of Schr\"odinger-invariant fields, and that a connection with actual physical models can eventually be established. Indeed, these representations
open the road to an explicit computation of $n$-point functions from the knowledge of the symmetries. In particular, some  three-point functions (which are known to depend on an arbitrary scaling function for massive $\sch_1$-covariant fields) are
computed here for a conjectural $\sv$-covariant massive field which must still be spelled out completely.

The paper is organized as follows.

Section 1 is introductory on the Schr\"odinger-Virasoro Lie algebra and its representations. Most of the material contained here is adapted from \cite{RogUnt06}. However, after developing the theory a while, it appeared necessary to deal with an {\it extended Schr\"odinger-Virasoro Lie algebra} denoted by
$\tilde{\sv}$ that is defined here for the first time. The extension of the results of \cite{RogUnt06}
to $\tilde{\sv}$ is more or less straightforward. The Lie algebra $\tilde{\sv}$ appears to have {\it three}
independent central extensions (in other terms,  three central charges), whereas $\sv$ admits only one
central extension. The centrally extended Lie algebra
is denoted by $\tilde{\sv}_{c,\kappa,\alpha}$ (see Lemma 1.2).

Section 2 deals mostly with the definition of $\sv$- and $\tilde{\sv}$-primary fields, see Definition 2.1.1. They depend
on the choice of a 'spin representation' $\rho$ of $\sv_0\cong\langle L_0\rangle\ltimes\langle Y_{\half},M_1\rangle\subset\sv$ or $\tilde{\sv}_0\cong \langle N_0\rangle\ltimes \sv_0\subset\tilde{\sv}$ (see below for a definition of $N_0$). It appears from the examples that $\tilde{\sv}$-primary fields are also characterized by a matrix $\Omega$ acting on the representation space of $\rho$, 
which is unexpected from a mathematical point of view.

Section 3 is devoted to the construction of the {\it $a\bar{b}$-theory}. The name refers to the
 fact that the  $\tilde{\sv}$ fields (see Definition 3.1.3) are built out of two independent fields of conformal field theory, namely a free boson $a(z)$ and a charged symplectic boson $\bar{b}(z)=(\bar{b}^+(z),\bar{b}^-(z)).$ Note that the complex variable $z$ becomes the real time variable $t$ in this theory and the conjugate variable $\bar{z}$ apparently leaves the picture. The so-called {\it polynomial fields} $\Phi_{j,k}$ and {\it generalized polynomial fields}
$_{\alpha} \Phi_{j,k}$, $j,k\in\N, \alpha\in\R$ -- all of them $\tilde{\sv}$-primary fields -- are constructed
(see Theorems 3.2.4 and 3.2.5)
as polynomials in the fields $a$, $\bar{b}$, the $_{\alpha} \Phi_{j,k}$ involving furthermore the 
vertex   operator $V_{\alpha}$ built from $a$. The space-dependence of the fields appears from the repeated application of the
generator $Y_{-\half}$, interpreted as a space-translation.

In Section 4, we compute the two- and three-point functions of the polynomial and generalized polynomial
 fields introduced in Section 3.

Finally, Section 5 conjectures the existence of {\it massive} fields, see Theorem 5.1 and Theorem 5.2 for a definition,  whose two-point and (at least in
one case) three-point
functions are explicitly computed.

\section{On the extended Schr\"odinger Lie algebra $\tilde{\sv}$ and its coinduced representations}

Recall from the Introduction the realization of $\sch_1$ as Lie symmetries of the free Schr\"odinger equation
\BEQ (2{\cal M}\partial_t-\partial_r^2)\psi(t,r)=0\EEQ (see formulas (\ref{massM1}) and (\ref{massM2}) above). Suppose now that the wave-function $\psi=\psi_{\cal M}(t,r)$ is indexed by the mass parameter.
Then a 'trick' first used in \cite{HenUnt03} (see also \cite{HenUnt06} for an application to the
Dirac-L\'evy-Leblond equation and \cite{Sto05} for other invariant equations), with far-reaching consequences, is to consider (formally)
a Laplace transform of the Schr\"odinger equation with respect to the mass: the Laplace transformed field
\BEQ \tilde{\psi}(t,r,\zeta):=\int \psi_{\cal M}(t,r)e^{{\cal M}\zeta} \ d{\cal M} \EEQ
satisfies the field equation $\Del \tilde{\psi}(t,r,\zeta)=0$, where
\BEQ \Delta:=2\partial_t\partial_{\zeta}-\partial_r^2 \EEQ
is formally equivalent to a Laplacian in three dimensions. Transforming accordingly the Lie symmetry
generators in $\sch_1$ is equivalent to 'replacing' ${\cal M}$ by $\partial_{\zeta}$ in (\ref{massM1},
\ref{massM2}). 

The difference with the usual fixed mass setting is that the new wave equation has
more symmetries (as well-known, the Laplacian in three dimensions is $\conf_3$-invariant, where $\conf_3\cong \so(4,1)$ is the Lie algebra of infinitesimal conformal transformations), including in particular
$N_0=-r\partial_r-2\zeta\partial_{\zeta}.$ This new generator of $\conf_3$ acts as a {\it derivation} on
$\sch_1$ in the above realization, namely
\BEQ [N_0,L_{0,\pm 1}]=0,\quad [N_0,Y_{\pm\half}]=Y_{\pm\half},\quad [N_0,M_0]=2M_0.\EEQ  One obtains
thus a 7-dimensional maximal parabolic Lie subalgebra of $\conf_3$ (see \cite{HenUnt03}), $\tilde{\sch}_1=\langle N_0\rangle\ltimes\sch_1$. Note that an embedding  of the Schr\"odinger algebra into the conformal
algebra (in $d=3$ space dimensions)  had been defined in a different context in \cite{Bur73}. 

{\bf Definition 1.1}

{\em
Let $\tilde{\sv}\supset\sv$ be the (abstract) Lie algebra generated by $L_n,M_n,N_n$ ($n\in\Z$) and
$Y_m$ ($m\in\half+\Z$) with the following additional brackets:
\BEQ
[L_n,N_p]=-pN_{n+p}, [N_n,N_p]=0, \quad
[N_n,Y_p]= Y_{n+p},\  [N_n,M_p]=2M_{n+p}
\EEQ
}

Note that the $N_n$, $n\in\Z$, may be interpreted as a second $L$-conformal current with
conformal weight $1$.

{\bf Lemma 1.2}

{\em
\begin{enumerate}
\item
Let \BEQ \h=\langle Y_m\ |\ m\in\half+\Z\rangle\oplus\langle M_p\ |\ p\in\Z\rangle\EEQ and
\BEQ \tilde{\h}=\langle N_n\ |\ n\in\Z\rangle\oplus\h. \EEQ
Then $\h$ and $\tilde{\h}$ are Lie subalgebras of $\tilde{\sv}$ and one has the following double semi-direct product structure:
\BEQ \tilde{\h}=\langle N_n\ |\ n\in\Z\rangle\ltimes\h,\quad \tilde{\sv}=\vir_0\ltimes\tilde{\h}.\EEQ

The Lie algebra $\tilde{\h}$ is solvable.
\item
The Lie algebra $\tilde{\sch}_1=\langle N_0\rangle\ltimes\sch_1$ is a maximal Lie subalgebra of $\wsv$.
\item
The Lie algebra $\wsv$ has three independent classes of central extensions
given by the cocycles
\BEQ
c_1(L_n,L_m)= \frac{1}{12} n(n^2-1)\del_{n+m,0};
\EEQ
\BEQ c_2(N_n,N_m)=n\del_{n+m,0};
\EEQ
\BEQ
 c_3(L_n,N_m)=n^2\del_{n+m,0}
\EEQ
(the zero components of the cocycles have been omitted).
\end{enumerate}
}

{\bf Proof.}

Points 1 and 2 are straightforward. Let us turn to the proof of point 3.

The Lie subalgebra $\sv$ is known (see \cite{Hen94} or \cite{RogUnt06})
to have only one class of central
extensions given by the multiples of the Virasoro cocycle $c_1$; it
extends straightforwardly by zero to $\wsv$. Then
any central cocycle $c$ of $\wsv$  which is non-trivial on the $N$-generators may be
decomposed by $L_0$-homogeneity (see \cite{GuiRog06}) into the following components 
\BEQ c(N_m,N_p)=a_m \del_{m+p},\ c(N_m,M_p)=b_m \del_{m+p},\ 
c(L_m,N_p)=c_m \del_{m+p} \EEQ
The $b_m$ are easily seen to vanish by applying the Jacobi  relation to 
$[N_n,[Y_m,Y_p]]$ where $n+m+p=0$. The same relation applied to
$[L_n,[N_m,N_p]]$, respectively $[L_n,[L_m,N_p]]$, 
yields $pa_m=ma_p$, viz. $(n+m)(c_n-c_m)=(n-m)c_{n+m}$,
 hence $a_m=\kappa m$ and $c_m=\alpha m^2+\beta m$ for some
coefficients $\kappa,\alpha,\beta$. The coefficient $\beta$ may be set to zero by adding a constant to $N_0$. Finally, the two remaining cocycles are easily
seen to be non-trivial and independent. \eop 

{\bf Definition 1.3}
{\em

Let $\tilde{\sv}_{c,\kappa,\alpha}$ be the central extension of $\sv$ corresponding to the cocycle
$c c_1+\kappa c_2+\alpha c_3$, i.e. such that
\BEQ [L_n,L_m]=(n-m)L_{n+m}+\frac{1}{12} cn(n^2-1)\del_{n+m,0};\quad [N_n,N_m]=\kappa n\del_{n+m,0};\quad
[L_n,N_m]=-m N_{n+m} + \alpha n^2 \del_{n+m,0}.\EEQ
}

We shall now define  a series of representations $\tilde{\rho}$ of
$\wsv$, that we call {\it coinduced representations}, which are the
analogues of the density modules or conformal currents of the Virasoro representation theory. They are indexed by a 'spin' parameter $\rho$ corresponding to the
choice of a class of equivalence of representations of the subalgebra
$\wsv_0\subset \wsv$ (see below for a definition of $\wsv_0$).

The Lie algebra $\wsv$ is provided with a graduation $\del$ defined by
\BEQ
\del(L_n)=nL_n,\  \del(N_n)=n, \ \del(Y_m)=(m-\half)Y_m,\ \del(M_n)=(n-1)M_n\quad (n\in\Z,m\in\half+\Z)
\EEQ Note that $\del=\ad(-\half N_0-L_0)=-\half [N_0,.]-[L_0,.].$

Set $\wsv_n=\{X\in\wsv\ |\ \del(X)=nX\}=\langle L_n, N_n, Y_{n+\half},
M_{n+1}\rangle$ for $n=0,1,2,\ldots$ and $\wsv_{-1}=\langle L_{-1},
Y_{-\half},M_0\rangle$. Note that we choose to exclude $N_{-1}$ from
$\wsv_{-1}$ although $\del(N_{-1})=-N_{-1}$. Then $\wfsv:=
\oplus_{n\ge -1} \wsv_n$ is a Lie subalgebra of $\wit{\sv}$. The
subspace $\wit{\sv}_{-1}$ is commutative and
the Lie subalgebra $\wsv_0:=\{X\in\wsv\ |\ \del(X)=0\}$ is
a double  extension of the commutative Lie algebra
$\langle Y_{\half},M_1\rangle\cong\R^2$ by $L_0$ and
$N_0$ as follows:
\BEQ
\wsv_0=(\langle L_0\rangle \oplus \langle N_0\rangle) \ltimes\langle
Y_{\half},M_1\rangle
\EEQ
Namely, one has
\BEQ
[L_0,Y_{\half}]=-\half Y_{\half},\ [L_0,M_1]=-M_1;\quad [N_0,L_0]=0,\ 
[N_0,Y_{\half}]=Y_{\half},\ [N_0,M_1]=2M_1. 
\EEQ
Note that $N_0$ acts by conjugation as $-2L_0$ on $\wsv_0$. Also, the adjoint action of $\wsv_0$ preserves $\wsv_{-1}$, so that
$\wit{\sv}_0\oplus\wsv_{-1}=\wsv_0\ltimes\wsv_{-1}$ is  a Lie algebra too.
Actually, $\wfsv$ appears to be the {\it Cartan prolongation} of 
$\wsv_0\ltimes\wsv_{-1}$ (see \cite{AlbMol84}): if one realizes
 $\wsv_0\ltimes\wsv_{-1}$ as the following polynomial vector fields
\footnote{Note that this realization was originally obtained in \cite{HenUnt03}, where the generator denoted
by $N$ coincides with  $L_0-\frac{N_0}{2}=-t\partial_t+\zeta\partial_{\zeta}$.}
\BEQ
L_{-1}=-\partial_t,\ Y_{-\half}=-\partial_r,\ M_0=-\partial_{\zeta}
\EEQ
\BEQ
L_0=-t\partial_t-\half r\partial_r,\ N_0=-r\partial_r-2\zeta\partial_{\zeta},
\ Y_{\half}=-t\partial_r-r\partial_{\zeta},\ M_1=-t\partial_{\zeta}
\EEQ
then the Lie algebra $\wsv_{-1}\oplus\wsv_0\oplus\wsv_{1}\oplus\ldots$ defined inductively by
\BEQ \wsv_n:=\{ {\cal X}\in{\cal P}_n\ |\ [{\cal X},\wsv_{-1}]\subset\wsv_{n-1}\}, \quad n\ge 1 \EEQ
(where ${\cal P}_n$ stands for the vector space of homogeneous polynomial vector fields on $\R^3$
of degree $n+1$) defines a vector field realization
 of $\wfsv$ which extends straightforwardly into a representation of $\wsv$. Namely, let $f\in\C[t,t^{-1}]$: then 

\BEQ
L_f=-f(t)\partial_t-\half f'(t)r\partial_r-{1\over 4}
f''(t) r^2 \partial_{\zeta}
\EEQ
\BEQ
N_f=-f(t)(r\partial_r+2\zeta\partial_{\zeta})-
\half f'(t) r^2\partial_{\zeta}
\EEQ
\BEQ
Y_f=-f(t)\partial_r-f'(t) r\partial_{\zeta}
\EEQ
\BEQ
M_f=-f(t)\partial_{\zeta}
\EEQ

The restriction to $\sv$ of the above realization of $\wsv$ extends (after a Laplace transform) the mass ${\cal M}$ realization of $\sch_1$, see formulas (\ref{massM1}), (\ref{massM2}) and was originally obtained in \cite{Hen94}.

Let us now find out the  {\it coinduced representations} of $\wfsv$. The work was done in \cite{RogUnt06} for
 the Lie algebra $\sv$. The generalization to $\tilde{\sv}$ is only a matter of easy computations. 
Hence we merely recall the definition and give the results.

Let $\rho$ be a representation of $\wsv_0=(\langle L_0\rangle\oplus \langle
N_0\rangle) \ltimes \langle Y_{\half},M_1\rangle$ into a vector space ${\cal H}_{\rho}$. Then
$\rho$ can be trivially extended to $\wsv_+=\oplus_{i\ge 0}\wsv_i$ by
setting $\rho(\sum_{i>0}\wsv_i)=0$. Standard examples are provided:
\begin{itemize}
\item[(i)] either  by  choosing a representation $\rho$ of the $(ax+b)$-Lie algebra $\langle L_0,Y_{\half}\rangle$ and  extending it to $\tilde{\sv}_0 $ by setting 
\BEQ \rho(N_0)=-2\rho(L_0)+\mu\Id \ (\mu\in\R),\quad \rho(M_1)=C\rho(Y_{\half})^2\ (C\in\R);\EEQ
\item[(ii)] or by  choosing a representation $\rho$ of the $(ax+b)$-Lie algebra $\langle L_0,M_1\rangle$ and  extending it to $\tilde{\sv}_0 $ by setting 
\BEQ \rho(N_0)=-2\rho(L_0)+\mu\Id \ (\mu\in\R),\quad \rho(Y_{\half})=0.\EEQ
\end{itemize}

Actually, one may show easily that finite-dimensional indecomposable representations of $\langle
L_0,Y_{\half}\rangle$ or $\langle L_0, M_1\rangle$ are given (up to the addition of a constant to $L_0$) by restricting any finite-dimensional representation of $\slin(2,\R)$ to its Borel subalgebra or traceless upper-triangular matrices.
(On the other hand, the classification  of all  indecomposable finite-dimensional representations
of $\tilde{\sv}_0$ is probably a very difficult task). It happens so that all examples considered in this
article are obtained  as in (i) or (ii).

Let us now define the  representation of $\wfsv$ coinduced from $\rho$.

{\bf Definition 1.4} (see \cite{RogUnt06})

{\em
 The {\em  $\rho$-formal density module} $(\tilde{\cal
H}_{\rho},\tilde{\rho})$ is the coinduced module
\BEA
\tilde{\cal H}_{\rho} &=& {\mathrm{Hom}}_{{\cal U}(\wsv_+)} ({\cal
U}(\wfsv),{\cal H}_{\rho}) \nonumber \\
&=& \{ \phi:\ {\cal U}(\wfsv)\to {\cal H}_{\rho} \ {\mathrm{linear}}\ |\
\phi(U_0 V)=\rho(U_0).\phi(V),
\ \ U_0\in{\cal U}(\wsv_+),V\in{\cal U}(\wfsv) \}
\} \nonumber\\
\EEA
with the natural action of ${\cal U}(\wfsv)$ on the right
\BEQ
(d\tilde{\rho}(U).\phi)(V)=\phi(VU),\quad U,V\in{\cal U}(\wfsv).
\EEQ}

These abstract-looking formal density modules may be identified with the
following representations by matrix first-order differential operators.

{\bf Theorem 1.5}

{\em

The $\wfsv$-module  $(\wit{\cal H}_{\rho},\tilde{\rho})$ of $\wfsv$  is isomorphic
to
the action of the
following matrix differential operators on functions:

\BEA
 && \tilde{\rho}(L_f)=\left( -f(t)\partial_t-\half f'(t)r\partial_r-{1\over 4}
f''(t) r^2 \partial_{\zeta}\right)\otimes {\mathrm{Id}}_{{\cal H}_{\rho}}
+f'(t) \rho(L_0)
+{1\over 2} f''(t) r \rho(Y_{\half})+{1\over 4} f'''(t) r^2 \rho(M_1); \nonumber \\
&& \tilde{\rho}(N_f)=\left(-f(t)(r\partial_r+2\zeta\partial_{\zeta})-
\half f'(t) r^2\partial_{\zeta}\right)\otimes 
 {\mathrm{Id}}_{{\cal H}_{\rho}}+f(t) \rho(N_0) \nonumber\\ && \quad \quad
  + f'(t) r \rho(Y_{\half})+(\half f''(t)
r^2+2\zeta f'(t))\rho(M_1); \nonumber\\
&& \tilde{\rho}(Y_f)=\left(-f(t)\partial_r-f'(t) r\partial_{\zeta}\right)
\otimes{\mathrm{Id}}_{{\cal H}_{\rho}}
+f'(t)
\rho(Y_{\half})+f''(t) r\  \rho(M_1); \nonumber\\
&&\tilde{\rho}(M_f)=-f(t)\partial_{\zeta}\otimes{\mathrm{Id}}_{{\cal H}_{\rho}}
    +f'(t) \ \rho(M_1). \nonumber\\
\EEA

It may be extended into a representation of $\wsv$ by simply extrapolating the above formulas
to $f\in\R[t,t^{-1}]$.

}

The representations of $\wsv$ thus obtained will be  called {\it coinduced representations}.


\section{The Schr\"odinger-Virasoro primary fields and the superfield interpretation of $\wsv$}


Just as conformal fields are given by quantizing density modules in the Virasoro 
representation theory, we shall define in this section $\tilde{\sv}$-primary fields
by quantizing the coinduced representations $\tilde{\rho}$ introduced
in the previous section.

\subsection{Definition of the Schr\"odinger-Virasoro primary fields}

Our foundamental hypothesis is that correlators of $\tilde{\sv}$-primary fields
$\langle \Phi_1(t_1,r_1,\zeta_1)\ldots\Phi_n(t_n,r_n,\zeta_n)\rangle$ should
be singular only when some of the time coordinates coincide; this is confirmed by the computations of two- and three-point functions for scalar massive Schr\"odinger-covariant fields (see \cite{Hen94} or \cite{HenUnt03}, or also Appendix A).  Hence one is led to the
following assumption:

A $\tilde{\sv}$-primary field $\Phi(t,r,\zeta)$ may be
written as $\Phi(t,r,\zeta)=\sum_{\mu} \Phi^{(\mu)}(t,r,\zeta)e_{\mu}$, where $(e_{\mu})_{\mu
=1,\ldots,\dim 
{\cal H}_{\rho}}$ is a basis of
the representation space ${\cal H}_{\rho}$ (see Section 1) and
\BEQ \Phi^{(\mu)}(t,r,\zeta):=\sum_{\xi} \Phi^{(\mu),\xi}(t,\zeta) r^{\xi} \EEQ
where $\xi$ varies in a denumerable set of real values which is bounded
below (so that it is possible to multiply two such formal series) and stable
with respect to  translations by positive integers.
It may have been more logical to decompose further $\Phi^{(\mu),\xi}(t,\zeta)$
as $\sum_{\sigma} \Phi^{(\mu),\xi,\sigma}(t) \zeta^{\sigma}$, as we shall
occasionally do (see subsection 3.2), but this leads to unncessarily complicated
notations and turns out to be mostly counter-productive. In any case, $\Phi^{(\mu),\xi}(t,\zeta)$ is to be seen as a $\zeta$-indexed quantum field in the variable $t$, the
latter
playing the same role as the complex variable $z$ of conformal field theory,
implying the possibility of defining normal ordering, operator product expansions
and so on. Note that the ${\cal H}_{\rho}$-components of the field $\Phi$ are written systematically
{\it inside parentheses} in order to avoid any possible confusion with other indices.

  Suppose now that $\wsv$ (or any of its central extensions) acts on $\Phi$ by the  coinduced representation $\tilde{\rho}$ of Theorem 1.5. 
This action decomposes naturally as an action on each field component $\Phi^{(\mu),\xi}$
as follows (where Einstein's summation convention is implied):
  \begin{eqnarray}
[L_m,\Phi^{(\mu),\xi}(t,\zeta)] &=&  -t^{m+1}\partial_t \Phi^{(\mu),\xi}(t,\zeta)-{\xi\over 2}
(m+1)t^m \Phi^{(\mu),\xi}(t,\zeta)-\frac{1}{4} (m+1)mt^{m-1} \partial_{\zeta} \Phi^{(\mu),\xi-2} \nonumber\\
&& +(m+1)t^m \rho(L_0)^{\mu}_{\nu} \Phi^{(\nu),\xi}(t,\zeta) \nonumber\\
&&+\half(m+1)mt^{m-1} \rho(Y_{\half})^{\mu}_{\nu} \Phi^{(\nu),\xi-1}(t,\zeta) \nonumber\\
&&+{1\over 4} (m+1)m(m-1)t^{m-2} \rho(M_1)^{\mu}_{\nu} \Phi^{(\nu),\xi-2}(t,\zeta); 
\end{eqnarray}
\begin{eqnarray}
[N_m,\Phi^{(\mu),\xi}(t,\zeta)]&=& -t^m (\xi+2\zeta\partial_{\zeta})
\Phi^{(\mu),\xi}(t,\zeta)-\frac{m}{2} t^{m-1} \partial_{\zeta} \Phi^{(\mu),\xi-2}(t,\zeta)+t^m \rho(N_0)^{\mu}_{\nu} \Phi^{(\nu),\xi}(t,\zeta) \nonumber\\
&+&  mt^{m-1} \rho(Y_{\half})^{\mu}_{\nu} \Phi^{(\nu),\xi-1}(t,\zeta)+\frac{m(m-1)}{2}
t^{m-2} \rho(M_1)^{\mu}_{\nu} \Phi^{(\nu),\xi-2}(t,\zeta) \nonumber\\
 &+& 2mt^{m-1}\zeta \rho(M_1)^{\mu}_{\nu}
\Phi^{(\nu),\xi}(t,\zeta);
\end{eqnarray}
 
\begin{eqnarray}
[Y_m,\Phi^{(\mu),\xi}(t,\zeta)]  &=& -t^{m+\half} (\xi+1)\Phi^{(\mu),\xi+1}(t,\zeta)-
(m+\half)t^{m-\half} \partial_{\zeta} \Phi^{(\mu),\xi-1}(t,\zeta) \nonumber\\
&& + (m+\half)t^{m-\half} \rho(Y_{\half})^{\mu}_{\nu} \Phi^{(\nu),\xi}(t,\zeta) \nonumber\\
&&+ (m+\half)(m-\half) t^{m-3/2} \rho(M_1)^{\mu}_{\nu} \Phi^{(\nu),\xi-1}(t,\zeta);
\end{eqnarray}
  
 \BEQ
[M_m,\Phi^{(\mu),\xi}(t,\zeta)]=-t^m \partial_{\zeta}\Phi^{(\mu),\xi}(t,\zeta)+mt^{m-1}
\rho(M_1)^{\mu}_{\nu} \Phi^{(\nu),\xi}(t,\zeta).
\EEQ

  In order to define $\tilde{\sv}_{c,\kappa,\alpha}$-primary fields, one needs first the following assumption: there
exist four mutually local fields $$L(t)=\sum_{n\in\Z} L_n t^{-n-2}\ ,\ Y(t)=\sum_{n\in
\Z+\half} Y_n t^{-n-3/2},\ M(t)=\sum_{n\in\Z} M_n t^{-n-1},\quad
N(t)=\sum_{n\in\Z} N_n t^{-n-1}$$ with the following OPE's:
  \BEQ
L(t_1)L(t_2)\sim \frac{\partial L(t_1)}{t_1-t_2}+ \frac{2L(t_2)}{(t_1-t_2)^2}+
\frac{c/2}{(t_1-t_2)^4}, \quad c\in\R \label{DPOLL}
\EEQ
so that $L$ is a Virasoro field with central charge $c$;
\BEQ
L(t_1)Y(t_2)\sim \frac{\partial Y(t_2)}{t_1-t_2}+\frac{\frac{3}{2}Y(t_2)}{(t_1-t_2)^2},
\quad 
L(t_1)M(t_2)\sim \frac{\partial M(t_2)}{t_1-t_2}+\frac{M(t_2)}{(t_1-t_2)^2} \label{DPOLYM}
\EEQ
and
\BEQ L(t_1) N(t_2)\sim  \frac{\partial M(t_2)}{t_1-t_2}+\frac{M(t_2)}{(t_1-t_2)^2} +
\frac{\alpha}{(t_1-t_2)^3}  \label{DPOLN} \EEQ
so that $Y$ (resp. $M$) is an $L$-primary field with conformal weight $\frac{3}{2}$
(resp. $1$) and $N$ is primary with conformal weight 1 up to the term $\frac{\alpha}{(t_1-t_2)^3}$ 
due to the central extension;
\BEQ
Y(t_1)Y(t_2)\sim \frac{\partial M}{t_1-t_2}+\frac{2M(t_2)}{(t_1-t_2)^2},\quad
Y(t_1)M(t_2)\sim 0,\quad
M(t_1)M(t_2)\sim 0 \label{DPOYM}
\EEQ
and
\BEQ N(t_1)M(t_2)\sim \frac{2M(t_2)}{t_1-t_2},\quad N(t_1)Y(t_2)\sim \frac{Y(t_2)}{t_1-t_2},\quad
N(t_1)N(t_2)\sim \frac{\kappa}{(t_1-t_2)^2} \label{DPONYM}  \EEQ
which all together yield in mode decomposition 
the centrally extended  Lie algebra $\wsv_{c,\kappa,\alpha}.$

  We may now define what a {\it ${\rho}$-$\tilde{\sv}$-primary field} is.
Note that  we leave aside for the time being the essential condition which states
 that the values of the index $\xi$ should be bounded from below; we shall
actually see in subsection 3.2 that our free field construction works only for fields
$\Phi^{(\mu)}=\sum_{\xi} \Phi^{(\mu),\xi} r^{\xi}$ such that $\Phi^{(\mu),\xi}=0$
for all negative indices $\xi$. For technical reasons that will be explained below, we shall also define $\sv$-primary fields
and $\langle N_0\rangle\ltimes\sv$-primary fields.

In the following definition, we call (following \cite{Kac97}) {\it mutually local fields} a set
$X_1,\ldots,X_n$ of operator-valued formal series in $t$ whose commutators $[X_i(t_1),X_j(t_2)]$ are distributions of finite order supported on the diagonal $t_1=t_2$. In other words, the fields $X_1,\ldots,X_n$ have meromorphic
operator-product expansions (OPE).

  {\bf Definition 2.1.1}

{\em
\begin{enumerate}
\item ($\sv$-primary fields)

Let $\rho:\sv_0\to {\cal L}({\cal H}_{\rho})$ be a finite-dimensional representation of $\sv_0$.
A {\em $\rho$-$\sv$-primary field}  $\Phi(t,r,\zeta)=\sum_{\mu} \Phi^{(\mu)}(t,r,\zeta)e_{\mu}$ is given (at least
in a formal sense) as an infinite series 
$$\Phi^{(\mu)}(t,r,\zeta)=\sum_{\xi} \Phi^{(\mu),\xi}(t,\zeta)r^{\xi}$$
where $\xi$ varies in a denumerable set of real values which is stable
with respect to integer translations, and  the $\Phi^{(\mu),\xi}(t,\zeta)$ are mutually local fields with respect to the time
variable $t$-- which are also mutually
local with the $\sv$-fields $L(t),Y(t),M(t)$ -- with the following OPE:
  \BEA
L(t_1)\Phi^{(\mu),\xi}(t_2,\zeta)\sim \frac{\partial\Phi^{(\mu),\xi}(t_2,\zeta)}{t_1-t_2} &+&
\frac{(\half\xi) \Phi^{(\mu),\xi}(t_2,\zeta)-\rho(L_0)^{\mu}_{\nu} \Phi^{(\nu),\xi}(t_2,\zeta)} {(t_1-t_2)^2}
\nonumber \\
&+& \frac{\half\partial_{\zeta}\Phi^{(\mu),\xi-2}(t_2,\zeta)-\rho(Y_{\half})^{\mu}_{\nu} \Phi^{(\nu),\xi-1}
(t_2,\zeta)}{(t_1-t_2)^3} \nonumber\\
&-& \frac{\frac{3}{2} \rho(M_1)^{\mu}_{\nu} \Phi^{(\nu),\xi-2}(t_2)}{(t_1-t_2)^4}  \label{DPOL}
\EEA

\BEA
Y(t_1)\Phi^{(\mu),\xi}(t_2,\zeta)\sim \frac{(1+\xi)\Phi^{(\mu),\xi+1}(t_2,\zeta)}{t_1-t_2} &+&
\frac{\partial_{\zeta}\Phi^{(\mu),\xi-1}(t_2,\zeta)-\rho(Y_{\half})^{\mu}_{\nu} \Phi^{(\nu),\xi}(t_2,\zeta)}
{(t_1-t_2)^2} \nonumber \\
&-& \frac{2\rho(M_1)^{\mu}_{\nu} \Phi^{(\nu),\xi-1}(t_2,\zeta)}{(t_1-t_2)^3} \label{DPOY} 
\EEA
\BEQ
M(t_1)\Phi^{(\mu),\xi}(t_2,\zeta)\sim \frac{\partial_{\zeta}\Phi^{(\mu),\xi}(t_2,\zeta)}{t_1-t_2}
-\frac{\rho(M_1)^{\mu}_{\nu} \Phi^{(\nu),\xi}(t_2,\zeta)}{(t_1-t_2)^2} \label{DPOM}
\EEQ

\item ($\langle N_0\rangle\ltimes\sv$-primary fields)
Let $\bar{\rho}:\tilde{\sv_0}=\langle N_0\rangle\ltimes \sv_0\to {\cal L}({\cal H}_{\rho})$ be a finite-dimensional representation of $\tilde{\sv}_0$, and $\rho$ be the restriction of $\bar{\rho}$ to $\sv_0$.
A {\em $\bar{\rho}$-$\langle N_0\rangle\ltimes\sv$-primary field}  $\Phi(t,r,\zeta)$
is a $\rho$-$\sv$-primary field such that
\BEQ [N_0,\Phi^{(\mu),\xi}(t,\zeta)]=(\xi+2\zeta\partial_{\zeta})\Phi^{(\mu),\xi}(t,\zeta)-\bar{\rho}(N_0)^{\mu}_{\nu} \Phi^{(\nu),\xi}(t,\zeta).\EEQ

\item ($\tilde{\sv}$-primary fields)
Let  $\bar{\rho}:\tilde{\sv_0}=\langle N_0\rangle\ltimes \sv_0\to {\cal L}({\cal H}_{\rho})$ be a finite-dimensional representation of $\tilde{\sv}_0$ and $\Omega:{\cal H}_{\rho}\to{\cal H}_{\rho}$ be a linear
operator such that $[\rho(L_0),\Omega]=\Omega$, $[\rho(Y_{\half}),\Omega]=[\rho(M_1),\Omega]=[\rho(N_0),\Omega]=0$. Then a {\em $(\bar{\rho},\Omega)$-$\tilde{\sv}$-primary field} is a $\bar{\rho}|_{\sv_0}$-$\sv$-primary field $\Phi(t,r,\zeta)$, local with $N$,  such that
\BEA
 && N(t_1)\Phi^{(\mu),\xi}(t_2,\zeta)\sim \frac{(\xi+2\zeta\partial_{\zeta})\Phi^{(\mu),\xi}(t_2,\zeta)-\rho(N_0)^{\mu}_{\nu} \Phi^{(\nu),\xi}(t_2,\zeta)}{t_1-t_2} \nonumber \\ && +
\frac{\half \partial_{\zeta}\Phi^{(\mu),\xi-2}(t_2,\zeta)-\rho(Y_{\half})^{\mu}_{\nu}\Phi^{(\nu),\xi-1}(t_2,\zeta)-2\zeta \rho(M_1)^{\mu}_{\nu} \Phi^{(\nu),\xi}(t_2,\zeta)-\Omega^{\mu}_{\nu} \Phi^{(\nu),\xi}(t_2,\zeta)}{(t_1-t_2)^2} \nonumber \\
&& -  \frac{\rho(M_1)^{\mu}_{\nu} \Phi^{(\nu),\xi-2}(t_2,\zeta)}{(t_1-t_2)^3} \label{DPON}
\EEA

In the case $\Omega=0$, we shall simply say that $\Phi$ is $\bar{\rho}$-$\tilde{\sv}$-primary.
\end{enumerate}
}

{\bf Remark :} Mind that in these OPE and in all the following ones, $\zeta$ is considered only 
as a parameter, as we mentioned earlier.

The operator $\Omega$ for $\tilde{\sv}$-primary fields does not follow from the coinduction method. However, it appears in all our examples, including for the superfield ${\cal L}$ with components
$L,Y,M,N$ with the adjoint action of $\tilde{\sv}$ on itself (see \textsection 2.2  below).

  {\bf Proposition 2.1.2}

{\em
Suppose $\Phi$ is a $(\rho,\Omega)$-$\tilde{\sv}$-primary field. Then the adjoint action of $\wsv$ on
$\Phi$ is given by the formulas of Theorem 1.1 except for the action of the $N$-generators which are
{\em twisted} as follows:
\BEA
 \tilde{\rho}(N_f) &=&\left(-f(t)(r\partial_r+2\zeta\partial_{\zeta})-
\half f'(t) r^2\partial_{\zeta}\right)\otimes 
 {\mathrm{Id}}_{{\cal H}_{\rho}}+f(t) \rho(N_0) \nonumber \\ &+&  f'(t) r \rho(Y_{\half})+(\half f''(t)
r^2+2\zeta f'(t))\rho(M_1)+f'(t) \Omega; \nonumber\\
\EEA
}

  {\bf Proof.} Straightforward computations. One may in particular check that the twisted representation is indeed a representation of $\tilde{\sv}$. \hfill\eop

Note that the usual conformal fields of weight $\lambda$ are a particular case of
this construction: they correspond to $\rho$-$\sv$-conformal fields $\Phi$ with
only one component $\Phi=\Phi^{(0)}(t)$, commuting with $N,Y,M$, such that $\rho$
is the one-dimensional character given by $\rho(L_0)=-\lambda$, $\rho(N_0)=\rho(Y_{\half})= \rho(M_1)=0$.

\subsection{A superfield interpretation}

Similarly to the case of superconformal field theory (see \cite{Kac97}, \textsection 5.9), one may consider the fields $L,Y,M,N$ as four
components of the same superfield $\cal L$. To construct ${\cal L}$, we first need to go over to the
'Heisenberg' point of view by setting
\BEQ
\bar{L}(t,r,\zeta):=e^{\zeta M_0} e^{rY_{-\half}} L(t) e^{-rY_{-\half}}e^{-\zeta M_0}
\EEQ
and similarly for $\bar{Y},\bar{M},\bar{N}$, the quantum generators $Y_{-\half}$, resp. $M_0$ corresponding to the infinitesimal generators of space, resp. $\zeta$-translations.

In the following, the sign $\partial$ alone always indicates a derivative with respect to time.
Differences of coordinates are abbreviated as $t_{12}=t_1-t_2$, $r_{12}=r_1-r_2$, $\zeta_{12}=\zeta_1-\zeta_2$.

{\bf Lemma 2.2.1}
{\em 

\begin{enumerate}
\item
The Heisenberg fields $\bar{L},\bar{Y},\bar{N},\bar{M}$ read
\BEA
\bar{L}(t,r)=L(t)+\half r\partial Y(t)+\frac{r^2}{4} \partial^2 M(t);\nonumber\\
\bar{Y}(t,r)= Y(t)+r\partial M(t);\nonumber\\
\bar{M}(t)=M(t);\nonumber\\
\bar{N}(t,r,\zeta)=N(t)-rY(t)-\frac{r^2}{2} \partial M(t)-2\zeta M(t).
\EEA

\item
Operator product expansions are given by the following formulas:
\BEA
&&\bar{L}(t_1,r_1) \bar{L}(t_2,r_2)\sim \frac{\partial_{t_2} \bar{L}(t_2,r_2)}{t_{12}}+\frac{2\bar{L}(t_2,r_2)-\frac{1}{4}r_{12} \partial_{t_2} \bar{Y}(t_2,r_2)}{t_{12}^2} 
-\frac{3}{2} \frac{r_{12}}{t_{12}^3} \bar{Y}(t_2,r_2) \\ && \quad \quad \quad \quad  +\frac{\frac{c}{2}+\frac{3}{2} r_{12}^2
\bar{M}(t_2)}{t_{12}^4}; \nonumber\\
&& \bar{L}(t_1,r_1) \bar{Y}(t_2,r_2)\sim \frac{\partial \bar{Y}(t_2,r_2)}{t_{12}} + 
\frac{\frac{3}{2} \bar{Y}(t_2,r_2)-\half r_{12} \partial_{r_2} \bar{Y}(t_2,r_2)}{t_{12}^2} 
 -2\frac{r_{12}}{t_{12}^3} \bar{M}(t_2) ; \nonumber\\
&& \bar{Y}(t_1,r_1)\bar{L}(t_2,r_2)\sim \frac{\partial_{r_2}\bar{L}(t_2,r_2)}{t_{12}}+\frac{3}{2}
\frac{\bar{Y}(t_2,r_2)}{t_{12}^2}-\frac{2r_{12}\bar{M}(t_2)}{t_{12}^3}; \nonumber\\
&& \bar{L}(t_1,r_1)\bar{M}(t_2)\sim \frac{\partial \bar{M}(t_2)}{t_{12}}+\frac{\bar{M}(t_2)}{t_{12}^2},
\quad \bar{M}(t_1)\bar{L}(t_2,r_2)\sim \frac{\bar{M}(t_2)}{t_{12}^2};\nonumber\\
&& \bar{L}(t_1,r_1) \bar{N}(t_2,r_2,\zeta_2)\sim \frac{\partial\bar{N}(t_2,r_2,\zeta_2)}{t_{12}}+
\frac{\bar{N}(t_2,r_2,\zeta_2)-\half r_{12} \partial_{r_2}\bar{N}(t_2,r_2,\zeta_2)}{t_{12}^2} \nonumber\\
&&\quad \quad + \half \frac{r_{12}^2 \partial_{\zeta}\bar{N}(t_2,r_2,\zeta_2)}{t_{12}^3}+
\frac{\alpha}{t_{12}^3};\nonumber\\
&& \bar{N}(t_1,r_1,\zeta_1)\bar{L}(t_2,r_2)\sim \left( - \frac{r_{12}\partial_{r_2} \bar{L}(t_2,r_2)}{t_{12}} +\frac{ -\frac{3}{2} r_{12} \bar{Y}(t_2,r_2)-2\zeta_{12} \bar{M}(t_2)}{t_{12}^2} +
\frac{r_{12}^2 \bar{M}(t_2)}{t_{12}^3} \right) \nonumber\\
&&  \quad \quad +\frac{\bar{N}(t_2,r_2,\zeta_2)}{t_{12}^2} ; \nonumber\\
&&\bar{Y}(t_1,r_1)\bar{Y}(t_2,r_2)\sim \frac{\partial \bar{M}(t_2)}{t_{12}}+\frac{2\bar{M}(t_2)}{t_{12}^2},
\quad \bar{Y}(t_1,r_1)\bar{M}(t_2)\sim \bar{M}(t_1)\bar{M}(t_2)\sim 0;\nonumber\\
&&\bar{N}(t_1,r_1,\zeta_1)\bar{Y}(t_2,r_2)\sim \frac{-r_{12}\partial_{r_2} \bar{Y}(t_2,r_2)+
\bar{Y}(t_2,r_2)}{t_{12}}-\frac{2r_{12}\bar{M}(t_2)}{t_{12}^2}; \nonumber\\
&& \bar{Y}(t_1,r_1)\bar{N}(t_2,r_2,\zeta_2)\sim \frac{-\bar{Y}(t_2,r_2)}{t_{12}}+\frac{2r_{12}\bar{M}(t_2)}{t_{12}^2}; \nonumber\\
&&\bar{N}(t_1,r_1,\zeta_1) \bar{M}(t_2)\sim \frac{2\bar{M}(t_2)}{t_{12}},\quad \bar{M}(t_1)
\bar{N}(t_2,r_2,\zeta_2) \sim \frac{\partial_{\zeta_2} \bar{N}(t_2,r_2,\zeta_2)}{t_{12}}; \nonumber\\
&& \bar{N}(t_1,r_1,\zeta_1)\bar{N}(t_2,r_2,\zeta_2)\sim \left( \frac{-(2\zeta_{12}\partial_{\zeta_2}
+r_{12}\partial_{r_2})\bar{N}(t_2,r_2,\zeta_2)}{t_{12}}+\half \frac{r_{12}^2 \partial_{\zeta_2}\bar{N}(t_2,r_2,\zeta_2)}{t_{12}^2} \right) + \frac{\kappa}{t_{12}^2}. \nonumber\\
\EEA

\item
A field $\Phi=\sum_{\mu} \Phi^{(\mu)} e_{\mu}$ is a $(\rho,\Omega)$-$\tilde{\sv}$-primary field if and only if
the following relations hold (we omit the argument $(t_2,r_2,\zeta_2)$ of the field $\Phi$ in the
right-hand side of the equations): 
\BEA
&&\bar{L}(t_1,r_1)\Phi^{(\mu)}(t_2,r_2,\zeta_2)\sim \frac{\partial_{t_2}\Phi^{(\mu)}}{t_{12}}-\half
\frac{r_{12}\partial_{r_2}\Phi^{(\mu)}}{t_{12}^2}-\frac{\rho(L_0)^{\mu}_{\nu}\Phi^{(\nu)}}{t_{12}^2}
\nonumber\\
&& +\frac{\half r_{12}^2 \partial_{\zeta}\Phi^{(\mu)}+r_{12}\rho(Y_{\half})^{\mu}_{\nu} \Phi^{(\nu)}}{t_{12}^3} -\frac{3}{2} \frac{r_{12}^2 \rho(M_1)^{\mu}_{\nu} \Phi^{(\nu)}}{t_{12}^4}; 
\EEA
\BEA
\bar{Y}(t_1,r_1)\Phi^{(\mu)}(t_2,r_2,\zeta_2)\sim \frac{\partial_{r_2}\Phi^{(\mu)}}{t_{12}}-\frac{r_{12}\partial_{\zeta}\Phi^{(\mu)}}{t_{12}^2} -\frac{\rho(Y_{\half})^{\mu}_{\nu} \Phi^{(\nu)}}{t_{12}^2}+
\frac{2r_{12}\rho(M_1)^{\mu}_{\nu}\Phi^{(\nu)}}{t_{12}^3}; \EEA
\BEQ \bar{M}(t_1) \Phi^{(\mu)}(t_2,r_2,\zeta_2)\sim \frac{\partial_{\zeta_2}\Phi^{(\mu)}}{t_{12}}-\frac{\rho(M_1)^{\mu}_{\nu} \Phi^{(\nu)}}{t_{12}^2}; \EEQ
\BEA && \bar{N}(t_1,r_1,\zeta_1) \Phi^{(\mu)}(t_2,r_2,\zeta_2)\sim \frac{-(r_{12}\partial_{r_2}+2\zeta_{12}
\partial_{\zeta_2})\Phi^{(\mu)}-\rho(N_0)^{\mu}_{\nu} \Phi^{(\nu)}}{t_{12}} \nonumber\\ && +
\frac{\half r_{12}^2 \partial_{\zeta_2} \Phi^{(\mu)} +r_{12} \rho(Y_{\half})^{\mu}_{\nu} \Phi^{(\nu)}+2\zeta_{12} \rho(M_1)^{\mu}_{\nu}\Phi^{(\nu)} + \Omega^{\mu}_{\nu} \Phi^{(\nu)} }{t_{12}^2}  - \frac{r_{12}^2 \rho(M_1)^{\mu}_{\nu} \Phi^{(\nu)}}{t_{12}^3}.
\EEA

\end{enumerate}
}

Putting all this together, one gets:

{\bf Theorem 2.2.2}
{\em

Set $c=\kappa=\alpha=0$. Then: 

\begin{itemize}
\item[(i)] The four-dimensional field 
\BEQ {\cal L}(t,r,\zeta)=\left(\begin{array}{c} \bar{L}\\
\bar{Y}\\ \bar{M}\\ \bar{N} \end{array}\right)(t,r,\zeta) \EEQ
is $\rho$-$\sv $-primary for the representation $\rho$ defined by:
\BEQ \rho(L_0)=\left(\begin{array}{cccc} -2&&&\\ & -\frac{3}{2} &&\\ && -1 & \\ &&& -1
\end{array}\right), \ \rho(Y_{\half})=\left(\begin{array}{cccc} 0 & -\frac{3}{2} & & \\
& 0 & -2 & \\ && 0 & \\ &&& 0\end{array}\right), \ \rho(M_1)=\left(\begin{array}{cccc} 0&0&-1&0\\
& 0 && \\ && 0 &\\ &&& 0\end{array}\right).\EEQ
\item[(ii)] It is {\em not} $\rho-\tilde{\sv}$-primary.
\end{itemize}
}

{\bf Proof.} Straightforward computations. Note that $\rho(M_1)$ is proportional to $\rho(Y_{\half})^2$, see 
the remarks preceding Definition 1.4.

So what happened ? Setting $\rho(N_0)=\left(\begin{array}{cccc} 0 &&& \\ & -1 && \\  && -2 &\\
&&& 0\end{array}\right)$, one gets a representation of $\tilde{\sv}_0=(\langle L_0\rangle\oplus
\langle N_0\rangle) \ltimes \langle Y_{\half},M_1\rangle$ and ${\cal L}$ looks
$\rho-\tilde{\sv}$-primary, except for the last term $\frac{\bar{N}}{t_{12}^2}$
in the above OPE $\bar{N}.\bar{L}$. Fortunately, a supplementary matrix $\Omega$ as in Definition
2.1.1. (3)  allows
to take into account this term:

 {\bf Theorem 2.2.3}
{\em

Set $c=\kappa=\alpha=0$. Then
${\cal L}$ is $(\rho,\Omega)$-$\tilde{\sv}$-primary if one sets
$\Omega=\left(\begin{array}{cccc} 0&0&0&-1\\ &0&&\\&&0&\\ &&&0\end{array}\right).$
}

{\bf Proof.} Straightforward computations.


  \section{Construction by  $U(1)$-currents or $a\bar{b}$-theory}


Now that the definition of what is intended by {\it $\wsv$-primary} has been completed, we proceed to
 give explicit examples. The rest of the article is devoted to the detailed analysis of a vertex algebra
constructed out of two bosons (called : {\it $a\bar{b}$-model}) containing a representation of $\tilde{\sv}$
and $\tilde{\sv}$-primary fields of any $L_0$-weight.

\subsection{Definition of the $\wsv$-fields}

  We shall use here a classical construction of current algebras given in all generality
in \cite{Kac97}. Let $V=V_{\bar{0}}\oplus V_{\bar{1}}$ be a (finite-dimensional) super-vector
space, with even generators $a^i$, $i=1,\ldots,N$ for $V_{\bar{0}}$ and odd generators $b^{+,i}$,
$b^{-,i}$, $i=1,\ldots,M$ for $V_{\bar{1}}$ (supposed to be even-dimensional).

  {\bf Definition 3.1.1}
{\em
\begin{enumerate}
\item
The {\em bosonic supercurrents} associated with $V$ (see \cite{Kac97}, section 3.5)
 are the mutually local $N$ bosonic
fields $a^i(z)=\sum_{n\in\Z} a^i_n z^{-n-1}$ and the $2M$ fermionic fields
$b^{\pm,i}(z)=\sum_{n\in\Z} b^{\pm,i}_n z^{-n-1}$ with the following non-trivial OPE's:
\BEQ
a^i(z)a^j(w)\sim \frac{\del^{i,j}}{(z-w)^2}
\EEQ
\BEQ
b^{\pm,i}(z)b^{\mp,j}(w)\sim \pm \frac{\del^{i,j}}{(z-w)^2}
\EEQ
or,equivalently, with the following non-trivial Lie brackets in mode decomposition
\BEQ
[a^i_n, a^j_m]_-=n\del^{i,j} \del_{n+m,0}
\EEQ
\BEQ
[b^{+,i}_n, b^{-,j}_m]_+=n\del^{i,j}\del_{n+m,0}.
\EEQ
  \item
The {\em fermionic supercurrents} associated with $V$ (see \cite{Kac97}, sections 2.5 and 3.6) are the mutually local $N$ fermionic
fields 
$\bar{a}^i(z)=\sum_{n\in\Z} \bar{a}^i_n z^{-n-\half}$ and the $(2M)$ bosonic fields
$\bar{b}^{\pm,i}(z)=\sum_{n\in\Z} \bar{b}^{\pm,i}_n z^{-n-\half}$ with the following non-trivial OPE's:
\BEQ
\bar{a}^i(z)\bar{a}^j(w)\sim \frac{\del^{i,j}}{z-w}
\EEQ
\BEQ
\bar{b}^{\pm,i}(z)\bar{b}^{\mp,j}(w)\sim \pm \frac{\del^{i,j}}{z-w}
\EEQ
or,equivalently, with the following non-trivial Lie brackets in mode decomposition
\BEQ
[\bar{a}^i_n, \bar{a}^j_m]_+=\del^{i,j} \del_{n+m,0}
\EEQ
\BEQ
[\bar{b}^{+,i}_n, \bar{b}^{-,j}_m]_-=\del^{i,j}\del_{n+m,0}.
\EEQ
\end{enumerate}
  }

{\bf Remark:} The bosonic supercurrents $\bar{b}^{{\pm},i}$ (with unusual parity considering
their half-integer weight) are sometimes called {\em symplectic bosons} in the physical literature, see for
instance \cite{FriMarShe86,BoerFeher97}. 

  {\bf Proposition 3.1.2} (see \cite{Kac97}, sections 3.5 and 3.6)

  {\em
Consider the canonical Fock realization of the superalgebra generated by $a^i, b^{i,\pm}$
(obtained by requiring that $a^i,b^{i,\pm}$, $i\ge 0$, vanish on the vacuum vector $|0\rangle$).
Then
\BEQ
\langle 0\ |\ a^{i}(z)a^j(w)\ |\ 0\rangle=\del^{i,j} (z-w)^{-2},\quad
\langle 0\ |\ b^{\pm,i}(z)b^{\mp,j}(w)\ |\ 0\rangle=\pm \del^{i,j} (z-w)^{-2}
\EEQ and
\BEQ
\langle 0\ |\ \bar{a}^{i}(z)\bar{a}^j(w)\ |\ 0\rangle=\del^{i,j} (z-w)^{-1},\quad
\langle  0\ |\ \bar{b}^{\pm,i}(z)\bar{b}^{\mp,j}(w)\ |\ 0\rangle=\pm \del^{i,j} (z-w)^{-1}.
\EEQ }
  One may build Virasoro fields out of these supercurrents, one for each type of currents:
\BEQ
L_a(t)=\half :a^2:(t), \quad L_b(t)=:b^+ b^-:(t) 
\EEQ with central charge $1$, viz. $-2$;
\BEQ
L_{\bar{a}}(t)=-\half :\bar{a}(\partial \bar{a}):(t),\quad 
L_{\bar{b}}(t)=\half \left( :\bar{b}^+ \partial \bar{b}^-:(t) - 
 :\bar{b}^- \partial \bar{b}^+:(t) \right)
\EEQ with central charge $\half$, viz. $-1$.

  For the appropriate Virasoro field, the bosonic supercurrents $a^i,b^i$ are primary
with conformal weight $1$, while the fermionic supercurrents $\bar{a}^i, \bar{b}^{\pm,i}$
are primary with conformal weight $\half$.
  The simplest way to construct a Lie algebra isomorphic to an appropriately centrally
extended $\tilde{\sv}$ with these generating
fields is the following:

  {\bf Definition 3.1.3}

{\em
Let $V=V_{\bar{0}}\oplus V_{\bar{1}}$ with $V_{\bar{0}}=\R a$ and $V_{\bar{1}}=\R b^+
\oplus \R b^-.$ Then $\wsv_{(0,-1,0)} $-fields $L$, $N$,  $Y$, $M$ may be defined as follows: 
\BEA
L=L_a+L_{\bar{b}} \quad \mathrm{with\ zero\ central\ charge};\\
N=-:\bar{b}^+ \bar{b}^-:  \quad \mathrm{with\  central\ charge}\ -1;\\
Y=:a\bar{b}^+:\ ;\\
M=\half :(\bar{b}^+)^2:
\EEA }
  Let us first check explicitly that one retrieves the OPE (\ref{DPOLL},\ref{DPOLYM},\ref{DPOLN},
\ref{DPOYM},\ref{DPONYM})
 with this definition:
 $$:a^2:(t_1):a\bar{b}^+:(t_2)\sim \frac{2:a(t_1)\bar{b}^+(t_2):}{(t_1-t_2)^2}
\sim \frac{2:a\bar{b}^+:(t_2)}{(t_1-t_2)^2}+\frac{2:\partial a\bar{b}^+:(t_2)}{t_1-t_2};$$
  $$:\bar{b}^+ \partial\bar{b}^-:(t_1) \ :a\bar{b}^+:(t_2) \sim \frac{:\bar{b}^+(t_1)
a(t_2):}{(t_1-t_2)^2}\sim \frac{:a\bar{b}^+:(t_2)}{(t_1-t_2)^2}+\frac{:a\partial\bar{b}^+:
(t_2)}{t_1-t_2};$$
  $$:\bar{b}^-\partial\bar{b}^+:(t_1)\ :a\bar{b}^+:(t_2)\sim -\frac{:\partial\bar{b}^+(t_1)
a(t_2):}{t_1-t_2}\sim - \frac{:\partial\bar{b}^+ a:(t_2)}{t_1-t_2}$$
  so $L(t_1)Y(t_2)\sim \frac{\partial Y(t_2)}{t_1-t_2}+\frac{\frac{3}{2}Y(t_2)}{(t_1-t_2)^2}$.

  Similarly,
\begin{eqnarray*}
:a^2:(t_1)\ :(\bar{b}^+)^2:(t_2)\sim 0; \\
:\bar{b}^+\partial\bar{b}^-:(t_1)\ :(\bar{b}^+)^2:(t_2)\sim \frac{2:\bar{b}^+(t_1)
\bar{b}^+(t_2):}{(t_1-t_2)^2}\sim \frac{2:(\bar{b}^+)^2(t_2)}{(t_1-t_2)^2} +
2\frac{:\partial\bar{b}^+ \bar{b}^+:(t_2)}{t_1-t_2};\\
:\bar{b}^- \partial\bar{b}^+:(t_1)\ :(\bar{b}^+)^2:(t_2)\sim -2
\frac{:\partial\bar{b}^+ \bar{b}^+:(t_2)}{t_1-t_2}
\end{eqnarray*}
so $L(t_1)M(t_2)\sim \frac{\partial M(t_2)}{t_1-t_2}+\frac{M(t_2)}{(t_1-t_2)^2}$.

  Finally,
  \BEA
Y(t_1)Y(t_2)&=& :a\bar{b}^+:(t_1) :a\bar{b}^+:(t_2) 
\sim \frac{:\bar{b}^+(t_1)\bar{b}^+(t_2):}{(t_1-t_2)^2} \nonumber\\
&\sim& \frac{:(\bar{b}^+)^2:(t_2)}{(t_1-t_2)^2}+\frac{:\bar{b}^+ \partial\bar{b}^+:(t_2)}
{t_1-t_2}
= \frac{2M(t_2)}{(t_1-t_2)^2}+\frac{\partial M(t_2)}{t_1-t_2}
\EEA
and $Y(t_1)M(t_2)\sim 0$, $M(t_1)M(t_2)\sim 0$, so one is done for the
$\sv$-fields $L,Y,M$.
Then
$$N(t_1)N(t_2)=:\bar{b}^+\bar{b}^-:(t_1) :\bar{b}^+\bar{b}^-:(t_2)
\sim -\frac{1}{(t_1-t_2)^2}$$
(the terms of order one cancel each other);
\begin{eqnarray*}
:\bar{b}^+\partial\bar{b}^-:(t_1):\bar{b}^+\bar{b}^-:(t_2)\sim
\frac{:\partial(\bar{b}^+\bar{b}^-):(t_2)}{t_1-t_2}+\frac{:\bar{b}^+\bar{b}^-:
(t_2)}{(t_1-t_2)^2} \\
:\bar{b}^-\partial\bar{b}^+:(t_1):\bar{b}^+\bar{b}^-:(t_2)\sim -
\frac{:\partial(\bar{b}^+\bar{b}^-):(t_2)}{t_1-t_2}-\frac{:\bar{b}^+\bar{b}^-:
(t_2)}{(t_1-t_2)^2}
\end{eqnarray*}
hence $L(t_1)N(t_2)\sim \frac{\partial N(t_2)}{t_1-t_2}+\frac{N(t_2)}{t_1-t_2};$
finally,
$$N(t_1)Y(t_2)=-:\bar{b}^+\bar{b}^-:(t_1):a\bar{b}^+:(t_2)\sim \frac{Y(t_2)}
{t_1-t_2}$$ and $$N(t_1)M(t_2)=-\half :\bar{b}^+\bar{b}^-:(t_1):(\bar{b}^+)^2:(t_2)\sim \frac{2M(t_2)}{t_1-t_2}.$$

  {\bf Definition 3.1.4}

{\em The {\em constrained 3D-Dirac equation} (or: constrained Dirac equation for short) is
the set of following equations for a spinor field $(\phi_1,\phi_2)=(\phi_1(t,r,\zeta),\phi_2(t,r,
\zeta))$ on $\R^3$:
\BEA
\partial_r \phi_0=\partial_t \phi_1 \\
\partial_r \phi_1=\partial_{\zeta}\phi_0 \\
\partial_{\zeta}\phi_1=0.
\EEA } 

  {\bf Theorem 3.1.5}

{\em
\begin{enumerate}
\item
The space of solutions of the constrained 3D-Dirac equation is in one-to-one correspondence
 with the
space of triples $(h_0^-,h_0^+,h_1)$ of functions of $t$ only: a natural bijection may be
obtained by setting
\BEA \phi_0(t,r,\zeta)=(h_0^-(t)+\zeta h_0^+(t))+rh_1(t)+\frac{r^2}{2} \partial h_0^+ (t)\\
 \phi_1(t,r,\zeta)=\int_0^t h_1(u)\ du+r h_0^+ (t)\EEA
\item
Put 
\BEQ
\Phi^{(0)}(t,r,\zeta)=(\bar{b}^-(t)+\zeta \bar{b}^+(t))+ra(t)+\frac{r^2}{2}
 \partial \bar{b}^+(t)
\EEQ
 and 
 \BEQ
\Phi^{(1)}(t,r,\zeta)=( \int a)(t) +r \bar{b}^+ (t)
\EEQ where 
\BEQ
\int a=-\sum_{n\not =0} a_n \frac{t^{-n}}{n}+a_0 \log t+\pi_0,\quad [a_0,\pi_0]=1 
\EEQ
 is the logarithmic bosonic field defined for instance in \cite{DiF97}. 
Then $\Phi:=\left(\begin{array}{c} \Phi^{(0)} \\ \Phi^{(1)} \end{array}\right)$ 
is a $\rho$-$\tilde{\sv}$-primary field, where $\rho$ is the two-dimensional
character defined by 
\BEQ \rho(L_0)=\left(\begin{array}{cc} -\half &\\ & 0\end{array}\right),\quad
\rho(N_0)=\left(\begin{array}{cc} 1 &\\ & 0\end{array}\right),\quad 
\rho(Y_{\half})=\rho(M_1)=0.\EEQ

\item
The two-point functions ${\cal C}^{\mu,\nu}(t_1,r_1,\zeta_1;t_2,r_2,\zeta_2):=
\langle 0|\ \Phi^{(\mu)}(t_1,r_1,\zeta_1)\Phi^{(\nu)}(t_2,r_2,\zeta_2)\ |0\rangle$,
$\mu,\nu=0,1$, are given by
\BEQ
{\cal C}^{0,0}=\frac{1}{t} (\zeta-\frac{r^2}{2t}),\ 
{\cal C}^{0,1}={\cal C}^{1,0}=r,\ {\cal C}^{1,1}=\ln t
\EEQ
where $t=t_1-t_2,r=r_1-r_2,\zeta=\zeta_1-\zeta_2.$

\end{enumerate}
}

{\bf Remark.} 

The free boson $\int a$ is not conformal in the usual sense since it contains a logarithmic term, contrary to
the vertex operators built as exponentials of $\int a$ that we shall use in the following sections. In this
very particular case, one needs to  consider $a_0$, $\pi_0$ as a couple of usual annihilation/creation operators in order for
 the scalar product $\langle 0|\ (\int a)(t_1) (\int a)(t_2)\ |0\rangle$ to make sense, so that $a_0$ and
$\pi_0$ are adjoint to each other. The usual normalization is quite different.

 {\bf Proof.}
\begin{enumerate}
\item
Let $(\phi,\psi)$ be a solution of the constrained Dirac equation. Then $\partial_r^2
\psi=\partial_{t}\partial_{\zeta} \psi=0$ so 
\BEQ \psi(t,r,\zeta)=\psi_0(t)+r\psi_1(t). \EEQ
On the other hand, $\partial_{\zeta}^2 \phi=\partial_{\zeta}\partial_r\psi=0$, 
$\partial_{\zeta}\phi=\psi_1$ and $\partial_r \phi=\partial_t\psi_0+r\partial_t \psi_1$,
hence, by putting together everything,
\BEQ \phi(t,r,\zeta)=\phi^{00}(t)+\psi_1(t)\zeta+\psi'_0(t)r+\psi'_1(t) \frac{r^2}{2}. \EEQ
Now one just needs to set $h_0^-:=\phi^{00}, h_0^+=\psi_1$ and $h_1=\psi'_0.$

\item
This follows directly from  Definition 2.1 once one has  established the following easy
relations
\BEA
L(t_1)\partial{\bar{b}^+}(t_2)\sim \frac{\partial^2\bar{b}^+(t_2)}{t_1-t_2}+\frac{3}{2}
\frac{\partial\bar{b}^+(t_2)}{(t_1-t_2)^2}+\frac{\bar{b}^+(t_2)}{(t_1-t_2)^3}\\
N(t_1)(\bar{b}^-(t_2)+\zeta\bar{b}^+(t_2))\sim \frac{-\bar{b}^-(t_2)+\zeta
\bar{b}^+(t_2)}{t_1-t_2} \\
N(t_1)a(t_2)\sim 0\\
N(t_1)\partial\bar{b}^+(t_2)\sim \partial_{t_2} \left(\frac{\bar{b}^+(t_2)}
{t_1-t_2}\right)=\frac{\partial\bar{b}^+(t_2)}{t_1-t_2}+\frac{\bar{b}^+(t_2)}
{(t_1-t_2)^2}\\
Y(t_1)(\bar{b}^-(t_2)+\zeta\bar{b}^+(t_2))\sim \frac{a(t_2)}{t_1-t_2} \\
Y(t_1) a(t_2)\sim \frac{\partial\bar{b}^+(t_2)}{t_1-t_2}+\frac{\bar{b}^+(t_2)}{
(t_1-t_2)^2} \\
Y(t_1)\bar{b}^+(t_2)\sim 0
\EEA
together with the fact that $\bar{b}^{\pm}$, resp. $a$, are $L$-conformal with conformal
weight $\half$ (resp. $1$).
\item Straightforward.
\end{enumerate}

In particular, one retrieves the fact that the classical constrained Dirac equation is $\sv$-invariant, see
\cite{RogUnt06} for a discussion and generalizations. Unfortunately, one can hardly say that this is
an interesting physical equation.

We give thereafter two other examples. They
exhaust all possibilities of $\tilde{\sv}$-primary linear fields of this model and
are only given for the sake of completeness.

{\bf Lemma 3.1.6}

{\em
\begin{enumerate}
\item
The {\em trivial}  field $\bar{b}^+(t)$ is  a $\rho$-Schr\"odinger-conformal field, where $\rho$ is the one-dimensional
character defined by 
\BEQ \rho(L_0)=-\half,\quad \rho(N_0)=-1,\quad 
\rho(Y_{\half})=\rho(M_1)=0.\EEQ 
The associated two-point function vanishes.

\item
Put $\Phi^{(0)}(t,r,\zeta)=a(t)+r\partial\bar{b}^+$ and $\Phi^{(1)}(t,r,\zeta)=-\bar{b}^+.$
Then  $\Phi=\left(\begin{array}{c} \Phi^{(0)} \\ \Phi^{(1)} \end{array}\right)$ 
is a $\rho$-$\tilde{\sv}$-primary field, where $\rho$ is the two-dimensional
representation defined by 
\BEQ \rho(L_0)=\left(\begin{array}{cc} -\half &\\ & -1\end{array}\right),\quad
\rho(N_0)=\left(\begin{array}{cc} -1 &\\ & 0\end{array}\right),\quad 
\rho(Y_{\half})=\left(\begin{array}{cc} 0 & 1 \\ 0& 0\end{array}\right)
,\quad \rho(M_1)=0.\EEQ
The two-point functions of the field $\Phi$ are given by
\BEQ
{\cal C}^{0,0}=t^{-2},\ {\cal C}^{0,1}={\cal C}^{1,0}={\cal C}^{1,1}=0.
\EEQ
\end{enumerate}
}
 
{\bf Proof.}
\begin{enumerate}
\item Straightforward.
\item Follows from preceding computations.
\end{enumerate}
 \eop

\subsection{Construction of the generalized polynomial fields $\Phia_{j,k}$}

We shall introduce in this paragraph  more general fields. Take any polynomial $P=P(\bar{b}^-,
\bar{b}^+,\partial\bar{b}^+,a,\int a)$ where 
\BEQ (\int a)(t):=-\sum_{n\not=0} a_n \frac{t^{-n}}{n}+a_0 \log t+\pi_0, \quad [a_0,\pi_0]=1 \EEQ
is the usual  logarithmic bosonic field of conformal field theory from which vertex operators
are built.  Since $[\bar{b}^+_n,\bar{b}^-_m]=0$
if $nm>0$ and similarly for the commutators of any of the fields $\bar{b}^-,
\bar{b}^+,\partial\bar{b}^+,a,\int a$, the normal ordering is {\it commutative}
and the field $:P:$ is well defined.

Let us introduce first for convenience the following notation for the coefficients of OPE of two mutually local fields.

{\bf Definition 3.2.1.}
{\em

Let $A,B$ be two mutually local fields: their OPE is given as
\BEQ A(t_1)B(t_2)\sim \sum_{k=0}^{\infty} \frac{C_k(t_2)}{t_{12}^k} \EEQ
for some fields $C_0(t),C_1(t),\ldots,C_p(t),\ldots$ which vanish for $p$ large enough. 

We shall denote by $A_{(k)}B$, $k=0,1,\ldots$ the field $C_k$.
}

{\bf Theorem 3.2.2.}

{\em

Let $P$ be any polynomial in the fields $\bar{b}^{\pm}$, $\partial \bar{b}^+$, $a$ and $\int a$. Then 
\BEA
L(t_1):P:(t_2)&\sim& \frac{:\partial P:(t_2)}{t_{12}}+\frac{\left( \half(\bar{b}^-
\partial_{\bar{b}^-}+\bar{b}^+\partial_{\bar{b}^+})+\frac{3}{2} \partial
\bar{b}^+ \partial_{\partial\bar{b}^+}+a\partial_a+\half\partial_{\int a}^2
\right)P:(t_2)}{t_{12}^2}  \nonumber \\
&+& \frac{:\left(\bar{b}^+\partial_{\partial\bar{b}^+}+\partial_{\int a}
\partial_a\right)P:(t_2)}{t_{12}^3}+\half \frac{:(\partial_a^2+\partial_{\bar{b}^-}\partial_{\partial\bar{b}^+})P:(t_2)}{t_{12}^4}  \label{LP}
\EEA
\BEA
N(t_1)P(t_2)&\sim& \frac{\left(\bar{b}^+\partial_{\bar{b}^+}+\partial \bar{b}^+
\partial_{\partial
\bar{b}^+}-\bar{b}^-\partial_{\bar{b}^-}\right)P:(t_2)}{t_{12}}+\frac{
\left(\bar{b}^+\partial_{\partial\bar{b}^+}+\partial_{\bar{b}^-}
\partial_{\bar{b}^+}\right)P:(t_2)}{t_{12}^2} \nonumber \\
&+& \frac{(\partial_{\bar{b}^-}\partial_{\partial\bar{b}^+})P:(t_2)}{t_{12}^3} \label{NP}
\EEA
\BEA
Y(t_1):P:(t_2)&\sim& \frac{:\left( a\partial_{\bar{b}^-}+\partial\bar{b}^+
\partial_a+\bar{b}^+ \partial_{\int a}\right)P:(t_2)}{t_{12}}+\frac
{:(\bar{b}^+\partial_a+\partial_{\bar{b}^-}\partial_{\int a})P:(t_2)}{t_{12}^2} \nonumber \\
&+& \frac{:\partial_a\partial_{\bar{b}^-}P:(t_2)}{t_{12}^3} \label{YP}
\EEA
\BEQ
M(t_1) :P:(t_2)\sim \frac{:\bar{b}^+\partial_{\bar{b}^-}P:(t_2)}{t_{12}}+
\half \frac{:\partial_{\bar{b}^-}^2 P:(t_2)}{t_{12}^2} \label{MP}
\EEQ }

{\bf Proof.}

Consider the monomial $P=P_{jklmn}=(\bar{b}^-)^j (\bar{b}^+)^k (\partial \bar{b}^+)^l
a^m (\int a)^n.$ Let us compute $L_{(n)}:P:$, $n\ge 0$ first. Apart from the contribution of
the logarithmic field $\int a $ which has special properties, one may deduce the coefficient of the terms of order $t_{12}^{-1}$ and $t_{12}^{-2}$ directly from general considerations
(see \cite{Kac97}): the field $(\bar{b}^-)^j(\bar{b}^+)^k (\partial\bar{b}^+)^l a^m$ is
quasiprimary with conformal weight $\frac{j+k}{2}+\frac{3l}{2}+m$. The contribution from
the field $\int a$ reads 
\BEA
L(t_1):P:(t_2)&\sim& \half :a^2:(t_1) :P_{jkl00}a^m(\int a)^n:+\ldots \nonumber\\
&\sim& n\frac{:P_{jkl00}a^{m+1}(\int a)^{n-1}:(t_2)}{t_{12}}+\frac{n(n-1)}{2}
\frac{:P_{jkl00}a^m (\int a)^{n-1}:(t_2)}{t_{12}^2} \nonumber
\\&+& mn\frac{:P_{jkl00}a^{m-1}(\int a)^{n-1}:(t_2)}
{t_{12}^3}+\ldots 
\EEA
in accordance with the Theorem. So, if we prove that $L_{(n)}:P_{jklm0}:$ agree with (\ref{LP})
for $n\ge 2$, we are done. One gets (leaving aside the poles of order $1$ or $2$)
$$
\half:a^2:(t_1)a^m(t_2)\sim \half m(m-1) \frac{(\bar{b}^-)^j (\bar{b}^+)^k (\partial\bar{b}^+)^l
a^{m-2}:(t_2)}{t_{12}^4}+\ldots \ {\mathrm{
(double\ contraction)}}; $$
$$
\half:\bar{b}^+\partial\bar{b}^-:(t_1) :(\partial\bar{b}^+)^l:(t_2)\sim l
\frac{:(\bar{b}^-)^j(\bar{b}^+)^k(\partial\bar{b}^+)^{l-1}a^m:(t_2)}{t_{12}^3}+\ldots
\ {\mathrm{(simple\ contraction)}}; $$
$$
\half:\bar{b}^+\partial\bar{b}^-:(t_1) :(\bar{b}^-)^j (\bar{b}^+)^k:(t_2)\sim \frac{jk}{2}
\frac{(\bar{b}^-)^{j-1}(\bar{b}^+)^{k-1}(\partial\bar{b}^+)^l a^m:(t_2)}{t_{12}^3}
\ {\mathrm{(double\ contraction)}}; $$
$$
\half:\bar{b}^+\partial\bar{b}^-:(t_1) :(\bar{b}^-)^j (\partial\bar{b}^+)^l:(t_2)\sim jl
\frac{:(\bar{b}^-)^{j-1} (\bar{b}^+)^k (\partial\bar{b}^+)^{l-1} a^m:(t_2)}{t_{12}^4}
\ {\mathrm{(double\ contraction)}}; $$
$$
-\half:\bar{b}^-\partial\bar{b}^+:(t_1) :(\bar{b}^-)^j (\bar{b}^+)^k:(t_2)\sim -\frac{jk}{2}
\frac{(\bar{b}^-)^{j-1}(\bar{b}^+)^{k-1}(\partial\bar{b}^+)^l a^m:(t_2)}{t_{12}^3}
\ {\mathrm{(double\ contraction)}};$$
$$
-\half:\bar{b}^-\partial\bar{b}^+:(t_1) :(\bar{b}^-)^j (\partial\bar{b}^+)^l:(t_2)\sim -\frac{jl}{2}
\frac{:(\bar{b}^-)^{j-1} (\bar{b}^+)^k (\partial\bar{b}^+)^{l-1} a^m:(t_2)}{t_{12}^4}
\ {\mathrm{(double\ contraction)}} $$ hence the result.

Let us consider now the OPE of $N$ with $P$. The fields $a $  and $\int a $ giving no contribution, one may just as well assume that $m=n=0$. Then
\begin{eqnarray*}
-:\bar{b}^+\bar{b}^-:(t_1) & :(\bar{b}^-)^j(\bar{b}^+)^k(\partial\bar{b}^+)^l:(t_2) \sim
\left( \frac{(k-j)(\bar{b}^-)^j(\bar{b}^+)^k (\partial\bar{b}^+)^l}{t_{12}}+l
\frac{:\bar{b}^+(t_1)(:(\bar{b}^-)^j(\bar{b}^+)^k(\partial\bar{b}^+)^{l-1}:)(t_2):}{t_{12}^2}\right)
\\ &+ \left(
\frac{jk:(\bar{b}^-)^{j-1} (\bar{b}^+)^{k-1} (\partial\bar{b}^+)^l:(t_2)}{t_{12}^2}+ jl
\frac{:(\bar{b}^-)^{j-1} (\bar{b}^+)^{k-1}(\partial\bar{b}^+)^{l-1}:(t_2)}{t_{12}^3}\right)
\end{eqnarray*}
adding the terms coming from a single contraction to the terms coming from a double
contraction. Hence the result. The OPE of $M$ with $P$ follows easily from the same rules. Finally, 
\begin{eqnarray}
Y(t_1):P:(t_2)&=& :a\bar{b}^+:(t_1) :(\bar{b}^-)^j(\bar{b}^+)^k (\partial\bar{b}^+)^l a^m (\int a)^n:(t_2) 
\\ &\sim& \left( j \frac{:(\bar{b}^-)^{j-1}(\bar{b}^+)^k (\partial\bar{b}^+)^l a^{m+1}
 (\int a)^n:(t_2) }{t_{12}} + n  \frac{:(\bar{b}^-)^{j}(\bar{b}^+)^{k+1} (\partial\bar{b}^+)^l a^{m}
 (\int a)^{n-1}:(t_2) }{t_{12}} \right. \nonumber\\
&+& \left.  m \frac{:\bar{b}^+(t_1)(:(\bar{b}^-)^{j}(\bar{b}^+)^k (\partial\bar{b}^+)^l a^{m-1}
 (\int a)^n:(t_2): }{t_{12}^2} \right) \nonumber \\
&+& \left( nj \frac{:(\bar{b}^-)^{j-1}(\bar{b}^+)^k (\partial\bar{b}^+)^l a^{m}
 (\int a)^{n-1}:(t_2) }{t_{12}^2}+mj 
  \frac{:(\bar{b}^-)^{j-1}(\bar{b}^+)^k (\partial\bar{b}^+)^l a^{m-1}
 (\int a)^{n}:(t_2) }{t_{12}^3} \right) \nonumber\\
\end{eqnarray}
(separating once more the terms coming from a single contraction from the terms with a double
contraction) hence (\ref{YP}). \hfill \eop

\vskip 1 cm

In any case, the $\wsv$-fields preserve this space of polynomial fields. The reason why we chose not to include powers of $\partial\bar{b}^-$ or $\partial a$ for instance, or higher derivatives of the field $\bar{b}^+$, will appear clearly in a moment. Take a $\rho$-Schr\"odinger-conformal field
$\Phi=(\Phi^{(\mu)})_{\mu}$ and suppose it has a formal expansion of the type $\sum_{\xi,\nu} \Phi^{(\mu),\xi,\sigma}(t)r^{\xi}\zeta^{\sigma}$ as in subsection 2.1, with $\sigma$ varying in a set of real values of the
same type as for $\xi$, while the $\Phi^{(\mu),\xi,\sigma}$ are polynomials in the variable $\int a$,
$\bar{b}^{\pm}$ and their derivatives of any order. Suppose $\Phi^{(\mu),\xi,\sigma}\not=0$ for
a negative value of $\xi$. Then
\BEQ \Phi^{(\mu),\xi,\sigma}=Y_{(0)} \frac{\Phi^{(\mu),\xi-1,\sigma}}{\xi}=\frac{1}{\xi} (a\partial_{\bar{b}^-}+\ldots)
\Phi^{(\mu),\xi-1,\sigma} \EEQ
hence $\Phi^{(\mu),\xi-1,\sigma}$ must include a monomial $P_{jklmn}$ with $m$ strictly less than for all the monomials in $\Phi^{(\mu),\xi,\sigma}$. But this argument can be repeated indefinitely,
going down one step $\xi\to\xi-1$ at a time, and one ends with a contradiction if negative
powers of $a$ are not allowed. The same goes for $\sigma$ since
\BEQ \Phi^{(\mu),\xi,\sigma}= M_{(0)}\frac{\Phi^{(\mu),\xi,\sigma-1}}{\sigma}=\frac{1}{\sigma} \bar{b}^+\partial\bar{b}^- \Phi^{(\mu),\xi,
\sigma-1}.\EEQ

A moment's thought proves then that if the $\Phi^{(\mu),\xi,\sigma}$ are to be polynomials, then the
indices $\xi$ and $\sigma$ should  be positive integers and all the terms $\Phi^{(\mu),\xi,\sigma}$ may
be obtained from the {\it lowest degree component fields} $\Phi^{(\mu),0,0}$ by using
Definition 2.1.1; in particular, $Y_{(0)}\Phi^{(\mu)}=\partial_r \Phi^{(\mu)}$ and $M_{(0)}\Phi^{(\mu)}=\partial_{\zeta}
\Phi^{(\mu)}$: by applying the operators $Y_{(0)}$ and $M_{(0)}$ to $\Phi^{(\mu),0,0}$, one retrieves
the whole series $\Phi^{(\mu)}=\sum\sum_{\xi,\sigma=0}^{\infty} \Phi^{(\mu),\xi,\sigma}(t) r^{\xi}\zeta^{\sigma}$.

Now $\Phi^{(\mu),0,0}$ may contain neither powers of $\partial\bar{b}^{\pm}$ (otherwise Theorem 3.2.1 gives
$L_{(2)}\Phi^{(\mu),0,0}\not=0$ and formula (\ref{DPOL}) proves that this is impossible) nor
powers of $a$, except, possibly, for fields of the type $(\bar{b}^+)^k a $ 
(otherwise Theorem 3.2.1 shows that  $Y_{(2)}\Phi^{(\mu),0,0}\not=0$ or $L_{(2)}\Phi^{(\mu),0,0}\not =0$ or
$L_{(3)}\Phi^{(\mu),0,0}
\not=0$,  and this is contradictory with formula (\ref{DPOL}) or (\ref{DPOY})).
Higher derivatives of the previous fields would yield higher order singularities in the
OPE with $L$ for instance.
Note also that powers of $\int a $ may be freely included under the previous conditions and entail
no supplementary constraint.

Hence (discarding fields such that $\Phi^{(\mu),0,0}$ is linear in $a$, which are not
very interesting, as one sees by considering the rather trivial action of the $\wsv$-fields
on them and their disappointingly simple 
$n$-point functions), one is led to consider the following family of fields,
where
we  make use of the {\it vertex operator} $V_{\alpha}:=
\exp \alpha \int a$ $(\alpha\in\R)$, see \cite{DiF97} for instance. Vertex
operators are known to be primary; with our normalization, $V_{\alpha}$
is $L$-primary with conformal weight $\frac{\alpha^2}{2}$.

{\bf Definition 3.2.3}

{\em  Set for $\al\in\C, j,k=0,1,\ldots$
\BEQ \phia_{j,k}(t)=:(\bar{b}^-)^j (\bar{b}^+)^k V_{\alpha}:(t) \EEQ
and 
\BEQ \phi_{j,k}(t) =\ _0 \phi_{j,k}(t)= :(\bar{b}^-)^j (\bar{b}^+)^k :(t). \EEQ
 }

All these fields appear to be the lowest-degree component fields of $\rho$-$\wsv$-primary
fields. The operator $\rho(Y_{\half})$ is trivial if $\al=0$; in the
contrary case, $\rho(M_1)$ may be  expressed as a coefficient times 
 $(\rho(Y_{\half}))^2$, in accordance with the discussion preceding Definition 1.4. Since
$\rho$ is quite different according to whether $\al\not=0$ or $\al=0$, and also for the sake
of clarity, we will state two different theorems. 

\bigskip

{\bf Theorem 3.2.4} (construction of the polynomial fields $\Phi_{j,k}$)

{\em

\begin{enumerate}
\item

Set 
\BEQ
\Phi^{(0),0}_{j,k}(t,\zeta)=\sum_{m=0}^j \left(\begin{array}{c} m\\ j\end{array}\right) \zeta^m :(\bar{b}^-)^{j-m} (\bar{b}^+)^{k+m}: (t)
\EEQ
and define inductively a series of fields $\Phi_{j,k}^{(\mu),\xi}$ ($\mu,\xi=0,1,2,\ldots$) by setting
\BEQ \Phi_{j,k}^{(\mu+1),\xi}(t,\zeta)=-\half :\partial_{\bar{b}^-}^2 \Phi_{j,k}^{(\mu),\xi}:(t,\zeta)
\label{ind1} \EEQ
and
\BEQ \Phi_{j,k}^{(\mu),\xi+1}(t,\zeta)=\frac{1}{1+\xi} :(a\partial_{\bar{b}}^- +\partial\bar{b}^+
\partial_a)\Phi_{j,k}^{(\mu),\xi}:(t,\zeta) \label{ind2} \EEQ
Then $\Phi_{j,k}^{(\mu)}=0$ for $\mu>[j/2]$ and 
\BEQ
\vec{\Phi}_{j,k}:=(\Phi_{j,k}^{(\mu)})_{0\le \mu\le[j/2]}, \quad \Phi_{j,k}^{(\mu)}(t,r,\zeta):=\sum_{\xi\ge 0} \Phi_{j,k}^{(\mu),\xi}(t,\zeta)
r^{\xi} \EEQ
defines a  $\rho$-$\langle N_0\rangle\ltimes\sv$-primary field,  $\rho$ being the representation of $\wsv_0$ defined by
\BEQ \rho(L_0)=-\left[ \frac{j+k}{2} \Id- \sum_{\mu=0}^{[j/2]} \mu E^{\mu}_{\mu} \right] \label{repL} \EEQ
\BEQ \rho(N_0)=-\left[ (k-j)\Id+ 2\sum_{\mu=0}^{[j/2]} \mu E^{\mu}_{\mu} \right] \label{repN} \EEQ
\BEQ \rho(Y_{\half})=0 \label{repY} \EEQ
\BEQ \rho(M_1)=\sum_{\mu=0}^{[j/2]-1} E^{\mu}_{\mu+1} \label{repM} \EEQ
where $E^{\mu}_{\nu}$ is the $([j/2]+1)\times ([j/2]+1)$ elementary matrix, with a single coefficient $1$
at the intersection of the $\mu$-th line and the $\nu$-th row. 

\item
Set $\vec{\Phi}=(\Phi_{j,k}^{(0)})_{j,k=0,1,\ldots}.$ Then $\vec{\Phi}$ is a $({\rho},\Omega)$-$\tilde{\sv}$-primary field if ${\rho},\Omega$ are defined as follows: 
\BEA
\rho(L_0)\Phi_{j,k}^{(0)}= -\frac{j+k}{2} \Phi_{j,k}^{(0)}; \nonumber\\
\rho(Y_{\half})\Phi_{j,k}^{(0)}=0;\quad \rho(M_1)\Phi_{j,k}^{(0)}=-\half j(j-1)\Phi_{j-2,k}^{(0)};\nonumber\\
\rho(N_0)\Phi_{j,k}^{(0)}=(j-k)\Phi_{j,k}^{(0)};\nonumber\\
\Omega\Phi_{j,k}^{(0)}=jk \Phi_{j-1,k-1}^{(0)}.
\EEA 

\end{enumerate}
}

{\bf Remarks.}

\begin{enumerate}
\item Both representations $\rho$ are of course the same; the passage from the first action on the $\Phi_{j,k}^{(\mu)}$ to the action on $\vec{\Phi}$ is given by the relation
\BEQ \Phi_{j,k}^{(\mu)}=(-\half)^{\mu} j(j-1)\ldots(j-2\mu+1)\Phi_{j-2\mu,k}^{(0)}.\EEQ
 The second case in the Theorem
 is an extension of the first one when one wants to consider covariance under all $N$-generators (not
only under $N_0$), which makes things more complicated.
\item
 Formally, one has
\BEQ \Phi_{j,k}^{(\mu)}=:\exp r(a\partial_{\bar{b}^-}+\partial\bar{b}^+
\partial_a).\Phi_{j,k}^{(\mu),0}: \EEQ
since $Y_{(0)}\equiv\partial_r\equiv a\partial_{\bar{b}^-}+\partial\bar{b}^+
\partial_a$ when applied to a polynomial $\sv$-primary  field of the form
$P(\bar{b}^{\pm},\partial\bar{b}^+,a).$ Hence, by the Campbell-Hausdorff formula
\BEQ \exp(A+B)=\exp \half [B,A] \exp A \exp B,\EEQ
valid if $[A,[A,B]]=[B,[A,B]]=0$, one
may also write 
\BEQ \Phi_{j,k}^{(\mu)}=:\exp ra\partial_{\bar{b}^-}  \exp \frac{r^2}{2} \partial\bar{b}^+
\partial_{\bar{b}^-} \ .\ \Phi_{j,k}^{(\mu),0}.\EEQ
\end{enumerate}

\vskip 1 cm

{\bf Proof.}

\begin{enumerate}

\item

First of all, $\Phi^{(\mu),\xi}$ is well-defined only because the operators $\partial_{\bar{b}^-}^2$
and $a\partial_{\bar{b}^-}+\partial\bar{b}^+\partial_a$ (giving the shifts $\mu\to\mu+1$ and
$\xi\to\xi+1$) commute. Let us check successively
the covariance under the action of $M,Y,N_0,L$. 

\begin{itemize}

\item

One finds from (\ref{MP})
\BEA
M_{(0)} \Phi^{(0),0}(t,\zeta) &=& \bar{b}^+\partial_{\bar{b}^-} \Phi^{(0),0}(t,\zeta)=
\sum_{m=0}^j \left(\begin{array}{c} j\\ m\end{array}\right) (j-m)\ : (\bar{b}^-)^{j-m-1}
 (\bar{b}^+)^{k+m+1}:\ (t)  \zeta^m  \nonumber\\
&=& \partial_{\zeta}\Phi^{(0),0}(t,\zeta); 
\EEA
\BEQ M_{(1)}\Phi^{(0),0}(t,\zeta)=\half\partial_{\bar{b}^-}^2 \Phi^{(0),0}(t,\zeta)=-\Phi^{(1),0}(t,\zeta)
\EEQ
which is coherent with formula (\ref{DPOM}) and the definition (\ref{repM}) of $\rho(M_1)$.
The field $\Phi_{j,k}$ is $M$-covariant if $\bar{b}^+\partial_{\bar{b}^-}\Phi^{(\mu),\xi}(t,\zeta)=
\partial_{\zeta} \Phi^{(\mu),\xi}(t,\zeta)$ for every $\mu,\xi\ge 0$. But this is true for $\mu,\xi=0$
and $[\bar{b}^+\partial_{\bar{b}^-},\partial_{\bar{b}^-}]=[\bar{b}^+\partial_{\bar{b}^-},
a\partial_{\bar{b}^-}+\partial\bar{b}^+ \partial_a]=0.$ Hence this is true for all values
of $\mu,\xi$ by induction.

\item

The action of $Y_{(0)}$ on $\Phi^{(\mu),\xi}$ is correct by definition - compare with  formulas (\ref{YP}) and (\ref{ind2}).
One has $Y_{(1)}\Phi^{(\mu),0}=0$ because $\partial_a \Phi^{(\mu),0}=0$, which is coherent with
(\ref{DPOY}) if one sets $\rho(Y_{\half})=0$. To prove that $Y_{(1)}\Phi^{(\mu),\xi}=\bar{b}^+
\partial_a \Phi^{(\mu),\xi}$ coincides with $\partial_{\zeta}\Phi^{(\mu),\xi-1}=\bar{b}^+\partial_{\bar{b}^-}\Phi^{(\mu),\xi-1}$, one uses induction on $\xi$ and the commutator relation
$[\bar{b}^+\partial_a,a\partial_{\bar{b}^-}+(\partial\bar{b}^+)\partial_a]=\bar{b}^+
\partial_{\bar{b}^-}.$ If this holds for some $\xi\ge 0$, then
\BEA
\bar{b}^+ \partial_a \Phi^{(\mu),\xi+1}&=& \frac{1}{1+\xi} (\bar{b}^+\partial_a)
(a\partial_{\bar{b}^-}+(\partial\bar{b}^+)\partial_a)\Phi^{(\mu),\xi} \nonumber\\
&=& \frac{1}{1+\xi} \left[ (a\partial_{\bar{b}^-}+(\partial\bar{b}^+)\partial_a)(\bar{b}^+
\partial_{\bar{b}^-})\Phi^{(\mu),\xi-1}+\bar{b}^+\partial_{\bar{b}^-}\Phi^{(\mu),\xi}\right] \nonumber\\
&=&  \frac{1}{1+\xi} \left[(\bar{b}^+
\partial_{\bar{b}^-}) (a\partial_{\bar{b}^-}+(\partial\bar{b}^+)\partial_a)
\Phi^{(\mu),\xi-1}+\bar{b}^+\partial_{\bar{b}^-}\Phi^{(\mu),\xi}\right] \nonumber\\
&=& \frac{1}{1+\xi} \left[ \xi \bar{b}^+ \partial_{\bar{b}^-} \Phi^{(\mu),\xi}+\bar{b}^+\partial_{\bar{b}^-} \Phi^{(\mu),\xi} \right] \nonumber\\
&=& \bar{b}^+ \partial_{\bar{b}^-}\Phi^{(\mu),\xi}
\EEA by (\ref{ind2}).

\item

One has  $Y_{(2)}\Phi^{(\mu),0}=0$ by (\ref{YP}) and, supposing that $Y_{(2)}\Phi^{(\mu),\xi}=
\partial_{\bar{b}^-}\partial_a\Phi^{(\mu),\xi}$ coincides with $-2\Phi^{(\mu+1),\xi-1}=-2\partial_{\zeta}
\Phi^{(\mu),\xi-1}= \partial_{\bar{b}^-}^2 \Phi^{(\mu),\xi-1}$ for some $\xi\ge 0$, then
\BEQ \partial_{\bar{b}^-}\partial_a \Phi^{(\mu),\xi+1}=\frac{1}{1+\xi} (\partial_{\bar{b}^-}
\partial_a)(a\partial_{\bar{b}^-}+\partial\bar{b}^+\partial_a)\Phi^{(\mu),\xi}=
\partial_{\bar{b}^-}^2 \Phi^{(\mu),\xi} \EEQ
by a proof along the same lines, since $[\partial_{\bar{b}^-}\partial_a,a\partial_{\bar{b}^-}
+\partial_{\bar{b}^+}\partial_a]=\partial_{\bar{b}^-}^2.$

\item

Since $N_{(0)}$ acts as $\bar{b}^+\partial_{\bar{b}^+}-\bar{b}^-\partial_{\bar{b}^-}$ on 
$\Phi^{(\mu),0}$, it simply measures the difference of degrees in $\bar{b}^+$ and $\bar{b}^-$
(for polynomial fields which depends only  on $\bar{b}^{\pm}$ and not on their derivatives). Hence one
sees easily that $N_{(0)}\Phi^{(\mu),0}=(2\zeta\partial_{\zeta}-j+k+2\mu)\Phi^{(\mu),0}$, which
is formula (\ref{DPON}). 
 Then $Y_{(0)}\equiv
a\partial_{\bar{b}^-}+(\partial\bar{b}^+)\partial_a$ increases by $1$ the eigenvalue
of $N_{(0)}$, see (\ref{NP}), which is also coherent with (\ref{DPON}). 

\item

There remains to check for the action of $L_{(i)}$, $i=2,3$. Supposing that $L_{(2)}\Phi^{(\mu),\xi}
=\bar{b}^+\partial_{\partial\bar{b}^+}\Phi^{(\mu),\xi}$ coincides with $\half\partial_{\zeta}
\Phi^{(\mu),\xi-2}=\half\bar{b}^+\partial_{\bar{b}^-}\Phi^{(\mu),\xi-2}$ for some $\xi\ge 0$,
then
\BEQ
\bar{b}^+\partial_{\partial\bar{b}^+}\Phi^{(\mu),\xi+1}=\frac{1}{1+\xi}\left[
\half(a\partial_{\bar{b}^-}+\partial\bar{b}^+\partial_a)(\bar{b}^+\partial_{\bar{b}^-})
\Phi^{(\mu),\xi-2}+\bar{b}^+\partial_a \Phi^{(\mu),\xi}\right];
\EEQ
since (as we just proved) $Y_{(1)}\Phi^{(\mu),\xi}=\bar{b}^+\partial_a \Phi^{(\mu),\xi}=\bar{b}^+\partial_{\bar{b}^-}
\Phi^{(\mu),\xi-1}$, one gets $L_{(2)}\Phi^{(\mu),\xi+1}=\half \partial_{\zeta}\Phi^{(\mu),\xi-1}$.

Finally, supposing that $2L_{(3)}\Phi^{(\mu),\xi}=(\partial_a^2+\partial_{\bar{b}^-}\partial_{\partial\bar{b}^+})\Phi^{(\mu),\xi}$ coincides with $-3\Phi^{(\mu)+1,\xi-2}=\frac{3}{2} \partial_{\bar{b}^-}^2
\Phi^{(\mu),\xi-2}$, then, using the commutator relation $[\partial_a^2+\partial_{\bar{b}^-}\partial_{\partial\bar{b}^+},a\partial_{\bar{b}^-}+(\partial\bar{b}^+)\partial_a]=3\partial_{\bar{b}^-}
\partial_a$ and the above equality $Y_{(2)}\Phi^{(\mu),\xi}=\partial_{\bar{b}^-}\partial_a
\Phi^{(\mu),\xi}=\partial_{\bar{b}^-}^2\Phi^{(\mu),\xi-1}$, one finds
\BEQ (\partial_a^2+\partial_{\bar{b}^-}\partial_{\partial\bar{b}^+})\Phi^{(\mu),\xi+1}=
\frac{1}{\xi+1}\left[ \frac{3}{2} (a\partial_{\bar{b}^+}+\partial\bar{b}^+\partial_a)
\partial_{\bar{b}^-}^2\Phi^{(\mu),\xi-2}+3\partial_{\bar{b}^-}\partial_a\Phi^{(\mu),\xi}\right]
=\frac{3}{2} \partial_{\bar{b}^-}^2 \Phi^{(\mu),\xi-1}.\EEQ

\end{itemize}

\item

First note that
\BEA
\partial_{\bar{b}^+}\partial_{\bar{b}^-}\Phi_{j,k}^{(0)}&=& \sum_{m\ge 0} (j-m)(k+m)
\left(\begin{array}{c} j\\ m \end{array}\right) (\bar{b}^-)^{j-m-1} (\bar{b}^+)^{k+m-1} \nonumber\\
&=& jk\sum_{m\ge 0} \left(\begin{array}{c} j-1\\ m\end{array}\right) (\bar{b}^-)^{j-m-1} (\bar{b}^+)^{k+m-1} \nonumber\\
&+& j(j-1)\sum_{m\ge 1} \left(\begin{array}{c} j-2\\ m-1\end{array}\right) (\bar{b}^-)^{j-m-1} 
(\bar{b}^+)^{k+m-1} \nonumber\\
&=& jk \Phi_{j-1,k-1}^{(0)} +\zeta j(j-1) \Phi_{j-2,k}^{(0)} \nonumber\\
&=& -\Omega\Phi_{j,k}^{(0)}-2\zeta \rho(M_1)\Phi_{j,k}^{(0)}
\EEA
by  Remark 1. following Theorem 3.2.4.

Hence one has identified the action of $N_{(1)}$ on $\Phi_{j,k}^{(0),0}$ as the correct one.
 Suppose now that
$N_{(1)}\Phi^{(0),\xi}=(\bar{b}^+\partial_{\partial\bar{b}^+}+\partial_{\bar{b}^+}
\partial_{\bar{b}^-})\Phi^{(0),\xi}$ coincides with $\half \partial_{\zeta}\Phi^{(0),\xi-2}+
\zeta j(j-1) \Phi_{j-2,k}^{(0),\xi}+jk \Phi_{j-1,k-1}^{(0),\xi}$ for some $\xi$. Then, by commuting
$\bar{b}^+ \partial_{\partial\bar{b}^+}+\partial_{\bar{b}^+}\partial_{\bar{b}^-}$ through
$Y_{(0)}=a\partial_{\bar{b}^-}+\partial\bar{b}^+ \partial_a$, one gets
\BEA && 
(\bar{b}^+\partial_{\partial_{\bar{b}^+}}+\partial_{\bar{b}^+}\partial_{\bar{b}^-})\Phi_{j,k}^{(0),\xi+1}=
\nonumber\\ && \frac{1}{1+\xi} \left[ (a\partial_{\bar{b}^-}+\partial\bar{b}^+\partial_a)(
 \half \partial_{\zeta}\Phi^{(0),\xi-2}+
\zeta j(j-1)\Phi_{j-2,k}^{(0),\xi}+jk\Phi_{j-1,k-1}^{(0),\xi})  + \bar{b}^+\partial_a \Phi^{(0),\xi}
\right] \nonumber\\ \EEA
 and $Y_{(1)}\Phi^{(0),\xi}=\bar{b}^+\partial_a \Phi_{j,k}^{(0),\xi}=\partial_{\zeta}
\Phi_{j,k}^{(0),\xi-1}$ as we have just proved, hence $N_{(1)}\Phi^{(0),\xi+1}$ is given by the
correct formula.

Finally, $N_{(2)}\Phi_{j,k}^{(0),\xi}=\partial_{\bar{b}^-}\partial_{\partial\bar{b}^+}\Phi^{(0),\xi}$
must be identified with $-\partial_{\zeta} \Phi_{j,k}^{(0),\xi-2}$ (which is certainly true for $\xi=0$). Supposing
this holds for some $\xi$,
\BEA
\partial_{\bar{b}^-}\partial_{\partial\bar{b}^+}\Phi^{(0),\xi+1}&=& \frac{1}{1+\xi}
(\partial_{\bar{b}^-}\partial_{\partial\bar{b}^+})(a\partial_{\bar{b}^-}+\partial\bar{b}^+
\partial_a)\Phi^{(0),\xi} \nonumber\\
&=& \frac{1}{1+\xi} \left[ -(a\partial_{\bar{b}^-}+\partial\bar{b}^+
\partial_a) \partial_{\zeta} \Phi^{(0),\xi-2}+\partial_{\bar{b}^-}\partial_a \Phi^{(0),\xi}\right]
\EEA
since $[\partial_{\bar{b}^-}\partial_{\partial\bar{b}^+},a\partial_{\bar{b}^-}+\partial\bar{b}^+
\partial_a]=\partial_{\bar{b}^-}\partial_a$; we now use the previous result $Y_{(2)}
\Phi^{(\mu),\xi}=\partial_{\bar{b}^-}\partial_a \Phi^{(\mu),\xi}=-2\partial_{\zeta} \Phi^{(0),\xi-1}$ and conclude by
induction.

\end{enumerate}

{\bf Theorem 3.2.5} (construction of the generalized polynomial fields $\Phia_{j,k}$)

{\em

\begin{enumerate}
\item

Set 
\BEQ
\Phia_{j,k}^{(0),0}(w,\zeta)=\sum_{m=0}^j \left(\begin{array}{c} j\\ m\end{array}\right) \zeta^m :(\bar{b}^-)^{j-m} (\bar{b}^+)^{k+m}
V_{\alpha}:
\EEQ
and define inductively a series of fields $\Phia^{(\mu),\xi}=
\Phia_{j,k}^{(\mu),\xi}$ ($\mu,\xi=0,1,2,\ldots$) by setting
\BEQ \Phia^{(\mu+1),\xi}(w,\zeta)=\frac{i}{\sqrt{2}} :\partial_{\bar{b}^-} \Phia^{(\mu),\xi}:(w,\zeta)
\label{ind1bis} \EEQ
and
\BEQ \Phia^{(\mu),\xi+1}(w,\zeta)=\frac{1}{1+\xi} :(a\partial_{\bar{b}}^- +\partial\bar{b}^+
\partial_a+\alpha\bar{b}^+)\Phia^{(\mu),\xi}:(w,\zeta) \label{ind2bis} \EEQ
Then 
\BEQ
\Phia_{j,k}:=(\Phia^{(\mu)})_{0\le i\le j}, \quad \Phia^{(\mu)}(t,r,\zeta)=\sum_{\xi\ge 0} \Phia^{(\mu),\xi}(w,\zeta)
r^{\xi} \EEQ
defines a  $\rho$-$\langle N_0\rangle\ltimes\sv$-primary  field,  $\rho$ being the representation of $\wsv_0$ defined by
\BEQ \rho(L_0)= - \left[ 
\frac{j+k+\alpha^2}{2} \Id -\half \sum_{\mu=0}^{j} \mu E_{\mu,\mu} \right]  \label{repLbis} \EEQ
\BEQ \rho(N_0)=-\left[ (k-j)\Id+\sum_{\mu=0}^{j} \mu E_{\mu,\mu} \right] \label{repNbis} \EEQ
\BEQ \rho(Y_{\half})=\II\alpha\sqrt{2} \sum_{\mu=0}^{j-1} E_{\mu,\mu+1} \label{repYbis} \EEQ
\BEQ \rho(M_1)=-\half (\frac{1}{\alpha} \rho(Y_{\half}))^2=
\sum_{\mu=0}^{j-2} E_{\mu,\mu+2} \label{repMbis} \EEQ

\item

 Set $\vec{\Phia}=(\Phia_{j,k}^{(0)})_{j,k=0,1,\ldots}.$ Then $\vec{\Phia}$ is a $({\rho},\Omega)$-$\tilde{\sv}$-primary field if ${\rho},\Omega$ are defined as follows: 
\BEA
\rho(L_0)\Phia_{j,k}^{(0)}= -\frac{j+k}{2} \Phia_{j-2,k}^{(0)}; \nonumber\\
\rho(Y_{\half})\Phi_{j,k}^{(0)}=-\alpha j\Phi_{j-1,k}^{(0)};\quad \rho(M_1)\Phia_{j,k}^{(0)}=-\half j(j-1)\Phia_{j-2,k}^{(0)};\nonumber\\
\rho(N_0)\Phia_{j,k}^{(0)}=(j-k)\Phia_{j,k}^{(0)};\nonumber\\
\Omega\Phia_{j,k}^{(0)}=jk \Phia_{j-1,k-1}^{(0)}.
\EEA 

\end{enumerate}
}

{\bf Remark.}
\begin{itemize}
\item The coherence between the two representations is given this time by:
\BEQ \Phia_{j,k}^{(\mu)}=\left(\frac{\II}{\sqrt{2}}\right)^k  j(j-1)\ldots(j-k+1)\Phi_{j-\mu,k}^{(0)}.\EEQ

\item  One may write formally 
\BEA \Phia_{j,k}^{(\mu)} &=& :\exp r\left( a\partial_{\bar{b}^-} +\partial \bar{b}^+ \partial_a+
\alpha\bar{b}^+ \right) \ . \Phi_{j,k}^{(\mu),0} : \nonumber\\
&=& :\exp \alpha r \bar{b}^+ \ . \ \exp ra\partial_{\bar{b}^-}\ .\ \exp \frac{r^2}{2} \partial\bar{b}^+
\partial_{\bar{b}^-}\Phi_{j,k}^{(\mu),0} \ : 
 \EEA
\end{itemize}

{\bf Proof.}

The proof is almost the same, with just a few modifications. We shall follow
the  proof of Theorem 3.2.4 line by line and rewrite only what has to be changed. 
\begin{itemize}
\item
One has $Y_{(1)}\Phia^{(\mu),0}(\zeta)=\alpha\partial_{\bar{b}^-}\Phia^{(\mu),0}(\zeta),$ 
to be identified with $-\rho(Y_{\half})^{\mu}_{\nu} \Phia^{(\nu),0}$. Hence one must set, in accordance
with (\ref{repYbis})
\BEQ \Phia^{(\mu)+1,0}=\frac{i}{\sqrt{2}} \partial_{\bar{b}^-} \Phia^{(\mu),0} \EEQ
so $\Phia^{(\mu)+2,0}=-\half\partial_{\bar{b}^-}^2 \Phia^{(\mu),0}$ as in Theorem 3.2.4, with a double
shift instead in the indices $i$. 

Suppose now $Y_{(1)}\Phia^{(\mu),\xi}=(\bar{b}^+\partial_a+\alpha\partial_{\bar{b}^-})\Phia^{(\mu),\xi}$
coincides with $\partial_{\zeta}\Phia^{(\mu),\xi-1}-i\alpha\sqrt{2}\Phia^{(\mu)+1,\xi}=
\bar{b}^+\partial_{\bar{b}^-}\Phia^{(\mu),\xi-1}-i\alpha\sqrt{2}\Phia^{(\mu)+1,\xi}:$ then
the commutator relation $[\bar{b}^+\partial_a+\alpha\partial_{\bar{b}^-},a\partial_{\bar{b}^-}
+\partial\bar{b}^+\partial_a+\alpha\bar{b}^+]=\bar{b}^+\partial_{\bar{b}^-}$ yields
\BEA
&& Y_{(1)}\Phia^{(\mu),\xi+1}=\\
&&  \frac{1}{\xi+1} \left\{ (a\partial_{\bar{b}^-}+(\partial\bar{b}^+)
\partial_a+\alpha\bar{b}^+)(\bar{b}^+\partial_{\bar{b}^-}\Phia^{(\mu),\xi-1}-i\alpha\sqrt{2}
\Phia^{(\mu)+1,\xi})+\bar{b}^+\partial_{\bar{b}^-}\Phia^{(\mu),\xi}\right\} \nonumber\\
&&= \bar{b}^+\partial_{\bar{b}^-}\Phia^{(\mu),\xi}-i\alpha\sqrt{2}\Phi^{(\mu)+1,\xi+1}
\EEA
Then $Y_{(2)}=\partial_{\bar{b}^-}\partial_a$ and $[\partial_{\bar{b}^-}\partial_a,Y_{(0)}]=
\partial_{\bar{b}^-}^2$ as in Theorem 3.2.4, so  covariance under $Y_{(2)}$ holds true.

\item
The action of $N_{(0)}, N_{(1)}, N_{(2)}$  on $\Phia^{(\mu)}$ or
$\Phia_{j,k}^{(0)}$ is exactly as in Theorem 3.2.4 since $N=-:\bar{b}^+\bar{b}^-:$ does not
involve neither the free boson $a$ nor its integral.

\item
One must still check for $L_{(2)}$ (nothing changes for $L_{(3)}$). Suppose that
$L_{(2)}\Phia^{(\mu),\xi}=\bar{b}^+\partial_{\partial\bar{b}^+}+\alpha\partial_a$ coincides
with $\half \partial_{\zeta}\Phia^{(\mu),\xi-2}-\II \alpha\sqrt{2} \Phia^{(\mu+1),\xi-1}=\half \bar{b}^+
\partial_{\bar{b}^-}\Phia^{(\mu),\xi-2}-\II \alpha\sqrt{2} \Phia^{(\mu)+1,\xi-1}$. Then, using
\BEQ [ \bar{b}^+\partial_{\partial\bar{b}^+}+\alpha\partial_a, a\partial_{\bar{b}^-}+(\partial\bar{b}^+)
\partial_a+\alpha\bar{b}^+]\Phia^{(\mu),\xi}= (\bar{b}^+\partial_a+\alpha\partial_{\bar{b}^-})
\Phia^{(\mu),\xi}=\partial_{\zeta}\Phia^{(\mu),\xi-1}- \II \alpha\sqrt{2} \Phi^{(\mu+1),\xi} \EEQ
(see computation of $Y_{(1)}\Phia^{(\mu),\xi}$ above) one gets
\BEA
(\bar{b}^+\partial_{\partial\bar{b}^+}+\alpha\partial_a)\Phia^{(\mu),\xi+1}&=&
\frac{1}{1+\xi} \big[ (a\partial_{\bar{b}^-}+(\partial\bar{b}^+)
\partial_a+\alpha\bar{b}^+)(\half \bar{b}^+
\partial_{\bar{b}^-}\Phia^{(\mu),\xi-2}- \II \alpha\sqrt{2} \Phia^{(\mu)+1,\xi-1}) \nonumber\\
&+&\partial_{\zeta}\Phi^{(\mu),\xi-1}- \II \alpha\sqrt{2} 
\Phia^{(\mu)+1,\xi}\big] \nonumber\\ &=& \half \bar{b}^+\partial_{\bar{b}^-}
\Phia^{(\mu),\xi-1}-\II \alpha \sqrt{2} \Phia^{(\mu)+1,\xi}.
\EEA

\end{itemize}

\hfill \eop

We shall now start computing explicitly the simplest $n$-point functions of the $\tilde{\sv}$-primary
fields we have just defined.


\section{Correlators of the 
polynomial and generalized polynomial  fields}


We obtain below the two-point functions of the generalized polynomial fields $_\alpha \Phi_{j,k}$ (see
Propositions 4.1 and 4.2) and the three-point functions in the case $\alpha=0$, see Proposition 4.3 (computations are much more involved in the case $\alpha\not=0$).

{\bf Proposition 4.1} (computation of the two-point functions when $\alpha=0$)

{\em
Set $t=t_1-t_2,r=r_1-r_2,\zeta=\zeta_1-\zeta_2$ for the differences of coordinates. Then
the two-point function $${\cal C}(t_1,r_1,\zeta_1;t_2,r_2,\zeta_2):=
 \langle 0\ |\ \Phi^{(0)}_{j_1,k_1}(t_1,r_1,\zeta_1) \Phi^{(0)}_{j_2,k_2}(t_2,r_2,\zeta_2)\ |\ 0
\rangle$$ is equal to
\BEQ {\cal C}(t,r,\zeta)=\del_{j_1+k_1,j_2+k_2} (-1)^{k_2} j_1! \ k_1!\ \left(\begin{array}{c} j_2\\ k_1 \end{array}\right)
t^{-(j_1+k_1)} (\zeta-\frac{r^2}{2t})^{j_1-k_2} \EEQ
if $j_1\ge k_2$ and $0$ else. }

{\bf Proof.}

We use the covariance of  $\cal C$ under the finite subalgebra ${\rho}_{j,k}(L_{\pm
1,0})$, ${\rho}_{j,k}(Y_{\pm\half})$, ${\rho}_{j,k}(N_0)$. In particular, $\cal C$ is
a function of the differences of coordinates $t,r,\zeta$ only. Covariance under
\BEQ 
{\rho}(L_0)=-\sum_{i=1}^2 (t_i\partial_{t_i}+\half r_i\partial_{r_i})
-\half \sum_{i=1}^2 (j_i+k_i),\quad {\rho}(Y_{\half})=-\sum_{i=1}^2 t_i\partial_{r_i}+
r_i\partial_{\zeta_i}, \EEQ
\BEQ  {\rho}(L_1)=-\sum_{i=1}^2 t_i^2\partial_{t_i}+t_i
r_i\partial_{r_i}+\frac{r_i^2}{2}\partial_{\zeta_i} - \sum_{i=1}^2 t_i(j_i+k_i)
\EEQ
 yields quite generally  (see \cite{Hen94},
\cite{HenUnt03})
\BEQ {\cal C}=C.\del_{j_1+k_1,j_2+k_2} t^{-(j_1+k_1)} f(\zeta-\frac{r^2}{2t}) \EEQ
for some function $f$.

Suppose $j_1+k_1=j_2+k_2.$ Assuming the extra covariance under $\tilde{\rho}_k(N_0)=-
\sum_{i=1}^2 (r_i\partial_{r_i}+2\zeta_i\partial_{\zeta_i})+(j_1-k_1)+(j_2-k_2)\equiv
-(r\partial_r+2\zeta\partial_{\zeta})+2(j_1-k_2),$ one gets $vf'(v)=(j_1-k_2)f$, hence
$f(v)=v^{j_1-k_2}$ up to a constant (see also Proposition A.1 in Appendix A). The coefficient $(-1)^{k_2} j_1 ! k_1 ! \left(\begin{array}{c} j_2\\ k_1 \end{array}\right)$
may be obtained from   the coefficient $C$ of the term of highest degree in $t$ (i.e. the least
singular term in $t$) in the formal series in $r_{1,2},\zeta_{1,2}$. Since 
$Y_{(0)}\equiv a\partial_{\bar{b}^-}+\partial\bar{b}^+\partial_a$ maps an 
$L$-quasiprimary field
of weight, say, $\lambda$ into an $L$-quasiprimary field of weight $\lambda+\half$, it is clear that
$C$ can be read from 
\BEA {\cal C}_0 &=& \langle 0\ |\ 
 \Phi^{(0),0}(t_1,r_1,\zeta_1)\Phi^{(0),0}(t_2,r_2,\zeta_2)\ |\ 0\rangle \nonumber\\
&=&\sum_{m_1,m_2} \zeta^{m_1+m_2}
 \left(\begin{array}{c} j_1 \\ m_1\end{array}\right) \left(\begin{array}{c}  j_2 \\ m_2\end{array}\right) \langle 0\ |\ :(\bar{b}^-)^{j_1-m_1} (\bar{b}^+)^{k_1+m_1}:(t_1)\ :
(\bar{b}^-)^{j_2-m_2} (\bar{b}^+)^{k_2+m_2}:(t_2) \ |\ 0\rangle  \nonumber\\ \EEA
For the same reason, ${\cal C}_0$ must be equal to $Ct^{-(j_1+k_1)}(\zeta_1-\zeta_2)^{j_1-k_2}$.
One gets immediately $C=0$ for $j_1<k_2$. In the contrary case, one gets $C$ by looking
for the coefficient of $\zeta_2^{j_1-k_2}$, which is given by
\BEA && (-1)^{j_1-k_2} \langle 0\ |\ \left( : (\bar{b}^-)^{j_1}(\bar{b}^+)^{k_1}:(t_1)\right) \left( 
\left(\begin{array}{c} j_2 \\j_1-k_2\end{array}\right) :
(\bar{b}^-)^{j_2-(j_1-k_2)} (\bar{b}^+)^{k_2+(j_1-k_2)}:(t_2)\right)\ |\ 0\rangle \nonumber\\  && =
(-1)^{k_2} t^{-(j_1+k_1)} \left(\begin{array}{c} j_2\\ k_1 \end{array}\right) j_1 ! k_1 !.
\nonumber\\
\EEA
\hfill\eop

{\bf Proposition 4.2} (computation of the two-point functions when $\alpha\not=0$)

{\em
Set $t=t_1-t_2,r=r_1-r_2,\zeta=\zeta_1-\zeta_2$ for the differences of coordinates. Write  
\BEQ {\cal C}_{(\alpha_1,j_1,k_1),(\alpha_2,j_2,k_2)}^{\mu_1,\mu_2}:=
 \langle 0\ |\  _{\alpha_1}\Phi^{(\mu_1)}_{j_1,k_1}(t_1,r_1,\zeta_1)\ _{\alpha_2}
 \Phi^{(\mu_2)}_{j_2,k_2}(t_2,r_2,\zeta_2)\ |\ 0
\rangle. 
\EEQ
 Then: 
\begin{itemize}
\item[(i)]
the two-point functions vanish unless  $\alpha_1=-\alpha_2$ and
$j_1\ge k_2$ and $j_2\ge k_1$;
\item[(ii)]
suppose that $j_1=j_2:=j,\ k_1=k_2=0$ and $\alpha:=\alpha_1=-\alpha_2$. Then
\BEQ
{\cal C}_{(\alpha,j,0),(-\alpha,j,0)}^{\mu_1,\mu_2}
= t^{-j-\alpha^2+\frac{\mu_1+\mu_2}{2}} \sum_{\del=\max(\mu_1,\mu_2)}^j
c_j^{\del} \frac{(i\alpha\sqrt{2})^{\del-\mu_1} (-i\alpha\sqrt{2})^{\del-\mu_2}}
{(\del-\mu_1)! (\del-\mu_2)!} (\frac{r^2}{t})^{\del-\frac{\mu_1+\mu_2}{2}} 
(\zeta-\frac{r^2}{2t})^{j-\del} \label{2ptfbis}
\EEQ
where $c_j^{\del}=(-1)^{\del}\frac{(j!)^2}{2^{\del}(j-\del)!}.$

\end{itemize}
 }

{\bf Remark.}
All the other cases may be deduced easily from formula (\ref{2ptfbis})
since, if $j_1\ge k_2$ and $j_2\ge k_1$ and (without loss of generality)
$(j_2+k_2)-(j_1+k_1)=\Del\ge 0$, 
\BEA
&& {\cal C}_{(\alpha,j_1,k_1),(-\alpha,j_2,k_2)}^{\mu_1,\mu_2} \nonumber\\
&& = \left( k_1! \left(\begin{array}{c} j_1+k_1\\ k_1\end{array}\right) \ .\ k_2!
 \left(\begin{array}{c} j_2+k_2 \\ k_2 \end{array}\right) \right)^{-1}
\langle 0\ |\ \left[ (\bar{b}^+\partial_{\bar{b}^-})^{k_1} 
\Phia_{j_1+k_1,0}^{(\mu_1)}\right] 
 \left[ (\bar{b}^+\partial_{\bar{b}^-})^{k_2} 
\Phima_{j_2+k_2,0}^{(\mu_2)}\right] \ |\  0 \rangle  \nonumber \\
&&=  \left( k_1!  \left(\begin{array}{c} j_1+k_1\\ k_1\end{array}\right)  \ .\
 k_2!     \left(\begin{array}{c} j_2+k_2 \\ k_2 \end{array}\right)   \ .\ \Del !
\left(\begin{array}{c} j_1+k_1+\Del \\ \Del\end{array}\right)
\right)^{-1} \nonumber \\ 
&& \langle 0\ |\ \left[ \partial_{\bar{b}^-}^{\Del} (\bar{b}^+\partial_{\bar{b}^-})^{k_1} 
\Phia_{j_2+k_2,0}^{(\mu_1)}\right] 
 \left[ (\bar{b}^+\partial_{\bar{b}^-})^{k_2} 
\Phima_{j_2+k_2,0}^{(\mu_2)}\right] \ |\  0 \rangle \nonumber\\
&&= 
\frac{(i\sqrt{2})^{\Del}}{
 k_1!  \left(\begin{array}{c} j_1+k_1 \\ k_1 \end{array}\right)
\ .\ k_2!   \left(\begin{array}{c} j_2+k_2 \\ k_2 \end{array}\right)  \ .\ 
 \Del !  \left(\begin{array}{c} j_1+k_1+\Del \\ \Del  \end{array}\right)
} 
\partial_{\zeta_1}^{k_1}\partial_{\zeta_2}^{k_2} {\cal C}_{(\alpha,
j_2+k_2,0),(-\alpha,j_2+k_2,0)}^{\mu_1+\Del,\mu_2} \nonumber\\
\EEA
thanks to the fact that $\partial_{\zeta} \Phia_{j,k}^{(\mu)}=\bar{b}^+
\partial_{\bar{b}^-} \Phia_{j,k}^{(\mu)}$ and 
$\Phia_{j,k}^{(\mu+1)}=-\frac{i}{\sqrt{2}} \partial_{\bar{b}^-} 
\Phia_{j,k}^{(\mu)}$. 

{\bf Proof.}

We only prove (ii) since (i) is clear from the preceding computations.
Applying Proposition B.1 from Appendix B, with $d=j+1$, $\lambda_{1,2}=
\frac{\alpha^2+j}{2}$,
$\alpha_{1,2}=\pm i\alpha\sqrt{2}$ and $\lambda'_{1,2}=-j$, one gets
(\ref{2ptfbis}). There remains to find the coefficients $c_{\del}^j$.
Let us first explain how to find $c_j^j$. One has
\BEA
{\cal C}^{j,j}&=&\langle 0\ |\ \Phia_{j,0}^{(j)} \Phima_{j,0}^{(j)} \ |
\ 0\rangle = c_j^j t^{-\alpha^2} \nonumber \\
&=& (\frac{i}{\sqrt{2}})^{2j} \langle 0\ |\ \left( (\partial_{\bar{b}^-})^j
\Phia_{j,0}^{(0)}\right) \left( (\partial_{\bar{b}^-})^j \Phima_{j,0}^{(0)}\right) \ |
\ 0\rangle \ {\mathrm{by\ Theorem\ 3.2.5}} \nonumber \\
&=& (-1)^j 2^{-j} (j!)^2 
\langle 0\ |\ 
\Phia_{0,0}^{(0)}  \Phima_{0,0}^{(0)} \ |
\ 0\rangle \nonumber \\
&=& (-1)^j  2^{-j} (j!)^2 
\langle 0\ |\ 
:\exp\alpha r_1\bar{b}^+ V_{\alpha}(t_1): \ :\exp -\alpha r_2\bar{b}^+ V_{-\alpha}(t_2): \ |
\ 0\rangle \nonumber \\
&=& (-1)^j 2^{-j} (j!)^2 t^{-\alpha^2}.
\EEA

By the same trick, one gets (by deriving $j-\eps$ times with respect to
$\bar{b}^-$)
\BEA
{\cal C}^{j-\eps,j-\eps}&=& (c_j^{j-\eps}(\zeta-\frac{r^2}{2t})^{\eps}+
O(r)) t^{-\eps-\alpha^2} \nonumber\\
&=& (-1)^{j-\eps} 2^{-(j-\eps)} (j(j-1)\ldots(\eps+1))^2 
\langle 0\ |\ 
\Phia_{\eps,0}^{(0)}  \Phima_{\eps,0}^{(0)} \ |
\ 0\rangle 
\EEA
and one may identify the lowest degree component in $r$ -- which does {\it not}
depend on $\alpha$, up to a multiplication by the factor $t^{-\alpha^2}$ -- by setting $r_1=r_2$,
\BEA
{\cal C}^{j-\eps,j-\eps}(r_1=r_2)&=& c_j^{j-\eps} \zeta^{\eps} t^{-\eps-\alpha^2} \nonumber\\
&=& (-1)^{j-\eps} 2^{-(j-\eps)} \left( \frac{j!}{\eps !}\right)^2 t^{-\alpha^2}
 \langle 0\ |\ 
\Phi_{\eps,0}^{(0)}  \Phi_{\eps,0}^{(0)} \ |
\ 0\rangle 
\nonumber\\ &=&  (-1)^{j-\eps} 2^{-(j-\eps)} \left( \frac{(j!)^2}{\eps !}\right) 
\zeta^{\eps} t^{-\eps-\alpha^2}
\EEA
by Proposition 4.1. \hfill \eop

{\bf Proposition 4.3} (computation of the three-point functions when $\alpha=0$)

{\em

The following formula holds:

\BEA && \langle \Phi^{(0)}_{j_1,0}(t_1,r_1,\zeta_1)\Phi^{(0)}_{j_2,0}(t_2,r_2,\zeta_2)\Phi^{(0)}_{j_3,0}(t_3,r_3,\zeta_3)\rangle 
= \frac{j_1 ! j_2 ! j_3 !}{(\half(j_1+j_3-j_2))! (\half(j_2+j_3-j_1))! (\half(j_1+j_2-j_3))!} \nonumber\\
&& \left(\frac{\xi_{12}}{t_{12}}\right)^{\half(j_1+j_2-j_3)} 
 \left(\frac{\xi_{13}}{t_{13}}\right)^{\half(j_1+j_3-j_2)} 
\left(\frac{\xi_{23}}{t_{23}}\right)^{\half(j_2+j_3-j_1)} 
\EEA

where $\xi_{ij}:=\zeta_{ij}-\frac{r_{ij}^2}{2t_{ij}}.$
}

{\bf Remark.}

All three-point correlators for the case $\alpha=0$ can be obtained easily from these results by
applying a number of times the operator $\bar{b}^+\partial_{\bar{b}^-}$ or equivalently $\partial_{\zeta}$.

{\bf Proof.}

Denote by ${\cal C}(t_i,r_i,\zeta_i)= \langle \Phi_{j_1}(t_1,r_1,\zeta_1)\Phi_{j_2}(t_2,r_2,\zeta_2)\Phi_{j_3}(t_3,r_3,\zeta_3)\rangle$ the three-point function. 
By Theorem A.3, 
\BEQ {\cal C}=C t_{12}^{-\alpha} t_{23}^{-\beta} t_{13}^{-\gamma} (\xi_{12}^{\alpha} \xi_{23}^{\beta} \xi_{31}^{\gamma}+\Gamma(\xi_{12},\xi_{13},\xi_{23})) \EEQ
where $C$ is a constant, $\alpha=\frac{j_1+j_2-j_3}{2},\beta=\frac{j_1+j_3-j_2}{2},\gamma=\frac{j_2+j_3-j_1}{2}$, and
 $\Gamma=\Gamma(\xi_{12},\xi_{23},\xi_{13})$ is any linear combination (with constant coefficients)
of monomials $\xi_{12}^{\alpha'} \xi_{23}^{\beta'}\xi_{31}^{\gamma'}$ with $(\alpha',\beta',\gamma')\not=(\alpha,\beta,\gamma)$ and $\alpha'+\beta'+\gamma'=\frac{J}{2}.$
Suppose $t_3\not=t_1,t_2$ and look at the degree of the pole in $\frac{1}{t_{12}}$ in ${\cal C}$ considered as a function of $t_1,t_2,t_3$ and $\zeta_1,\zeta_2,\zeta_3$. Each term in the asymptotic expansion
of $\Phi_{j_i}$ in powers of $r_i,\zeta_i$ is a polynomial of degree $j_i$ in the fields $a,\bar{b}^-,
b^+,\partial_{\bar{b}^+}$. The covariance ${\cal C}=\langle \Phi_1 \Phi_2\Phi_3\rangle$ may be computed
as any polynomial of Gaussian variables by using Wick's theorem; calling $a_{ij}$ the number of
couplings of $\Phi_i$ with $\Phi_j$, an easy argument yields $j_1=a_{12}+a_{13},j_2=a_{12}+a_{23},
j_3=a_{13}+a_{23}$, hence in particular $a_{12}=\alpha.$ 
Hence ${\cal C}$ has a pole in $\frac{1}{t_{12}}$ of degree at most $2\alpha$ and $\Gamma$ may not
contain any term of the type   $\xi_{12}^{\alpha'} \xi_{23}^{\beta}\xi_{31}^{\gamma'}$ with $\alpha'>\alpha.$ By taking into
account the poles in $\frac{1}{t_{23}}$ and $\frac{1}{t_{13}}$, one sees that $\Gamma=0$.

There remains to compute the coefficient $C$. By rewriting ${\cal C}$ as
\BEQ {\cal C}=\sum_{\alpha'+\beta'+\gamma'=J/2} C_{\alpha',\beta',\gamma'} \xi_{12}^{\alpha'}
\xi_{13}^{\beta'} (\xi_{12}+\xi_{23}+\xi_{31})^{\gamma'},\EEQ
and using $\xi_{31}=(\xi_{12}+\xi_{23}+\xi_{31})-\xi_{12}-\xi_{23}$, one sees that
$C=C_{\alpha,\beta,\gamma}.$
 Now a minute's thought shows that the coefficient of 
\BEQ (\zeta_1^0 \zeta_2^{\beta} \zeta_3^{\gamma})(r_1^0 r_2^0 r_2^{2\gamma}) t_{23}^{-2\gamma}
 t_{12}^{-\alpha} t_{13}^{-\beta}  \EEQ
 in ${\cal C}$ is equal to $(-1)^{J/2} 2^{-\gamma} C_{\alpha,\beta,\gamma}$.
hence (using the asymptotic expansion of $\Phi_1$, $\Phi_2$ and $\Phi_3$ in powers of $\zeta_i,r_i$)
$C_{\alpha,\beta,\gamma}$ may be computing by extracting the coefficient of $t_{23}^{-2\gamma} t_{12}^{-\alpha} t_{13}^{-\beta}$
in $$\langle 0\ |\ :(\bar{b}^-)^{j_1}:(t_1) : \left(\begin{array}{c} j_2\\ \alpha\end{array}\right)
(\bar{b}^-)^{j_2-\alpha} (\bar{b}^+)^{\alpha}:(t_2) :\frac{(a\partial_{\bar{b}^-}+\partial\bar{b}^+ \partial_a)^{2\gamma}}{(2\gamma)!} \left(\begin{array}{c} j_3 \\ \beta\end{array}\right)
(\bar{b}^-)^{j_3-\beta} (\bar{b}^+)^{\beta} : (t_3) \ |0\rangle,$$
and multiplying by $(-1)^{J/2} 2^{-\gamma}$. Now the coefficient of $r^{2\gamma}$ in
$\exp rY_{(0)}\ .\ \left( (\bar{b}^-)^{j_3-\beta} (\bar{b}^+)^{\beta} \right)$ is equal to

\BEA
&&\sum_{i+2j=2\gamma} r^{2\gamma} \frac{(a\partial_{\bar{b}^-})^i}{i!} \left[ 2^{-j}
 \left(\begin{array}{c} j_3-\beta \\ j \end{array}\right)
(\bar{b}^-)^{j_3-\beta-j} (\partial\bar{b}^+)^{j} (\bar{b}^+)^{\beta} \right] \nonumber\\
&&=\sum_{i+2j=2\gamma} r^{2\gamma} 2^{-j} \frac{(j_3-\beta)!}{j!i!(j_3-\beta-i-j)!} (\partial\bar{b}^+)^j
a^i (\bar{b}^+)^{\beta}.
\EEA

The terms with $i>0$ do not contribute to $C_{\alpha,\beta,\gamma}$ since $a$ can only
be found in the field with the variable $t_3$ and does not couple to the other fields. Hence 
\BEA
C_{\alpha,\beta,\gamma}  t_{23}^{-2\gamma} t_{12}^{-\alpha} t_{13}^{-\beta} &=&
(-1)^{J/2}
\left(\begin{array}{c}  j_3 \\ \beta \end{array}\right)  \frac{(j_3-\beta)!}{\gamma!(j_3-\beta-\gamma)!}
 \left(\begin{array}{c} j_2 \\ j_2-\alpha\end{array}\right) 
\nonumber \\ &&
\langle 0 \ |\ :(\bar{b}^-)^{j_1}:(t_1) :(\bar{b}^-)^{j_2-\alpha} (\bar{b}^+)^{\alpha}:(t_2)
:(\partial\bar{b}^+)^{J/2-\alpha-\beta} (\bar{b}^+)^{\beta}:(t_3)\ |\ 0\rangle. \nonumber\\
\EEA

The first field $(\bar{b}^-)^{j_1}$ couples $\alpha$ times (resp. $\beta$ times)  with the second (resp. third) fields, yielding $(t_{12})^{-\alpha}(t_{13})^{-\beta}$ times
\BEQ
\left(\begin{array}{c} j_1 \\ \alpha \end{array}\right) \alpha! (-1)^{\alpha} \beta! (-1)^{\beta} .\EEQ
There remains the  coupling of the second and third fields, namely,
\BEQ \langle 0\ |\ :(\bar{b}^-)^{j_2-\alpha}(t_2) :(\partial\bar{b}^+)^{J/2-\alpha-\beta}:(t_3)\EEQ
which yields $t_{23}^{-2\gamma}$ times 
$\left(\begin{array}{c} j_2-\alpha \\ J/2-\alpha-\beta \end{array}\right) (J/2-\alpha-\beta)!
(-1)^{j_2-\alpha}.$

All together one gets 
\BEQ C_{\alpha,\beta,\gamma}= \frac{j_1 ! j_2 ! j_3 !}{(\half(j_1+j_3-j_2))! (\half(j_2+j_3-j_1))! (\half(j_1+j_2-j_3))!}.\EEQ
Hence the result.

\hfill \eop


\section{Construction of the massive fields}


All the fields constructed until now involve only {\it polynomials} in 
the unphysical variable $\zeta$. Inverting the Laplace transform
${\cal L}: f_{\cal M} \to {\cal L}f(\zeta)=\int_0^{\infty} f_{\cal M} e^{{\cal M}\zeta}\
 d{\cal M}$ is a priori impossible since polynomials in $\zeta$
only give derivatives of the delta-function $\del_{\cal M}$; one may say that these fields
represent singular zero-mass fields, which are a priori irrelevant from a physical point of view.

  However, we believe it is possible to construct {\it massive fields} by combining the above
polynomial fields into a formal series depending on a parameter $\Xi$ 
and taking an analytic continuation, whose status is yet unclear. Let us formalize
this as a conjecture:

{\bf Conjecture:}

{\it Massive fields may be obtained as an analytic continuation for $\Xi\to 0$ of  series 
in $\Phi_{j,k}$, $\Phia_{j,k}$ of the form
\BEQ  \Xi^{\lambda} \sum_{j,k\ge 0} a_{j,k} \Xi^{-\frac{j+k}{2}} \Phi_{j,k}(t,r,\zeta)
\quad {\mathrm{or}}\   \Xi^{\lambda} \sum_{j,k\ge 0} a_{j,k} \Xi^{-\frac{j+k}{2}} \Phia_{j,k}(t,r,\zeta)
\EEQ
for some $\lambda$, with a non-zero radius of convergence in $\Xi^{-1}$.
}

The idea lying behind this is that the discrepancy in the scaling behaviours in $t$ of the fields
$\Phi_{j,k}$ (namely, $\Xi^{-\frac{j+k}{2}} \Phi_{j,k}(t)$ behaves as $(\Xi t)^{-\frac{j+k}{2}}$ when
$t\to\infty$ since $\Phi_{j,k}$ has $L_0$-weight $\frac{j+k}{2}$) disappears in the above sums in the 
limit $\Xi\to 0$. As for the appearance of a {\it massive} behaviour in the limit $\Xi\to 0$, it is reminiscent of the construction of the coherent state $e^{{\cal M} a^{\dag}}|0\rangle$, an eigenvector of the annihilation
operator $a$ in the theory of the harmonic oscillator. We hope to make this analogy more precise in the future.

We introduce in Theorem 5.1 and Theorem 5.2 below good potential candidates for massive fields. Theorems 5.1, 5.2 and 5.3 show that
all two-point functions and (at least) some three-point functions may indeed be analytically
extended, and give  explicit expressions for the corresponding $n$-point functions of the
{\it would-be} massive field. The missing part in the picture is a formal proof that all $n$-point
functions have an analytic extension to $\Xi\to 0$. An encouraging fact is that the limit for $\Xi\to 0$ does not seem
to depend (up to a physically irrelevant overall coefficient depending only on the mass) on the
precise asymptotic series.  

We made some attempts to prove the existence of the desired  analytic extension by constructing the  $n$-point functions as solutions of differential equations coming from the symmetries  (for instance, the two-point
function $\langle \psi_{-1}^{\Xi} \psi_{-1}^{\Xi}\rangle$, see below, may be computed -- up to a constant --
by using the covariance under $\sch_1$ and under $N_1$, and it should be possible to compute more generally
$\langle \psi_{d_1}^{\Xi} \psi_{d_2}^{\Xi}\rangle$ in the same way by induction in $d_1,d_2$). This scheme
 may work, at least for the three-point functions,  but it looks like a difficult task in general, involving
a precise analysis of the singularities at $\Xi=0$ of differential operators with regular singularities.

In the case of the polynomial fields $\Phi_{j,k}$, one obtains (up to an irrelevant function of ${\cal M}$) the heat kernel in any even dimension (this is impossible for odd dimensions because the heat
kernel then involves a square root of $t_1-t_2$ and one should use instead non-local  conformal fields in the
first place instead of the bosons). In the case of the generalized polynomial fields $\Phia_{j,k}$, 
the two-point function is non-standard, which is not surprising since the $\Phia_{j,k}$ are themselves
non-scalar. The exact form is new and involves a Bessel function. There are (to the best of our knowledge)
no known examples at the moment of a physical model with a two-point function of this form.

{\bf Theorem 5.1} (polynomial fields $\Phi_{j,k}$) 

{\em
\begin{enumerate}
\item
Let $d=-1,0,1,\ldots$ and $\Xi>0$. Set 
\BEQ \phi_{d}^{\Xi}:=
\sum_{j=0}^{\infty} \frac{\II^{j+d}} {\Xi^{\frac{j+1}{2}}}
\frac{\sqrt{j!}}{(j+d+1)!}  \Phi_{j+d+1,d+1}^{(0)}.
\EEQ
Then the inverse Laplace transform of the two-point function
$${\cal C}^{\Xi}(t,r,\zeta)=\langle 0\ |\ \phi_d^{\Xi}(t_1,r_1,\zeta_1)
\phi_d^{\Xi}(t_2,r_2,\zeta_2)
\ |\  0 \rangle,$$ defined a priori for $\Xi \gg 1$, may be analytically
extended to the following function:
\BEQ
({\cal L}^{-1}{\cal C}^{\Xi})({\cal M};t,r)= e^{{\cal M} \Xi t} t^{-2d-1} e^{-{\cal M}\frac{r^2}{2t}}.
\EEQ
When $\Xi\to 0$, this goes to  the standard heat kernel $K_{4d+2}(t,r)=t^{-2d-1}e^{-{\cal M}\frac{r^2}{2t}}.$
\item
Let $d=0,1,\ldots$ and $\Xi>0$.  Set
\BEQ \tilde{\phi}_d^{\Xi}:=
\sum_{j=1}^{\infty} \frac{\II^{j+d}} {\Xi^{\frac{j+1}{2}}}
\frac{\sqrt{j!}}{(j+d)!}  \Phi_{j+d,d+1}^{(0)}.
\EEQ
Then the inverse Laplace transform of the two-point function $\langle
\tilde{\phi}_d^{\Xi}\tilde{\phi}_d^{\Xi}\rangle$ 
may be analytically extended into the function  $
{\cal M} e^{{\cal M} \Xi t} t^{-2d} e^{-{\cal M}\frac{r^2}{2t}}.$
When $\Xi\to 0$, this goes to ${\cal M}$ times  the standard heat kernel $K_{4d}(t,r)=t^{-2d}
e^{-{\cal M}\frac{r^2}{2t}}.$

\item
(same hypotheses) Set
\BEQ {\psi}_{2d}^{\Xi}:=
\sum_{j=0}^{\infty} \II^j \frac{\Xi^{-j-d-\frac{3}{2}}}{j!}  \Phi_{2j+2d+1,2d+1}^{(0)}.
\EEQ
Then  the two-point function
 $\langle{\psi}_d^{\Xi}{\psi}_d^{\Xi}\rangle$ has an analytic continuation to
small $\Xi$. The inverse Laplace transform of its value for $\Xi=0$ is
equal (up to a constant) to ${\cal M}^{2d+2} K_{4d-2}(t,r)$.

\item
(same hypotheses) Set
\BEQ \tilde{\psi}_{2d}^{\Xi}:=
\sum_{j=0}^{\infty} \II^j  \frac{\Xi^{-j-d-\half}}{j!}  \Phi_{2j+2d,2d+1}^{(0)}.
\EEQ
Then  the two-point function
 $\langle\tilde{\psi}_d^{\Xi}\tilde{\psi}_d^{\Xi}\rangle$ has an analytic continuation to
small $\Xi$. The inverse Laplace transform of its value for $\Xi=0$ is
equal (up to a constant) to ${\cal M}^{2d+1} K_{4d}(t,r)$.

\end{enumerate}

}

{\bf Remark.} One may also define 
\BEQ \psi_d^{\Xi}:= \sum_{j=0}^{\infty} \II^j \frac{\Xi^{-j-\frac{d}{2}-1}}{j!} \Phi^{(0)}_{2j+d+1,d+1} \EEQ
for $d$ odd, but similar computations (using a different connection formula for the hypergeometric
function though, see proof below) show that its two-point function is equal (up to a constant) to that of
$\psi^{\Xi}_{d+1}$, i.e. (up to a polynomial in ${\cal M}$) to $K_{2d}$. (Note however the
strange-looking but necessary shift by $\half$ in the powers of $\Xi$ in the expression of the $\psi_{d}^{\Xi}$
with 
odd index $d$  with respect to those with an even index).  Hence the need for $\tilde{\psi}_{2d}^{\Xi}.$

{\bf Proof.}

Note first quite generally that, if $K_d({\cal M};t,r):=\frac{e^{-{\cal M}\frac{r^2}{2t}}}{t^{d/2}}$ is
the standard heat kernel in $d$ dimensions, then
\BEQ {\cal L}({\cal M}^n K_d({\cal M};t,r))=\partial_{\zeta}^n \left(t^{-d/2} \left(\frac{r^2}{2t}-\zeta\right)^{-1}\right)=(-1)^{n+1} n! t^{-d/2} \left(\zeta-\frac{r^2}{2t}\right)^{-n-1}.\EEQ

We shall use the notation $\xi:=\zeta-r^2/2t$ in the proof.

\begin{enumerate}
\item
The Laplace transform of the function
$g^{(d)}_{\Xi}({\cal M};t,r):=
 e^{{\cal M}\Xi t} t^{-2d-1} e^{-{\cal M}\frac{r^2}{2t}}$ is equal to
$$({\cal L}g^{(d)})(t,r,\zeta)=-t^{-2d-1} \frac{1}{\Xi t+(\zeta-\frac{r^2}{2t})}=-
\sum_{j=0}^{\infty} (-1)^j \Xi^{-j-1} t^{-2(d+1)-j}
(\zeta-\frac{r^2}{2t})^{j}$$
(provided that the series converges, or taken in a formal sense). Then
Proposition 4.1 shows that the two-point function of the field $\phi_d^{\Xi}$ defined above
is equal to this series.
\item
Set $\tilde{g}^{(d)}_{\Xi}({\cal M};t,r):={\cal M}e^{{\cal M}\Xi t} t^{-2d} e^{-{\cal M} \frac{r^2}{2t}}:$
then $$({\cal L}\tilde{g}_{\Xi}^{(d)})(t,r,\zeta)=\partial_{\zeta} ({\cal L}g_{\Xi}^{(d-\half)})(t,r,\zeta)=-
\sum_{j=1}^{\infty} j(-1)^j t^{-2d-j-1} (\zeta-\frac{r^2}{2t})^{j-1}$$ is easily checked to be equal to the two-point function $\langle \tilde{\phi}_d^{\Xi}  \tilde{\phi}_d^{\Xi} \rangle$. 

\item
Set 
$$I^{\Xi}:=\Xi^{2d+3} \partial_t^{-(2d+1)} (t^{2d} \langle\psi_d^{\Xi} \psi_d^{\Xi}\rangle)$$ where
$\partial_t^{-1}=\int_0^t dt$ is the integration operator from $0$ to $t$.
Then Proposition 4.1, together with the duplication formula for the Gamma function,  yield
\BEA
I^{\Xi} &=& \sum_{j\ge 0} (2j+2d+1)!\  t^{-2j-1} \frac{(-\xi^2/\Xi^2)^j}{(j!)^2} \nonumber\\
&=& \frac{1}{t} \ .\ \frac{2^{2j+2d+1}}{\sqrt{\pi}} \sum_{j\ge 0}
\frac{\Gamma(j+d+1)\Gamma(j+d+\frac{3}{2})}{\Gamma(j+1)} \frac{ \left(-\left(\frac{\xi}{\Xi t}\right)^2
\right)^j}{j!} \nonumber\\
&=& \frac{1}{t} \frac{2^{2d+1}}{\sqrt{\pi}} \Gamma(d+1)\Gamma(d+\frac{3}{2}) \ 
_2 F_1(d+1,d+\frac{3}{2};1;-\left(\frac{2\xi}{\Xi t}\right)^2)
\EEA
which is defined for $\Xi \gg 1$. The connection formula (see \cite{Abra84}, 15.3.3) for the  Gauss hypergeometric function
$_2 F_1$
\BEQ _2 F_1(a,b,c;z)=(1-z)^{c-a-b} \ _2 F_1(c-a,c-b;c;z) \EEQ  yields
\BEQ
_2 F_1(d+1,d+\frac{3}{2};1;-(\frac{2\xi}{\Xi t})^2) =\left[
1+\left(\frac{2\xi}{\Xi t}\right)^2\right]^{-2d-\frac{3}{2}}  \ 
_2 F_1(-d,-d-\half;1;-\left( \frac{2\xi}{\Xi t}\right)^2).
\EEQ
The hypergeometric function on the preceding line is simply a polynomial
in $\Xi^{-1}$ since $-d$ is a negative integer. By extracting the most singular
term  in $\Xi^{-1}$, one sees that
$$I^{\Xi}\sim_{\Xi\to 0} (-1)^d \frac{d!}{4\pi} \Gamma(d+\frac{3}{2})^2   \Xi^{2d+3}
\xi^{-2d-3} t^{2d+2}.$$ 
Hence $$\langle \psi_d^{\Xi}\psi_d^{\Xi}\rangle\to_{\Xi\to 0} (-1)^d
\frac{(2d+2)! d!}{4\pi} \Gamma(d+\frac{3}{2})^2 \xi^{-2d-3} t^{1-2d}=(-1)^{d+1}
\frac{d!}{4\pi} \Gamma(d+\frac{3}{2})^2 {\cal L}({\cal M}^{2d+2} K_{4d-2}({\cal M};t,r)).$$
\item
Same method. 

\end{enumerate}
 \hfill \eop

{\bf Theorem 5.2} (generalized polynomial fields $\Phia_{j,k}$)

{\em
Let $\alpha\in\R$ and  $\Xi>0$.
\begin{enumerate}
\item

 Set 
\BEQ \phia^{\Xi}:=
\sum_{j=0}^{\infty} \frac{\II^{j}} {\Xi^{\frac{j+1}{2}}}
\frac{1}{\sqrt{ j!}}  \Phia_{j,0}^{(0)}.
\EEQ 

Then the  two-point function
$${\cal C}^{\Xi}(t,r,\zeta)=\langle 0\ |\ \phia^{\Xi}(t_1,r_1,\zeta_1)\phima^{\Xi}(t_2,r_2,\zeta_2)
\ |\  0 \rangle$$ has an analytic continuation to small $\Xi$, and its inverse Laplace
transform at $\Xi=0$  is equal to

\BEQ
\widetilde{\cal C}_{\cal M}(t,r)= -t^{1-\alpha^2} 
 e^{-{\cal M}r^2/2t}
I_0(2|\alpha|\sqrt{{\cal M}r^2/t})
\EEQ
where $I_0$ is the modified Bessel function of order 0.

\item

Set
\BEQ \psia^{\Xi}:=
\sum_{j=0}^{\infty}   (-1)^j 
\frac{\Xi^{-j-\half}}{ j!}  \Phia_{2j,0}^{(0)}.
\EEQ 
Then the same results hold for the two-point function $\langle\psia^{\Xi}\psima^{\Xi}\rangle$
(up to an overall multiplicative constant).
\end{enumerate}
}

{\bf Remark:} if one replaces $\alpha$ with $\II \alpha$, then the two-point function involves this
time the Bessel function $J_0$.

{\bf Proof.}  

\begin{enumerate}

\item

By applying Proposition 4.2, one gets
\BEA
{\cal C}(t,r,\zeta)&=& \sum_{j\ge 0} \frac{(-1)^j}{\Xi^{j+1}} 
t^{-j-\alpha^2} \sum_{\del=0}^j (-1)^{\del} \frac{j!}{2^{\del}(j-\del)!}
\frac{(2\alpha^2)^{\del}}{(\del!)^2} \left( \frac{r^2}{t}\right)^{\del}
(\zeta-\frac{r^2}{2t})^{j-\del}  \nonumber\\
&=& \frac{t^{-\alpha^2}}{\Xi} \sum_{n=0}^{\infty} \left(- \frac{\zeta-
r^2/2t}{\Xi t}
\right)^n \sum_{\del=0}^{\infty} \frac{(n+\del)!}{n! (\del!)^2}
\left( \frac{\alpha^2 r^2}{\Xi t^2}\right)^{\del}.
\EEA

The function
$$f(y)=\sum_{\del=0}^{\infty} \frac{\left(\begin{array}{c} n+\del\\
n \end{array}\right)}{\del!} y^{\del}=\sum_{\del=0}^{\infty} \frac{
(n+\del)!}{n! (\del!)^2} y^{\del}$$
is entire and admits a Laplace transform
$$h(\lambda)={\cal L}f(\lambda)=\int_0^{\infty} f(y)e^{-\lambda y}\ dy
=\sum_{\del=0}^{\infty} \left(\begin{array}{c} n+\del\\ n\end{array}\right)
\lambda^{-\del-1}=\frac{1}{\lambda} (1-1/\lambda)^{-n-1}=\lambda^{n}
(\lambda-1)^{-n-1}$$
which is given by a converging series for $\lambda>1$; by inverting the
Laplace transform, one gets
$$f(y)=\partial_y^{n} \left( \frac{(-1)^{n} y^{n}}{n!} e^y\right).$$
An application of Leibniz formula gives $$\partial_y^n (\frac{y^n}{n!} e^y)=
(\sum_{k=0}^n \left(\begin{array}{c} n\\ k\end{array}\right) \frac{y^k}{k!}
)e^y.$$
By putting everything together and setting $y=\frac{\alpha^2 r^2}{\Xi t^2}$,
one gets
\BEA {\cal C}(t,r,\zeta)&=& \frac{t^{-\alpha^2}}{\Xi} \sum_{n=0}^{\infty}
\left( \frac{\zeta-r^2/2t}{\Xi t}\right)^n \left( \sum_{k=0}^n
\frac{\left(\begin{array}{c} n\\ k\end{array}\right)}{k!} \left(
\frac{\alpha^2 r^2}{\Xi t^2}\right)^k \right) e^{\frac{\alpha^2 r^2}
{\Xi t^2}} \nonumber\\
&=&  \frac{t^{-\alpha^2}}{\Xi} \sum_{k=0}^{\infty} \frac{1}{k!} \left[
\sum_{n=k}^{\infty} \left(\begin{array}{c} n\\ k\end{array}\right)
 \left(  \frac{\zeta-r^2/2t}{\Xi t}\right)^n
 \right]
\left(
\frac{\alpha^2 r^2}{\Xi t^2}\right)^k  e^{\frac{\alpha^2 r^2}
{\Xi t^2}}.
\EEA
By comparing with the generating series
$$\sum_{k=0}^{\infty} \left( \sum_{n=k}^{\infty}  \left(\begin{array}{c} n\\ k\end{array}\right)
a^n \right) x^k= \sum_{n=0}^{\infty} a^n \sum_{k=0}^n \left(\begin{array}{c} n\\ k\end{array}
\right) x^k=\sum_{n=0}^{\infty} \left[ a(1+x)\right]^n=\frac{1}{1-a}
\sum_{k=0}^{\infty} \left(\frac{a}{1-a}\right)^k x^k,$$
one gets
\BEQ
{\cal C}(t,r,\zeta) =- \frac{ t^{1-\alpha^2}}{\zeta-r^2/2t-\Xi t}
\exp\left[- \alpha^2 \frac{r^2}{t} \left( \frac{1}{\zeta-r^2/2t-\Xi t}\right)
\right].
\EEQ
One finds in \cite{Erd54} 
${\cal L}^{-1} (\lambda^{-1} e^{a/\lambda})(t)=I_0(2\sqrt{at})$, $a>0$ (mind our unusual
convention for the Laplace transform with respect to the mass ${\cal M}$!), where 
 $I_0$ is the  modified Bessel function of order 0.
Hence $${\cal C}_{\cal M}(t,r)=-t^{1-\alpha^2} e^{{\cal M}\Xi t} e^{-{\cal M}r^2/2t}  I_0(2|\alpha| \sqrt{{\cal M}
r^2/t}).$$
\item

The method is the same but computations are considerably more involved. Set  $y:=
\frac{\alpha^2 r^2}{\Xi t^2}$ and $x=-\frac{4\xi}{\Xi t}$.  Applying
Proposition 4.2 yields this time
$${\cal C}(t,r,\zeta)
= \frac{t^{-\alpha^2}}{\Xi}\sum_{n=0}^{\infty} \left(- \frac{\zeta-
r^2/2t}{\Xi t}
\right)^n \sum_{\del\ge 0, \del+n\equiv  0 [2]} (-1)^{\frac{n+\del}{2}} \frac{((n+\del)!)^2}
{n! [ (\half(n+\del))!]^2  (\del!)^2}
\left( \frac{\alpha^2 r^2}{\Xi t^2}\right)^{\del}.$$

Let $h(\lambda)$ be the Laplace transform of ${\cal C}$ with respect to $y$. Formally, 
this is equivalent to replacing
$y^{\del}/ \del !$ by $\lambda^{-\del-1}$. Separating the cases $n,\del$
even, resp. odd, and using the duplication formula for the Gamma function, one
gets
\BEA
h(\lambda)&=& \frac{1}{\Xi\sqrt{\pi}\lambda t^{\alpha^2}}\left[
\sum_{n=0}^{\infty} (-1)^n \frac{x^{2n}}{(2n)!} \sum_{m=0}^{\infty} (-1)^m
\frac{\Gamma(n+m+\half)^2}{\Gamma(m+\half) m!} (\frac{2}{\lambda})^{2m}  \right.\nonumber\\ &-& \left.
\sum_{n=0}^{\infty} (-1)^n \frac{x^{2n+1}}{(2n+1)!} \sum_{m=0}^{\infty} (-1)^m
\frac{\Gamma(n+m+\frac{3}{2})^2}{\Gamma(m+1)\Gamma(m+\frac{3}{2})} 
\left( \frac{2}{\lambda}\right)^{2m+1}\right]  \nonumber\\
&=&  \frac{1}{\Xi\sqrt{\pi}\lambda t^{\alpha^2}}\left[
\sum_{n=0}^{\infty} (-1)^n  \frac{x^{2n}}{(2n)!}\  _2 F_1(n+\half,n+\half;\half;-
\frac{4}{\lambda^2}) \frac{\Gamma(n+\half)^2}{\Gamma(\half)}  \right.\nonumber\\ &-& \left.
\sum_{n=0}^{\infty} (-1)^n \frac{x^{2n+1}}{(2n+1)!}\  _2 F_1(n+\frac{3}{2},
n+\frac{3}{2};\frac{3}{2};-
\frac{4}{\lambda^2}) \frac{2}{\lambda}
 \frac{\Gamma(n+\frac{3}{2})^2}{\Gamma(\frac{3}{2})}\right]. \nonumber\\
\EEA

Hence $h(\lambda)=\frac{1}{\Xi\sqrt{\pi}\lambda t^{\alpha^2}} (T_1(\lambda)+T_2
(\lambda))$ where (using once more the duplication formula and connection formulas for the
hypergeometric function)
$$T_1(\lambda)=\sqrt{\pi} (1+\frac{4}{\lambda^2})^{-\half} \sum_{n=0}^{\infty}(-1)^n 
\frac{ (\half)_n}{n!} \left( \frac{x}{2}\right)^{2n}
 (1+\frac{4}{\lambda^2})^{-2n}\  _2 F_1(-n,-n;
\half;-\frac{4}{\lambda^2})$$ and
$$T_2(\lambda)=-\sqrt{\pi} \frac{x}{\lambda} (1+\frac{4}{\lambda^2})^{-\frac{3}{2}} 
\ .\ \sum_{n=0}^{\infty} (-1)^n \frac{(\frac{3}{2})_n}{n!}
 ( \frac{x}{2})^{2n}  (1+\frac{4}{\lambda^2})^{-2n} \ 
 _2 F_1(-n,-n;\frac{3}{2};-\frac{4}{\lambda^2})
.$$

These hypergeometric functions are simple polynomials since they
 have negative integer arguments; however, the sum obtained
by expanding these polynomials looks hopelessly intricate. We use instead
the following formula
\BEA
&& \sum_{n=0}^{\infty} \frac{(c)_{2n}}{n! (a)_n} (-v^2)^n\  _2 F_1(-n,1-a-n;b;
u^2)  \nonumber\\ 
&& = 2^{a+b-c-2} u^{1-b}v^{-c} \frac{\Gamma(a)\Gamma(b)}{\Gamma(c)} 
\int_0^{\infty} J_{a-1}(w) J_{b-1}(uw) \exp(-\frac{w}{2v}) w^{c-a-b+1}
dw, \nonumber \\
 \EEA
(see \cite{Hans75}, formula (65.3.11)), valid if $\Re(c)>0,\  \Re(v)>0,\ 
\Re(\frac{1}{2v}\pm iu)>0,\ \Re(a+b+c)>0$.

Hence
\BEQ T_1(\lambda)=-2x^{-1}\sqrt{\pi} (1+\frac{4}{\lambda^2})^{\half}
\int_0^{\infty} dw J_0(w) \cosh(\frac{2w}{\lambda}) \exp\left(
-2wx^{-1}(1+\frac{4}{\lambda^2})\right)\EEQ and
\BEQ T_2(\lambda)=-2x^{-1}\sqrt{\pi} (1+\frac{4}{\lambda^2})^{\half}
\int_0^{\infty} dw J_0(w) \sinh(\frac{2w}{\lambda}) \exp\left(
-2wx^{-1} (1-\frac{4}{\lambda^2})\right).\EEQ

By applying the following formula \cite{Grad80}
$$\int_0^{\infty}\ dw \ e^{-\beta w} J_0(w)=\frac{1}{\sqrt{1+\beta^2}}$$
(Laplace transform of the Bessel function) and expanding the cosinh
and sinh functions into exponentials, one gets
\BEQ T_1=-\sqrt{\pi}x^{-1} (1+\frac{4}{\lambda^2})^{\half}
\left[ \left(1+\left(-\frac{2}{\lambda}+\frac{2}{x} (1+\frac{4}{\lambda^2})\right)^2
\right)^{-\half}+ \left(1+\left(\frac{2}{\lambda}+\frac{2}{x} (1+\frac{4}{\lambda^2})\right)^2
\right)^{-\half} \right] \EEQ
and
\BEQ  T_2=-\sqrt{\pi}x^{-1} (1+\frac{4}{\lambda^2})^{\half}
\left[ \left(1+\left(-\frac{2}{\lambda}+\frac{2}{x} (1+\frac{4}{\lambda^2})\right)^2
\right)^{-\half}  -  \left(1+\left(\frac{2}{\lambda}+\frac{2}{x} (1+\frac{4}{\lambda^2})\right)^2
\right)^{-\half} \right] \EEQ

Using 
\BEQ
 (1+\frac{4}{\lambda^2})^{\half}
\left[ 1+\left(-\frac{2}{\lambda}+\frac{2}{x} (1+\frac{4}{\lambda^2})\right)^2\right]^{-\half}
=\frac{\lambda}{\sqrt{(1+\frac{4}{x^2})\lambda^2-\frac{8}{x}\lambda+\frac{16}
{x^2}}}) 
\EEQ and the inverse Laplace transform
\BEQ
{\cal L}^{-1}\left(\frac{1}{\sqrt{a\lambda^2+2b\lambda+c}}\right)(y)=\frac{1}
{\sqrt{a}}e^{-\frac{b}{a} y} J_0(y\sqrt{\frac{c}{a}-\frac{b^2}{a^2}} )
\EEQ
one gets 
$$h(\lambda)=\frac{-2}{\Xi x t^{\alpha^2}} \frac{1}{\sqrt{(1+\frac{4}{x^2})\lambda^2-\frac{8}{x}
\lambda+\frac{16}{x^2}}}$$
hence
\BEA
C(t,r,\zeta)&=&\frac{-2}{\Xi t^{\alpha^2} \sqrt{\frac{16\xi^2}{\Xi^2 t^2}+4}} 
\exp\left( -\frac{16\xi/\Xi t}{16\xi^2/\Xi^2 t^2+4} \frac{\alpha^2 r^2}{\Xi t^2}\right)
J_0\left( \frac{8\alpha^2 r^2/\Xi t^2}{16\xi^2/\Xi^2 t^2+4}\right) \nonumber\\
&\sim_{\Xi\to 0}& -\frac{1}{2\xi} t^{1-\alpha^2} \exp(-\frac{\alpha^2 r^2}{\xi t}) 
\EEA
which is (up to a constant) exactly the same expression we got for the two-point function
$\langle\phia^{\Xi}\phima^{\Xi}\rangle$.

\end{enumerate}
\hfill\eop

We did not manage to compute explicitly the three-point functions $\langle \psi_{d_1}^{\Xi} \psi_{d_2}^{\Xi} \psi_{d_3}^{\Xi}\rangle$ except in the simplest case $d_1=d_2=d_3=-1$ (see the remark after Theorem 4.1 for the
definition of $\psi_{-1}^{\Xi}$). One obtains:

{\bf Theorem 5.3} 

{\em
Let 
\BEQ {\cal C}^{\Xi}:= \langle \psi_{-1}^{\Xi}(t_1,r_1,\zeta_1)\psi_{-1}^{\Xi}(t_2,r_2,\zeta_2)\psi_{-1}^{\Xi}(t_3,r_3,\zeta_3)\rangle \EEQ
be the three point-function of the massive field
$$\psi_{-1}^{\Xi}:=\sum_{j=0}^{\infty} \II^j \frac{\Xi^{-j-\half}}{j!} \Phi_{2j,0}^{(0)}$$
defined in Theorem 5.1. Then
\BEQ {\cal C}^{\Xi}\to_{\Xi\to 0} C. \left(\frac{t_{12}t_{23}t_{31}}{\xi_{12}\xi_{23}\xi_{31}}\right)^{\half}\EEQ
where $\xi_{ij}:=\zeta_{ij}-\frac{r^2_{ij}}{2t_{ij}}.$

An  inverse Laplace transform of ${\cal C}^{\Xi}$ with respect to the $\zeta$-parameters yields the following
three-point function in terms of the masses :

\BEQ 
\EEQ

}

{\bf Proof.}

Let $x_1=\II \frac{\xi_{12}\xi_{13}t_{23}}{t_{12}t_{13}\xi_{23}\Xi}$, $x_2=\II \frac{\xi_{23}\xi_{21}t_{31}}{t_{23}t_{21}\xi_{31}\Xi}$ and $x_3=\II\frac{\xi_{31}\xi_{32}t_{12}}{t_{31}t_{32}\xi_{12}\Xi}$. Then Proposition 4.3.3 yields
\BEQ {\cal C}^{\Xi}:=\Xi^{-\frac{3}{2}}\sum_{j_1,j_2\ge 0} \frac{x_1^{j_1}x_2^{j_2}}{j_1 ! j_2 !} (2j_1)! (2j_2)! \sum_{|j_1-j_2|\le j_3
\le j_1+j_2} \frac{x_3^{j_3}}{j_3 !} \frac{ (2j_3)!}{(j_1+j_3-j_2)! (j_1+j_2-j_3)! (j_2+j_3-j_1)!}.\EEQ

Write
\BEQ \frac{1}{(j_1+j_2-j_3)!} = (-1)^{j_3-j_1-j_2} \lim_{\eps\to 0} \eps \Gamma(j_3-j_1-j_2+\eps). \EEQ
This form of the complement formula for the Gamma function is valid whatever the argument. Then 
\BEQ {\cal C}^{\Xi}= \Xi^{-\frac{3}{2}} \sum_{j_1,j_2\ge 0} \frac{x_1^{j_1}x_2^{j_2}}{j_1 ! j_2 !} (2j_1)! (2j_2)! \  I_3(j_1,j_2;x_3)\EEQ
where (by using also the duplication formula for the Gamma function)
\BEA
&& I_3(j_1,j_2;x_3)=(-1)^{j_1+j_2} \lim_{\eps\to 0} \eps \sum_{j_3=|j_1-j_2|}^{\infty} (-4x_3)^{j_3} \frac{\Gamma(j_3+\half)}{\sqrt{\pi}} \frac{\Gamma(j_3-j_1-j_2+\eps)}{\Gamma(j_3+(j_1-j_2)+1)\Gamma(j_3+(j_2-j_1)+1)} \nonumber\\
&&= \frac{(-1)^{j_1+j_2}}{\sqrt{\pi}} (-4x_3)^{|j_1-j_2|} \sum_{j=0}^{\infty} \frac{(-4x_3)^j}{j!} \frac{\Gamma(j+|j_1-j_2|+\half)\Gamma(j-2\min(j_1,j_2)+\eps)}{\Gamma(j+2|j_1-j_2|+1)} \nonumber\\
&=& \frac{(-1)^{j_1+j_2}}{\sqrt{\pi}} (-4x_3)^{|j_1-j_2|}  \ _2 \bar{F}_1(|j_1-j_2|+\half,\eps-2\min(j_1,j_2);2|j_1-j_2|+1;-4x_3) \nonumber\\
\EEA
The symbol $_2 \bar{F}_1$ stands for Gauss' hypergeometric function apart from a different normalization, namely,
\BEQ _2 \bar{F}_1(a,b,c;z)=\frac{\Gamma(a)\Gamma(b)}{\Gamma(c)}\ _2 F_1(a,b,c;z)=\sum_{n\ge 0} 
\frac{\Gamma(a+n)\Gamma(b+n)}{\Gamma(c+n)} \frac{z^n}{n!}. \EEQ
 Now the well-known formula connecting
the behaviour around 0 with the behaviour around infinity of the hypergeometric function, see \cite{Abra84} for instance, yields
\BEA
&& _2 \bar{F}_1(|j_1-j_2|+\half,\eps-2\min(j_1,j_2);2|j_1-j_2|+1;-4x_3) \nonumber\\
&& =\Gamma(\eps-j_1-j_2-\half) \left(\frac{1}{4x_3}
\right)^{|j_1-j_2|+\half} \ _2 F_1(|j_1-j_2|+\half,\half-|j_1-j_2|;j_1+j_2+\frac{3}{2};-\frac{1}{4x_3})  \nonumber\\
&&+ \frac{\Gamma(\eps-2\min(j_1,j_2))\Gamma(j_1+j_2+\half)}{\Gamma(2\max(j_1,j_2)+1-\eps)} \left(\frac{1}{4x_3}\right)^{-2\min(j_1,j_2)} \nonumber\\ &&  _2 F_1(-2\min(j_1,j_2),-2\max(j_1,j_2);\half-j_1-j_2;-\frac{1}{4x_3}).
\EEA

In the limit $\eps\to 0$, only the second term in the right-hand side has a pole,
 $\Gamma(\eps-2\min(j_1,j_2))\sim_{\eps\to 0} \frac{1}{\Gamma(1+2\min(j_1,j_2))} \frac{1}{\eps},$ hence

\BEQ
I_3=\frac{(4x_3)^{j_1+j_2}}{\sqrt{\pi}} \frac{\Gamma(j_1+j_2+\half)}{\Gamma(1+2j_1)\Gamma(1+2j_2)} \ _2 F_1(-2j_1,-2j_2;
\half-j_1-j_2;-\frac{1}{4x_3})
\EEQ

and

\BEQ
{\cal C}^{\Xi}=\frac{\Xi^{-\frac{3}{2}}}{\sqrt{\pi}}  \sum_{j_1,j_2\ge 0} \frac{(4x_1 x_3)^{j_1} (4x_2 x_3)^{j_2}}{j_1 ! j_2 !}
\Gamma(j_1+j_2+\half)\ _2 F_1(-2j_1,-2j_2;\half-j_1-j_2;-\frac{1}{4x_3}).\EEQ

Kummer's quadratic transformation formulas for the hypergeometric functions  give (see \cite{Abra84}, 15.3.22)
\BEQ 
 _2 F_1(-2j_1,-2j_2;\half-j_1-j_2;-\frac{1}{4x_3})=\ _2 F_1(-j_1,-j_2;\half-j_1-j_2;1-(1+\frac{1}{2x_3})^2).
\EEQ

Now for any $\beta$
\BEQ
\sum_{j_1,j_2\ge 0} \frac{y_1^{j_1}}{j_1 !} \frac{y_2^{j_2}}{j_2 !} \Gamma(j_1+j_2+\beta)=\sum_{j_1 \ge 0}
\frac{y_1^{j_1}}{j_1 !} \Gamma(j_1+\beta) (1-y_2)^{-j_1-\beta}=\Gamma(\beta) (1-y_1-y_2)^{-\beta}.\EEQ

Applying this formula to each term in the series expansion of the above hypergeometric function yields

\BEA 
{\cal C}^{\Xi}&=&\frac{\Xi^{-\frac{3}{2}}}{\sqrt{\pi}} \sum_{k\ge 0} \frac{ \left( 16x_3^2 x_1 x_2 ((1+\frac{1}{2x_3})^2-1)\right)^k}{k!}
\sum_{l_1,l_2\ge 0} \frac{(4x_1x_3)^{l_1} (4x_2 x_3)^{l_2}}{l_1 ! l_2 !} \Gamma(l_1+l_2+(k+\half)) \nonumber\\
&=& \frac{\Xi^{-\frac{3}{2}}}{\sqrt{\pi}} \sum_{k\ge 0}  \frac{ \left( 16x_3^2 x_1 x_2 ((1+\frac{1}{2x_3})^2-1)\right)^k}{k!}
\Gamma(k+\half) (1-4x_1 x_3-4x_2 x_3)^{-k-\half} \nonumber\\
&=& \Xi^{-\frac{3}{2}} \left(1-4x_3(x_1+x_2)+16x_3^2 x_1 x_2 (1-(1+\frac{1}{2x_3})^2)\right)^{-\half} \nonumber\\
&=& \Xi^{-\frac{3}{2}} \left(1-4x_1 x_2 -4x_1 x_3-4x_2 x_3-16x_1 x_2 x_3\right)^{-\half}
\EEA
hence the limit when $\Xi\to 0$.

Casting this result into the usual coordinates $({\cal M},t,r)$ (i.e. taking an inverse Laplace transform) is a
technical task, although some partial results are available through the usual results in conformal field theory
(see Remark after Theorem A.3 in the Appendix).

\underline{Acknowledgements}

Special thanks go to M. Henkel for patiently re-reading the manuscript and giving a statistical physicist's view
on the subject.


\newpage


\appsection{A}{Two- and three-point functions for general coinduced fields}


Let $\Phi_i$, $i=1,2,\ldots$ be $\tilde{\sch}_1$-quasi-primary fields. The general problem
we address in this Appendix is: what is
the most general $n$-point function $\langle 0\ |\ \Phi_1(t_1,r_1,\zeta_1)
\ldots\Phi_n(t_n,r_n,\zeta_n)\ |\ 0\rangle$ compatible with the constraints
coming from symmetries ? 

It has been solved in general (see \cite{Hen94,HenUnt03,HenUnt06} and \cite{BauStoHen05}, Appendix B) for scalar massive
$\sch_1$-quasi-primary fields, i.e. for fields such that the representation $\rho$ of $\sv_0=\langle
L_0\rangle\ltimes\langle Y_{\half},M_1\rangle$ is one-dimensional, namely $\rho(L_0)=-\lambda$
(where $\lambda$ is the scaling exponent \footnote{Physicists usually call 'scaling exponent' $2\lambda=:x$ instead of $\lambda$. For instance, the Schr\"odinger field defined in the Introduction has scaling exponent $\lambda=\frac{1}{4}$
or $x=\half$  depending on the convention.}   of the field) and $\rho(Y_{\half})=0$. Note 
that in the whole discussion, the value of $\rho(M_1)$ is irrelevant since $M_1$ does not belong
to $\tilde{\sch}_1$. Let us recall the results for two- and three-point functions. In the
following proposition, we also consider the natural extension to scalar $\tilde{\sch}_1$-quasi-primary
fields:

{\bf Definition A.1}

{\em
A  {\em scalar $(\lambda,\lambda')$-quasi-primary field} is a $\rho$-$\tilde{\sch}_1$-quasi-primary
field for which $\rho$ is scalar, with $\rho(L_0)=-\lambda$, $\rho(N_0)=-\lambda'$
and $\rho(Y_{\half})=0$.
}

When speaking of two-point functions, we shall generally use the notation $t=t_1-t_2$, $r=r_1-r_2$,
$\zeta=\zeta_1-\zeta_2$. The notations $u=r^2/2t$, $\xi=\zeta-u=\zeta-r^2/2t$ will also show up
frequently.

Note quite generally that the Bargmann superselection rule (due to the covariance under the phase shift $M_0$), as
mentioned in the Introduction, forbids scalar massive fields $\Phi_1,\ldots,\Phi_n$ with total
mass ${\cal M}={\cal M}_1+\ldots+{\cal M}_n$ different from $0$ to have a non-zero $n$-point function.

{\bf Proposition A.1}
{\em

\begin{itemize}
\item[(i)] Let $\Phi_{1,2}$ be two scalar $\sch_1$-quasi-primary fields with scaling exponents
$\lambda_{1,2}$. Then their two-point function ${\cal C}=\langle \Phi_1(t_1,r_1,\zeta_1)\Phi_2(t_2,r_2,\zeta_2)\rangle$ vanishes except if $\lambda_1=\lambda_2=:\lambda$, in which case it is equal to 
\BEQ {\cal C}=t^{-2\lambda} f(\zeta-\frac{r^2}{2t}) \EEQ
where $f$ is an arbitray scaling function. The inverse Laplace transform with respect to $\zeta$ gives
(up to the multiplication by an arbitrary function of the mass) 
for fields with the same mass ${\cal M}$ a generalized heat kernel,
\BEQ {\cal C}=g({\cal M}) t^{-2\lambda} e^{-{\cal M}\frac{r^2}{2t}}.\EEQ

\item[(ii)] Suppose furthermore that $\Phi_{1,2}$ are $(\lambda_i,\lambda'_i)-\tilde{\sch}_1$-quasi-primary, $i=1,2$, with $\lambda_1=\lambda_2=:\lambda$ (otherwise $\cal C$ vanishes). Then the
two-point function is fixed (up to a constant),
\BEQ {\cal C}=t^{-2\lambda} (\zeta-\frac{r^2}{2t})^{-\frac{\lambda'_1+\lambda'_2}{2}}.\EEQ
The inverse Laplace transform with respect to $\zeta$ of this function yields (up to a constant)
\BEQ {\cal C}={\cal M}^{\frac{\lambda'_1+\lambda'_2}{2}-1} t^{-2\lambda} e^{-{\cal M}\frac{r^2}{2t}}.\EEQ
\item[(iii)]
Let $\Phi_{1,2,3}$ be three scalar $\sch_1$-quasi-primary fields with scaling exponents $\lambda_{1,2,3}$.
Then 
\BEA {\cal C}&=&\langle \Phi_1(t_1,r_1;{\cal M}_1)\Phi_2(t_2,r_2;{\cal M}_2)\Phi_3(t_3,r_3;{\cal M}_3)\rangle \nonumber\\
&=& \del({\cal M}_1+{\cal M}_2+{\cal M}_3) t_{12}^{\lambda_3-\lambda_1-\lambda_2} t_{23}^{\lambda_1-\lambda_2-\lambda_3} t_{31}^{\lambda_2-\lambda_3-\lambda_1} \nonumber\\
&& \exp\left[ -\frac{{\cal M}_1}{2} \frac{r_{13}^2}{t_{13}}-\frac{{\cal M}_2}{2} \frac{r_{23}^2}{t_{23}}
\right] \  F\left( \frac{(r_{13}t_{23}-r_{23}t_{13})^2}{t_{12}t_{23}t_{13}}\right) \nonumber\\
\EEA
where $F$ is an arbitray scaling function.

\end{itemize}

}

Note that the $N_0$-symmetry constraint  is necessary to fix (up to a constant) even the two-point function
in the variables $(t,r,\zeta)$, contrary to the more rigid case of conformal invariance which fixes
two- and three-point functions. That is the reason why we  consider fields that are covariant
under the extended Schr\"odinger or Schr\"odinger-Virasoro algebra.

The non-scalar fields considered below are actually the most general possible for finite-dimensional
representations $\rho$ (see discussion before Definition 1.4), since one does not consider 
$\rho(M_1)$.

{\bf Theorem  A.2.} (two-point functions for non-scalar fields)

{\em
Consider two  $d$-dimensional representations $\rho_i$, $i=1,2$ of
$\wsv_0=(\langle L_0\rangle \oplus\langle N_0\rangle)\ltimes
\langle Y_{\half},M_1\rangle$ indexed by the parameters $\lambda_{1,2},
\lambda'_{1,2},\alpha_{1,2}$  such that
\BEA
\rho_i(L_0)=-\lambda_i\Id+\half\sum_{\mu=0}^{d-1} \mu E_{\mu,\mu},\ \rho_i(N_0)
=-\lambda'_i \Id-\sum_{\mu=0}^{d-1}\mu  E_{\mu,\mu}\\
\rho_i(Y_{\half})=\alpha_i \sum_{\mu=0}^{d-2} E_{\mu,\mu+1}
\EEA
Let $\Phi_i=(\Phi_i^{\mu}(t,r,\zeta))_{\mu=0,\ldots,d-1}$,  be 
${\rho}_i$-quasiprimary fields, $i=1,2$. Then their two-point functions 
${\cal C}^{\mu,\nu}=\langle \Phi_1^{(\mu)}
(t_1,r_1,\zeta_1)\Phi_2^{\nu}(t_2,r_2,\zeta_2)\rangle$ vanish unless $2(\lambda_1-
\lambda_2)$ is an integer. Supposing that $\lambda_1=\lambda_2$, they may be expressed in terms of
 $d$ arbitrary parameters $c_0,\ldots,c_{d-1}$ as follows:
\BEQ
{\cal C}^{\mu,\nu}=t^{-\lambda+\frac{\mu+\nu}{2}} \sum_{\del=\max(\mu,\nu)}^{d-1}
c_{\del} \frac{\alpha_1^{\del-\mu}\alpha_2^{\del-\nu}}{(\del-\mu)!(\del-\nu)!}
(\frac{r^2}{t})^{\del-\frac{\mu+\nu}{2}} (\zeta-\frac{r^2}{2t})^{-(\frac{\lambda'}{2}+\del)} \label{AppAth1}
\EEQ
where $\lambda=\lambda_1+\lambda_2$ ($=2\lambda_1$ here) and $\lambda'=
\lambda'_1+\lambda'_2$.
}

{\bf Remark.} The assumption $\lambda_1=\lambda_2$ is no restriction of
generality: supposing that $\Del:=2(\lambda_1-\lambda_2)$ is  (say) a positive
 integer 
implies a shift in the index $\mu$ with respect to $\nu$ in formula
(\ref{AppAth1}) and restricts the number of unknown constants. By working 
through the proof of this Theorem, it is possible to see that 
the ${\cal C}^{\mu,\nu}$ vanish for $\max(\mu,\nu)>d-1-\Del$ (hence
all of them vanish if $\Del\ge d$) and that the other components depend
on $d-\Del$ coefficients.

{\bf Proof.}

First of all, invariance under translations $\rho(L_{-1})=-\partial_t,
\rho(Y_{-\half})=-\partial_r,\rho(M_0)=-\partial_{\zeta}$
 implies that ${\cal C}^{\mu,\nu}$
is a function of the differences of coordinates $t,r,\zeta$ only. Set
$\lambda=\lambda_1+\lambda_2,\lambda'=\lambda'_1+\lambda'_2$; we do not assume anything
on $\lambda_1-\lambda_2$ for the moment. Let us write the action of $\rho(L_0)
= -t\partial_t-\half r\partial_r+\rho_1(L_0)\otimes\Id+\Id\otimes
\rho_2(L_0)$ on ${\cal C}^{\mu,\nu}$. By definition, one has
\BEQ (t\partial_t+\half r\partial_r){\cal C}^{\mu,\nu}=(-\lambda+\frac{\mu+\nu}{2}){\cal C}^{\mu,\nu} \EEQ
hence 
\BEQ
{\cal C}^{\mu,\nu}=f^{\mu,\nu}(\zeta,u)t^{-\lambda+\frac{\mu+\nu}{2}} \label{AppAeq0}
\EEQ
where $u:=\frac{r^2}{2t}.$ Then invariance under $\rho(Y_{\half})=
-t\partial_r-r\partial_{\zeta}+\rho_1(Y_{\half})\otimes\Id+\Id\otimes
\rho_2(Y_{\half})$ implies
\BEA
(t\partial_r+r\partial_{\zeta}){\cal C}^{\mu,\nu} &=& \rho_1(Y_{\half})^{\mu}_l
{\cal C}^{l,\nu}+\rho_2(Y_{\half})^{\nu}_l {\cal C}^{\mu,l} \\
&=& \alpha_1 {\cal C}^{\mu+1,\nu}+\alpha_2 {\cal C}^{\mu,\nu+1}
\EEA
hence 
\BEQ \sqrt{2u} (\partial_u+\partial_{\zeta})f^{\mu,\nu}(\zeta,u)=
\alpha_1 f^{\mu+1,\nu}(\zeta,u)+\alpha_2 f^{\mu,\nu+1}(\zeta,u) .
\label{AppAeq1} \EEQ

The solutions of the homogeneous equation associated with (\ref{AppAeq1})
are the functions of $\xi:=\zeta-u$. In the new set of coordinates $(\xi,u)$,
equation (\ref{AppAeq1}) reads as
\BEQ \sqrt{2u} \partial_u f^{\mu,\nu}(\xi,u)=\alpha_1 f^{\mu+1,\nu}(\xi,u)
+\alpha_2 f^{\mu,\nu+1}(\xi,u).\label{AppAeq1bis}\EEQ

These coupled equations are easily solved. First, $\partial_u f^{d-1,d-1}(\xi,u)=0$, hence $g^{d-1,d-1}:=f^{d-1,d-1}$ is a function of $\xi$ only. It
is clear by decreasing induction on $\mu$ and $\nu$ that the general solution
may be expressed in terms of $d^2$ undetermined functions $g^{\mu,\nu}(\xi)$,
$0\le \mu,\nu\le d-1$, through the relations
\BEQ f^{\mu,\nu}(\xi,u)=\alpha_1 \int \frac{f^{\mu+1,\nu}(\xi,u)}{\sqrt{2u}}\
du\ +\ \alpha_2 \int  \frac{f^{\mu,\nu+1}(\xi,u)}{\sqrt{2u}}\
du\ +g^{\mu,\nu}(\xi) \label{AppAeq2} \EEQ

Let us now use covariance under $\rho(N_0)\equiv -r\partial_r-2
\zeta\partial_{\zeta}+\rho_1(N_0)\otimes\Id +\Id\otimes \rho_2(N_0)$: one gets
$$2(u\partial_u+\xi\partial_{\xi})f^{\mu,\nu}(\xi,u)=-(\lambda'+\mu+\nu)f^{\mu,\nu}(\xi,u) $$ hence $$f^{\mu,\nu}(\xi,u):=\xi^{-\frac{\lambda'+\mu+\nu}{2}}
f_0^{\mu,\nu}(\frac{\xi}{u});$$ this implies immediately $g^{d-1,d-1}(\xi)=
\xi^{-(\frac{\lambda'}{2}+d-1)}$ up to a multiplicative constant. Then \\
$\int \frac{\xi^{-\frac{\lambda'+\mu+\nu+1}{2}} f_0^{\mu+1,\nu}(\frac{\xi}{u})}
{\sqrt{2u}}\ du$ is homogeneous of degree $-(\lambda'+\mu+\nu)$ with respect
to $2(u\partial_u+\xi\partial_{\xi})$, hence the defining relations
(\ref{AppAeq2}) are compatible with covariance under $\rho(N_0)$,
provided that $g^{\mu,\nu}(\xi)=\xi^{-\frac{\lambda'+\mu+\nu}{2}}$ up
to a constant.

Covariance under $\rho(L_1)= -\sum_{i=1}^2 
(t_i^2 \partial_{t_i}+t_i r_i\partial_{r_i}+\half r_i^2 \partial_{\zeta_i})
+(2t_1\rho_1(L_0)+r_1\rho_1(Y_{\half}))\otimes\Id+\Id\otimes
(2t_2\rho_2(L_0)+r_2\rho_2(Y_{\half}))$ is seen to be equivalent (after
some easy computations) to the coupled equations
\BEQ
(t^2\partial_t+tr\partial_r+\half r^2\partial_{\zeta}){\cal C}^{\mu,\nu}(t,r,\zeta)=2t\rho_1(L_0)^{\mu}_l {\cal C}^{l,\nu}+r\rho_1(Y_{\half})_l^{\mu}
{\cal C}^{l,\nu} 
\EEQ
Using the above Ansatz (\ref{AppAeq0}) yields 
\BEQ \left[ (\frac{\nu-\mu}{2}+\lambda_1-\lambda_2)+u\partial_u\right]
f^{\mu,\nu}(\xi,u)=\alpha_1 \sqrt{2u} f^{\mu+1,\nu}(\xi,u). \label{AppAeq3}
\EEQ

Applying this relation to $f^{d-1,d-1}=c_{d-1}. \xi^{-(\frac{\lambda'}{2}+d-1)}$
gives $c_{d-1}=0$ unless $\lambda_1=\lambda_2$, which we assume from now on.
Let us compute $f^{d-2,d-1}$ and $f^{d-1,d-2}$ before we cope with the
general case; one may  set $c_{d-1}=1$ for the moment.
Then $f^{d-2,d-1}(\xi,u)=(\alpha_1\sqrt{2u}+c\sqrt{\xi})\xi^{-(\frac{\lambda'}{2}+d-1)}$ must satisfy $(\half+u\partial_u)f^{d-2,d-1}(\xi,u)=\alpha_1
\sqrt{2u} \xi^{-(\frac{\lambda'}{2}+d-1)}$. The function $\mu_1 \sqrt{2u}
\xi^{-(\frac{\lambda'}{2}+d-1)}$ is indeed a solution of this equation, and
any other solution will be a linear combination of this with some function
$u^{-\half} h(\xi)$, hence $c=0$ and $f^{d-2,d-1}$ is totally determined
by $f^{d-1,d-1}$.
On the other hand,
$f^{d-1,d-2}(\xi,u)=(\alpha_2 \sqrt{2u}+c\sqrt{\xi})\xi^{-(\frac{\lambda'}{2}
+d-1)}$ must satisfy $(-\half+u\partial_u)f^{d-1,d-2}(\xi,u)=\alpha_1
\sqrt{2u} \xi^{-(\frac{\lambda'}{2}+d-1)}$. The general solution
of this equation is 
\BEQ f^{d-1,d-2}(\xi,u)=\sqrt{2u} h(\xi)+\alpha_1 \sqrt{2u}
(\ln u) \xi^{-(\frac{\lambda'}{2}+d-1)}. 
\EEQ 
 Both Ans\"atze are clearly compatible if and
only if $c=0$.

Let us now prove the general case by decreasing induction
on $\max(\mu,\nu)$. Assume formula (\ref{AppAth1}) of
the Theorem has been proved for $\max(\mu,\nu)>M$. Then 
formula (\ref{AppAeq1bis})  gives $f^{M,M}$
up to an undetermined function  $g^{M,M}(\xi)$ which is proportional
to $\xi^{-\lambda'-M}$ due to covariance with respect to $\rho(N_0)$; it is compatible with
formula (\ref{AppAth1}) and formula (\ref{AppAeq3}). One may now go down
or left along a line or a row: if for instance all $f^{M-i,M}$, $i< I$ have been found to agree with
(\ref{AppAth1}), then formula (\ref{AppAeq1bis}) again gives $f^{M+I,M}$,
in accordance with (\ref{AppAth1}), up to an undetermined function
$g^{M+I,M}(\xi)$. Compatibility with covariance under $\rho(L_1)$ (formula
(\ref{AppAeq3})) gives $(\frac{I}{2}+u\partial_u)g^{M+I,M}(\xi)=0$, hence
$g^{M+I,M}=0$ as soon as $I>0$.   \hfill \eop

\vskip 3cm

Let us now turn to the computation of the general three-point function for {\it scalar} quasi-primary 
fields.

{\bf Theorem A.3}

{\em
Let $\Phi_i$, $i=1,2,3$ be $(\lambda_i,\lambda'_i)$-quasi-primary fields. Then their general three-point
function ${\cal C}(t_i,r_i,\zeta_i)=\langle \Phi_1(t_1,r_1,\zeta_1)\Phi_2(t_2,r_2,\zeta_2)
\Phi_3(t_3,r_3,\zeta_3)\rangle$  may be written as
\BEQ
{\cal C}= t_{12}^{-\lambda_1-\lambda_2+\lambda_3} t_{23}^{-\lambda_2-\lambda_3+\lambda_1}
t_{13}^{-\lambda_1-\lambda_3+\lambda_2} \xi_{12}^{\half(-\lambda'_1-\lambda'_2+\lambda'_3)}
\xi_{23}^{\half(-\lambda'_2-\lambda'_3+\lambda'_1)} \xi_{13}^{\half(-\lambda'_1-\lambda'_3+\lambda'_2)}
\ .\ F(\xi_{12},\xi_{13},\xi_{23})
\EEQ
where $F$ is any function of $\xi_{12}:=\zeta_{12}-r_{12}^2/2t_{12}, \xi_{13}:=\zeta_{13}-r_{13}^2/2t_{13}, \xi_{23}:=\zeta_{23}-r_{23}^2/2t_{23}$ which is homogeneous of degree zero, i.e.
\BEQ (\xi_{12}\partial_{\xi_{12}}+\xi_{13}\partial_{\xi_{13}}+\xi_{23}\partial_{\xi_{23}})F=0.\EEQ
}

{\bf Remark.}

In the case $\lambda'_i=2\lambda_i$, $F$ constant, one retrieves the standard result for the three-point function
in  3d conformal field theory, with a Lorentzian pseudo-distance given (in light-cone coordinates) by
$d^2((t_i,r_i,\zeta_i),(t_j,r_j,\zeta_j))=t_{ij}\zeta_{ij}-r^2_{ij}/2.$ The explicit connection between the $n$-point
functions in the Schr\"odinger/conformal cases has been made in \cite{HenUnt03} and in \cite{HenPicPle04}. In the last
reference, an explicit computation of the three-point function in the  dual {\it mass} coordinates ${\cal M}_i$, $i=1,2,3$  is given -- assuming
covariance under the whole conformal group -- in the case when $\lambda_1=\lambda_2$, ${\cal M}_1={\cal M}_2$, $r_1=r_2$.
The general result is a combination of two  confluent hypergeometric functions. Note that in the present case, $\lambda'_i\not=2\lambda_i$
in general, but this leads simply to a different time-dependent pre-factor. The general conformally invariant solution
in coordinates ${\cal M},t,r$  (after 
removing the restriction on $\lambda_1,{\cal M}_1,r_1$) might be given by a generalized hypergeometric function of
two variables, see \cite{Hen?}. 

{\bf Proof.}

Set $r=r_1-r_3$, $r'=r_2-r_3$ and similarly for $t,t'$ and $\zeta,\zeta'$. The covariance under
the action of $L_0, Y_{\half}, N_0$ and $L_1$ yields respectively
\BEA
(\sum_i t_i \partial_{t_i}+\half\sum_i r_i \partial_{r_i}+\lambda){\cal C}=0,  \label{cov31}\\
(\sum_i t_i \partial_{r_i}+r_i \partial_{\zeta_i}){\cal C}=0 \label{cov32},\\
(\sum_i r_i \partial_{r_i}+2\zeta_i \partial_{\zeta_i}+\lambda'){\cal C}=0, \label{cov33}\\
(\sum_i t_i^2 \partial_{t_i}+t_i r_i \partial_{r_i}+\half r_i^2 \partial_{\zeta_i}+2\lambda_i t_i){\cal C}=0  \label{cov34}
\EEA
with $\rho_i(L_0)=-\lambda_i, \rho_i(N_0)=-\lambda'_i$, and $\lambda:=\sum_i \lambda_i, \lambda'=\sum_i
\lambda'_i.$ The function
\BEQ {\cal C}= t_{12}^{-\lambda_1-\lambda_2+\lambda_3} t_{23}^{-\lambda_2-\lambda_3+\lambda_1}
t_{13}^{-\lambda_1-\lambda_3+\lambda_2} \xi_{12}^{\half(-\lambda'_1-\lambda'_2+\lambda'_3)}
\xi_{23}^{\half(-\lambda'_2-\lambda'_3+\lambda'_1)} \xi_{13}^{\half(-\lambda'_1-\lambda'_3+\lambda'_2)}
\EEQ
is a particular solution of this system of equations. Hence the general solution is given by
${\cal C}_0(t_i,r_i,\zeta_i){\cal C}(t_i,r_i,\zeta_i)$, where ${\cal C}_0$ is any solution of the homogeneous
system obtained by setting $\lambda_i,\lambda'_i=0$. By taking an appropriate linear combination
of (\ref{cov31}) and (\ref{cov33}), one gets
\BEQ (E+E'){\cal C}_0=0,\EEQ
where $E=\sum_i t_i\partial_{t_i}+r_i\partial_{r_i}+\zeta_i\partial_{\zeta_i}$ is the Euler operator
in the variables $t_i,r_i,\zeta_i$ and similarly for $E'$. An appropriate linear combination of
(\ref{cov31}), (\ref{cov32}) and (\ref{cov34}) gives
\BEQ (t^2 \partial_t+t'^2\partial_{t'}+tr\partial_r+t' r'\partial_{r'}+\half r^2 \partial_{\zeta}+
\half r'^2 \partial_{\zeta'}){\cal C}_0=0 \label{cov35} \EEQ

Equation (\ref{cov32}) is equivalent to saying that
\BEQ {\cal C}_0:={\cal C}_1(t,t',\rho,\xi,\xi') \EEQ
where 
\BEQ \rho=\frac{r}{t}-\frac{r'}{t'}, \xi=\zeta-r^2/2t, \xi'=\zeta'-r'^2/2t'. \EEQ
Equation (\ref{cov31}) may then be rewritten
\BEQ (t\partial_t+t'\partial_{t'}-\half \rho\partial_{\rho}){\cal C}_1=0 \EEQ
hence 
\BEQ {\cal C}_1:={\cal C}_2(\tau,\tau',\xi,\xi')\EEQ
where \BEQ \tau=\rho^2 t,\tau'=\rho^2 t'.\EEQ
Equation (\ref{cov35}) reads  now simply
\BEQ (\tau^2 \partial_{\tau}+\tau'^2 \partial_{\tau'}){\cal C}_2=0,\EEQ
with general solution
\BEQ {\cal C}_2:={\cal C}_3(\frac{1}{\tau}-\frac{1}{\tau'},\xi,\xi').\EEQ

Set $\xi_{ij}:=\zeta_{ij}-r_{ij}^2/2t_{ij}.$ Then
\BEQ \half (\frac{1}{\tau}-\frac{1}{\tau'})=\xi_{12}+\xi_{23}+\xi_{31} \EEQ

Hence the final result by taking into account the equation $(E+E'){\cal C}_0=0.$ \hfill \eop

\end{document}